\documentclass[
12pt, 
english, 
onehalfspacing, 
headsepline, 
]{MastersDoctoralThesis} 

\usepackage[utf8]{inputenc} 
\usepackage[T1]{fontenc} 

\usepackage[backend=bibtex,style=alphabetic,natbib=true,firstinits=true,maxcitenames=5,maxbibnames=9,doi=false,url=false]{biblatex} 

\usepackage[autostyle=true]{csquotes} 
\usepackage{amsthm}
\usepackage{amssymb}
\usepackage{amsmath}
\usepackage{mathrsfs}
\usepackage{physics}
\usepackage{mathtools}
\usepackage{array}
\usepackage{booktabs}
\usepackage{enumitem}
\usepackage{MnSymbol}
\usepackage{tablefootnote}
\usepackage{setspace}
\usepackage{longtable}
\usepackage{needspace}
\usepackage{makeidx}

\newcommand{\ZZ}{{\mathbb Z}}

\theoremstyle{plain}
\newtheorem{theorem}{Theorem}[chapter]

\newtheorem{lemma}[theorem]{Lemma}
\newtheorem{corollary}[theorem]{Corollary}
\newtheorem{conjecture}[theorem]{Conjecture}

\theoremstyle{definition}

\theoremstyle{remark}
\newtheorem*{remark}{Remark}

\allowdisplaybreaks

\newbibmacro{string+doiurlisbn}[1]{%
  \iffieldundef{doi}{%
    \iffieldundef{url}{%
      \iffieldundef{isbn}{%
        \iffieldundef{issn}{%
          #1%
        }{%
          \href{http://books.google.com/books?vid=ISSN\thefield{issn}}{#1}%
        }%
      }{%
        \href{http://books.google.com/books?vid=ISBN\thefield{isbn}}{#1}%
      }%
    }{%
      \href{\thefield{url}}{#1}%
    }%
  }{%
    \href{http://dx.doi.org/\thefield{doi}}{#1}%
  }%
}

\DeclareFieldFormat{title}{\usebibmacro{string+doiurlisbn}{\mkbibemph{#1}}}
\DeclareFieldFormat[article,incollection]{title}%
    {\usebibmacro{string+doiurlisbn}{\mkbibquote{#1}}}

\thesistitle{Coupled Free Fermion Conformal Field Theory and Representations} 
\supervisor{Prof. Peter \textsc{Bouwknegt}} 
\examiner{} 
\degree{Doctor of Philosophy} 
\author{Bolin \textsc{Han}} 
\addresses{} 

\subject{Mathematical Physics} 
\university{\href{https://www.anu.edu.au}{The Australian National University}} 
\department{\href{https://science.anu.edu.au/}{Joint College of Science}} 
\group{\href{https://maths.anu.edu.au}{Mathematical Sciences Institute}} 

\AtBeginDocument{
\hypersetup{pdftitle=\ttitle} 
\hypersetup{pdfauthor=\authorname} 
\hypersetup{pdfkeywords=\keywordnames} 
}
\makeindex

\begin{document}

\frontmatter 

\pagestyle{plain} 


\begin{titlepage}
\begin{center}

\vspace*{.06\textheight}
{\scshape\LARGE \univname\par}\vspace{1.5cm} 
\textsc{\Large Doctoral Thesis}\\[0.5cm] 

\HRule \\[0.4cm] 
{\huge \bfseries \ttitle\par}\vspace{0.4cm} 
\HRule \\[1.5cm] 
 
\begin{minipage}[t]{0.4\textwidth}
\begin{flushleft} \large
\emph{Author:}\\
{\authorname} 
\end{flushleft}
\end{minipage}
\begin{minipage}[t]{0.4\textwidth}
\begin{flushright} \large
\emph{Supervisor:} \\
{\supname} 
\end{flushright}
\end{minipage}\\[3cm]
 
\vfill

\large \textit{A thesis submitted in fulfillment of the requirements\\ for the degree of \degreename}\\[0.3cm] 
\textit{in the}\\[0.4cm]
\groupname\\\deptname\\[2cm] 
 
\vfill

{\large \today}\\[4cm] 

\vfill
\end{center}
\end{titlepage}


\begin{declaration}
\noindent I, \authorname, declare that this thesis titled, \enquote{\ttitle} and the work presented in it are my own. I confirm that:

\begin{itemize} 
\item This work was done wholly or mainly while in candidature for a research degree at this University.
\item Where any part of this thesis has previously been submitted for a degree or any other qualification at this University or any other institution, this has been clearly stated.
\item Where I have consulted the published work of others, this is always clearly attributed.
\item Where I have quoted from the work of others, the source is always given. With the exception of such quotations, this thesis is entirely my own work.
\item I have acknowledged all main sources of help.
\item Where the thesis is based on work done by myself jointly with others, I have made clear exactly what was done by others and what I have contributed myself.\\
\end{itemize}
 
\noindent Signed: \includegraphics[scale=0.2,trim=0 2cm 0 0]{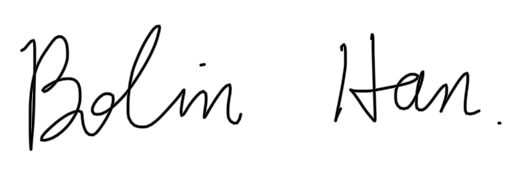}\\
\rule[0.5em]{25em}{0.5pt} 
 
\noindent Date: \today\\
\rule[0.5em]{25em}{0.5pt} 
\end{declaration}

\cleardoublepage






\begin{abstract}
Conformal field theory (CFT) has become an active area of research beyond its origins in statistical physics and attracted much attention due to its intrinsic mathematical interest, which reveals deep connections with other diverse branches of mathematics. We study a specific subclass of CFTs that involve either uncoupled or coupled free fermions. Coupled free fermion CFTs arise from parafermion CFTs and lattice constructions. We analyse their representation spaces and reveal the exclusion statistics of coupled free fermions with universal chiral partition functions under specific bases. We explicitly decompose their modules into Virasoro modules of minimal models. We reveal an unexpected connection that integrates the coset construction, lattice construction, and orbifold construction, which is supported by proving a range of character identities within the context of coupled free fermions. Simultaneously, we obtain explicit expressions of certain string functions in terms of Dedekind eta functions.
\end{abstract}


\begin{acknowledgements}
I would like to express my heartfelt gratitude to the people who have provided invaluable support and guidance throughout my academic journey and the completion of this thesis.

First and foremost, I would like to thank my supervisor Peter Bouwknegt, for his exceptional guidance, patience, and encouragement throughout my research. His insightful feedback and constructive criticism have been instrumental in shaping the direction and quality of this thesis.  I am incredibly fortunate to have had him as my supervisor, as he has always been there for me and pushed me to achieve my best. The meetings and conferences that he encouraged and supported me to attend have broadened my horizons and changed my attitude towards academia. 

I am also grateful to the faculty members and colleagues in Mathematical Science Institute, for their invaluable lectures, seminars, and discussions that have expanded my knowledge and understanding of the subject matter.

Furthermore, I would like to extend my sincere thanks to the university and Tim \& Margaret Bourke, for their generous financial support that has enabled me to pursue my academic goals.

I am grateful to the participants of my study, for their willingness to share their valuable time and insights that have contributed significantly to the findings of this research. I would like to give special thanks to Shane Chern who led the project to a level that I could not even have imagined with his exceptional skills and talents, David Ridout who had discussions with me on this project and gave valuable insights, Yiming Xu who provided crucial suggestions on some of the proofs in the thesis, and Xilin Lu and Chris Raymond who kindly offer to proofread the thesis. 

My thanks also extend to my colleagues, friends, and family, for their support, encouragement, and understanding throughout this journey. I sincerely thanks Shiqiu for his accompany during all the hard time and my dear sister-like roommate for made sure I was fed and motivated me to enjoy life. I am also grateful for Micheal, Emmanouil and Daniel - the friends I made in my very first conferences -  who welcomed me so warmly and helped me fit in with the group.  

Thank you all for your unwavering support, guidance, and inspiration.
\end{acknowledgements}


\tableofcontents 


\listoftables 

\dedicatory{To everyone who sees the world as it is and loves it} 


\mainmatter 

\pagestyle{thesis} 



\chapter{Introduction} 

\label{Chapter1} 


\section{Background}

The concept of scale invariance is a central aspect of critical phenomena in statistical physics. Specifically, it has been established that all ferromagnet-paramagnet critical points are described by scale invariant theories, such as the two-dimensional Ising model \parencite{Ons1944Ising}. This perspective has been incorporated into the renormalization group (RG) theory of phase transitions introduced in \parencite{Wil1983RG}, which asserts that the fixed points of RG flows describe continuous phase transitions, and are therefore scale invariant.

An extension of scale invariance is conformal invariance, which is the invariance under conformal transformations, that is, transformations preserving angles. It was conjectured in \parencite{Polyakov:1970xd} that scale invariant theories describing critical points possess full conformal invariance, and since then, several arguments have been given for the conditions of scale invariance implying conformal invariance. It has been gradually understood that most physically relevant scale invariant theories are in fact conformally invariant \parencite{henkel1999conformal,NAKAYAMA2015ScaleConformal}, and hence referred to the so-called ``Conformal Field Theories" (CFTs). \index{Conformal Field Theory (CFT)}

The study of formal properties of CFTs was initiated in the late 1960s \parencite{MACK1969174,polyakov1974nonhamiltonian,Ferrara:1971vh,Parisi:1972zm,Caianiello1978NonPerturbativeMI}. Early work was done in general dimension $d\ge 2$, where the Lie algebra of the group of conformal transformations is almost always finite-dimensional, except when $d=2$ where the group is infinite-dimensional. The infinite-dimensional symmetry group in 2D CFTs allows us to define the theories in a rather abstract way via operator algebras and their representation theory. 

Operator product expansion (OPE)\index{operator product expansion (OPE)} is a powerful concept going back to \parencite{Wil1969OPE,Kad1969OPE} which defines the product of a set of fields as a sum over the same set of fields. The conformal bootstrap approach, introduced in \parencite{MIGDAL1971386,FERRARA1973161}, combines conformal invariance with the existence of the OPE and leads to consistency conditions on the CFT parameters \parencite{Rattazzi_2008,Poland2019}. Following this approach, the exact solutions to the 2D minimal models, an infinite sequence of CFTs describing the critical points of lattice models such as the 2D Ising model and the 3-state Potts model \parencite{Baxter:1982zz}, were recovered using the constraints of conformal symmetry alone, with minimal or no microscopic input \parencite{BELAVIN1984333}.

With these developments, CFT has become an active area of research beyond its origins in statistical physics. In the 1980s, it was realised that CFTs provide a natural non-perturbative framework for studying string theory \parencite{FRIEDAN198693}, which is a theoretical framework for describing the behaviour of strings - one-dimensional objects moving in a higher-dimensional space \parencite{becker2006string}. In particular, the classical solutions to the spacetime equations of string theory correspond to CFTs \parencite{verlinde1988conformal,MS1989CFT}. Another important connection between CFT and string theory is the AdS/CFT correspondence, which is a duality between the large $N$ limit of $SU(N)$ superconformal field theories and string theory in a higher dimensional anti-de Sitter spacetime \parencite{Maldacena_1999,witten1998anti}. The discovery of string duality, which relates apparently different string theories, led to a better understanding of the underlying CFTs.

CFT has also attracted much attention for its intrinsic mathematical interest, which reveals deep connections with other diverse branches of mathematics including but not limited to non-commutative and/or non-associative algebras \parencite{kac1998vertex}, group theory \parencite{gannon_2006,Bae2021}, number theory \parencite{andrews1984hecke}, category theory \parencite{huang2013tensor}, differential geometry \parencite{CHE2022}, and harmonic analysis \parencite{Benjamin_2021}.

One of the most influential examples is the discovery of vertex operator algebras (VOAs), which were constructed to confirm the Monstrous Moonshine conjecture - an unexpected connection between the monster group and modular functions - and soon realised as a mathematical formulation of CFT \parencite{lepowsky1983vertex,Bor1986VOA,frenkel1989vertex,gannon_2006}. Subsequent work on VOAs, such as the geometric interpretation by \parencite{Huang1991} and Zhu's algebra by \parencite{zhu1996modular}, has deepened our understanding of the connection between VOAs and CFTs. More recent developments on VOA include framed VOAs \parencite{DongFVOA}, generalised VOAs \parencite{dong2012generalized}, and tensor categories \parencite{huang2013tensor}.

Besides the VOA formulation, other mathematical formulations of CFT were also proposed. One of them went back to the 1950s when the Wightman axioms for (QFT and) CFT were proposed in \parencite{Wightman1956QuantumFT}. Modern frameworks include the conformal net approach that axiomatises algebras of local observables \parencite{Kawahigashi_2001} and Segal's CFT which is a functorial field theory \parencite{Segal1988}. The relationship between the formulations is investigated with various techniques \parencite{Carpi_2018,tener2019geometric,Tener_2019,Raymond_2022}.

The quest to classify and solve CFTs (and super CFTs \parencite{NAHM1978149,GGRS2001SCFT}) is a major goal of modern theoretical physics. A complete understanding of the space of CFTs is not yet available. However, a particular subclass called rational CFTs - theories whose quantum state spaces are completely reducible into finitely many
irreducible representations of the infinitesimal symmetry algebra - is believed to be classifiable and investigated intensively. By contrast, the logarithmic CFTs, whose quantum state spaces are not completely reducible and do not necessarily have finitely many irreducible representations, are actively being researched in recent years \parencite{Gaberdiel_1999,Creutzig_2013,Fuchs_2019}. 

With a series of papers \parencite{Fuchs_2002,Fuchs_2004,Fuchs2004,Fuchs_2005,fjelstad2006tft,fjelstad2008uniqueness}, the authors had made certain progress on the classification of all rational full 2D CFTs, defined on all 2-dimensional cobordisms. There is also extensive literature on the classification of 2d CFTs defined on the torus using the modular differential equations (MDEs) satisfied by their module characters - formal power series that counts the tower of states created by a primary field. The efforts of classifying CFTs by MDEs started from late 1980s \parencite{MATHUR198915,MATHUR1988303,MATHUR1989483} and continue since then \parencite{Gaberdiel_2016,Chandra_2019,bae2019monster}. 

It is not a coincidence that classifying characters becomes crucial considering the physical perspective of CFT regarding critical exponents. Systems with the same critical exponents tend to have the same set of partition functions that are built out of characters. Thus, a classification of characters is equivalent to the classification of various classes of critical behaviours \parencite{hr2016classification}. For this reason, the study of character identities has been given prominence as well.

The famous Weyl-Kac character formula allows us to compute the characters of Kac-Moody algebras \parencite{KAC197885}, which have an inevitable connection to CFTs via WZW models \parencite{GepnerWitten}. It was noticed in \parencite{FEINGOLD1978271} that one can gain insights of certain Rogers-Ramanujan identities \parencite{andrews1974general} from the characters of modules of certain Kac-Moody algebras and the idea was further developed in \parencite{LEPOWSKY198221,Lepowsky1984}. Following these observations, it was then noticed that some character identities may have important physics interpretations \parencite{FRENKEL1981259,Gepner_1995}. Conversely, more and more character identities can be revealed by looking at the physical perspective and different frameworks of CFT, especially the Bose-Fermi correspondence and the Bethe ansatz \parencite{KUNIBA_1993,Melzer_1994,kedem1995sums,BERKOVICH1997621,BelGep}. 

The investigation of character identities grew slowly until it was revisited in \parencite{BelGep,genish2014level,Gepner2014yva}, based on the establishment of a direct relation between the characters of Kac-Moody algebras and the characters of parafermion CFTs presented in \parencite{GEPNER87}. We believe that there are still many mysteries regarding character identities that are worth exploring and our goal is to contribute to this area by studying a specific subclass of CFTs that involve either uncoulped or coupled free fermions. The later ones naturally arise from parafermion CFTs in \parencite{GEPNER87}. It is worth mentioning that some character identities that were derived from the algebraic structure may be
directly deduced from the classical theory of q-series as done in \parencite{andrews1984hecke}. We will pursue such deductions for our character identities as well. 


\section{Preliminaries}

 In two dimensions, the infinite-dimensional conformal algebra can be thought of as being composed of two commuting Virasoro algebras\index{Virasoro algebra} each of the form 
\begin{equation}\label{eq:Virasoro}
    [L_m,L_n]=(m-n)L_{m+n}+\frac{1}{12}cm(m^2-1)\delta_{m+n,0},
\end{equation}
where $m,n\in\mathbb{Z}$ and $c$ is the central element\footnote{However, with abuse of notation, we will often use it to mean the scalar by which the central element acts on modules of the Virasoro algebra.}, also known as the central charge\index{central charge} of the CFT. The algebra is related to (the holomorphic component of) the energy-momentum tensor\index{energy-momentum tensor} $L(z)$  of the CFT in the way that
\[L_n=\dfrac{1}{2\pi i}\oint dz z^{n+1}L(z).\]
The OPE of $L(z)$ with itself is given by
\[L(z)L(w)=\dfrac{c/2}{(z-w)^4}+\dfrac{2L(w)}{(z-w)^2}+\dfrac{\partial_{w}L(w)}{z-w}.\]
The anti-holomorphic component $\overline{L}(\overline{z})$ gives rise to a similar set of Virasoro operators $\overline{L}_n$, satisfying the same algebra \eqref{eq:Virasoro}. We will mostly focus on the holomorphic component only. 

The Virasoro operators $L_n$ and $\overline{L}_n$ act in the vector space (typically a Fock space) $\mathcal{H}$ of the CFT. A representation of the Virasoro algebra is called  unitary\index{Virasoro algebra!unitary representation} if  all generators $L_n$ are realized as operators acting on a Fock space with the condition that
\[L^{\dagger}_n=L_{-n}.\]
In that case, the states in $\mathcal{H}$ can be arranged into irreducible representations of the two Virasoro algebras. The vacuum state
$\ket{0}$ is the unique state annihilated by all Virasoro operators $L_n$ with $n \ge -1$. The other states $\ket{\phi}$ are obtained by the state-field correspondence:
\begin{equation}\label{eq:statefield}
    \ket{\phi}=\lim_{(z,\overline{z})\rightarrow(0,0)}\phi(z,\overline{z})\ket{0}.
\end{equation}
Here $\phi(z,\overline{z})$ denotes generically any field that might appear.

The operator $L_0$ is bounded below, so that 
one is only interested in highest weight representations\index{Virasoro algebra!highest weight !representation} whose representation space can be built up from a highest weight state\index{Virasoro algebra!highest weight !state} $\ket{h}$, satisfying 
\begin{align}\label{eq:Virprifield}
    &L_0\ket{h}=h\ket{h},\quad h\ge0, &L_n\ket{h}=0,\quad \forall n>0.
\end{align}
By the state-field correspondence \eqref{eq:statefield}, there is a corresponding field $\phi_h(z)$ for each highest weight state $\ket{h}$. Such fields are known as (Virasoro) primary fields. In terms of OPEs, we have (for the holomorphic part)
\[L(z)\phi_h(w)\sim \dfrac{h\phi_h(w)}{(z-w)^2}+\dfrac{\partial_w\phi_h(w)}{z-w},\]
where $h$ is also known as the conformal dimension
\index{conformal dimension} of $\phi_h(z)$.

The (algebraic Virasoro) character $\chi(\tau)$ of a highest weight representation $\mathcal{L}$ of the CFT is defined as 
\[\chi(\tau):=\Tr_{\mathcal{L}}e^{2\pi i\tau\left(L_0-\frac{c}{24}\right)}=\Tr_{\mathcal{L}}q^{L_0-\frac{c}{24}}.\]

The following will mainly focus on the holomorphic part of rational CFTs and their unitary representations\footnote{Mathematically, this simply means that we are studying VOAs and their representations.  However, we will keep using CFT terminologies (even though we are not working with full CFTs) but not VOA formalisation throughout the thesis to avoid too much technical discussion on locality issues in terms of VOA for coupled free fermion theories.}. Comprehensive reviews on 2D CFT can be found in literature, for instance,  \parencite{francesco1997conformal,SchCFT,Poland2019,cardy1988conformal,gaberdiel2000introduction}.


\subsection{Constructions of conformal field theories}\label{sec:CFTs}

\subsubsection*{Minimal models}
It has been long known that the CFTs containing finitely many irreducible representations of Virasoro algebras can only take certain values of the central charge \parencite{FQSminmodel,Belavin:1984vu,kacWang1994vertex}. Such CFTs are the so-called rational minimal models\index{minimal model}, among which the unitary ones form a nice discrete series, indexed by one parameter, say $m\in\mathbb{Z}_{>0}$, which reads 
\begin{equation}\label{eq:mincc}
    c(m)=1-\dfrac{6}{(m+2)(m+3)},
\end{equation}
and the conformal dimensions of their representations\index{minimal model!representation} are 
\begin{equation}\label{eq:minwt}
    h^{(m)}_{r,s}=\dfrac{[(m+3)r-(m+2)s]^2-1}{4(m+2)(m+3)},\qquad 1\le r\le m+1, \; 1\le s\le m+2.
\end{equation}
The irreducible modules of a minimal model with central charge $c(m)$ are characterised by the values of $r$ and $s$ as in \eqref{eq:minwt}. Therefore, we denote the modules of minimal models by $\mathcal{L}\left(c(m),h^{(m)}_{r,s}\right)$ hereafter\footnote{Note that this notation does not distinguish isomorphic modules with different values of $r$ and $s$}. The character formula\index{minimal model!character} of modules of minimal models was discovered in \parencite{rocha1985vacuum} as follows:
\begin{equation}
    \mathrm{ch}\left[\mathcal{L}\left(c(m),h^{(m)}_{r,s}\right)\right](\tau)=\dfrac{1}{\eta(\tau)}\sum_{k\in\mathbb{Z}}\left(q^{a_{+}(k,r,s)}-q^{a_{-}(k,r,s)}\right)
\end{equation}
where
\[a_{\pm}(k,r,s)=\dfrac{\left(2(m+2)(m+3)k\pm(m+3)r+ms\right)^2}{4(m+2)(m+3)}.\]

Minimal models are the simplest cases of CFTs and have been fully classified \parencite{cappelli1987ade}. Due to this fact, when studying a CFT, it is sometimes helpful, if possible, to decompose the theory in terms of minimal models, as we will do in Section \ref{sec:cosetexp}.

\subsubsection*{Lattice construction}
\index{lattice construction}
Given an even Euclidean lattice $\Gamma$ of dimension $d$, one can introduce orthonormal states $\ket{\gamma}$ for $\gamma\in\Gamma$ with inner product $\bra{\;}\ket{\;}$ and generate the fock space\index{fock space} $\mathcal{H}(\Gamma)$ by the action of $d$ copies of the Heisenberg Lie algebra with a basis $\{a^j_n|n\in\mathbb{Z},1\le j\le d\}$. Then, for each state $\ket{\psi}=\left(\prod_{m=1}^M a^{j_m}_{-n_m}\ket{\gamma}\right)\in\mathcal{H}(\Gamma)$, one associates it with the field
\begin{equation*}
    \mathcal{V}(\ket{\psi},z)=:\left(\prod_{m=1}^N\dfrac{i}{(n_m-1)!}\dfrac{d^{n_m}X^{j_m}(z)}{dz^{n_m}}\right)e^{i\gamma\cdot X(z)}: \sigma_{\gamma},
\end{equation*}
where $:A(z)B(z):$ denotes the normal ordering of $A(z)B(z)$ and
\[X^{j}(z)=q^j-ia_0^j\ln{z}+i\sum_{n\ne0}\dfrac{a^{j}_n}{n}z^{-n},\]
where $q$ is the position operator which only appears in the form $e^{i\gamma\cdot q}$ defined by
\[e^{i\gamma\cdot q}\ket{\mu}=\ket{\gamma+\mu},\] 
and $\sigma_{\gamma}$ is a cocycle operator such that
\[e^{i\gamma\cdot q}\sigma_{\gamma}e^{i\gamma'\cdot q}\sigma_{\gamma'}=(-1)^{\bra{\gamma}\ket{\gamma'}}e^{i\gamma'\cdot q}\sigma_{\gamma'}e^{i\gamma\cdot q}\sigma_{\gamma}\]
for any $\gamma,\gamma'\in\Gamma$. It is proven in \parencite{Dolan1996} that this structure forms a CFT\footnote{We should mention that more precisely speaking, this is a VOA of CFT type.} with central charge $d$, which we shall denote as $V_{\Gamma}$. 

The conformal state (i.e. the state corresponding to the energy-momentum tensor) is taken to be
\[\ket{\psi_{L}}=\frac{1}{2}\sum_{j}a_{-1}^ja_{-1}^j\ket{0}\]
and the conformal dimension\index{lattice construction!conformal dimension} (i.e. $L_0$-eigenvalue) of a general state $\ket{\psi}$ is given by
\[h_{\psi}=\frac{1}{2}|\gamma|^2+\sum_{m=1}^M n_m.\]

It is proved in \parencite{DONG1993245} that the irreducible modules \index{lattice construction!module}of $V_{\Gamma}$ are in one-to-one correspondence with the cosets in $\Gamma^*/\Gamma$, where $\Gamma^*=\{v\in\mathbb{Q}\otimes_{\mathbb{Z}}\Gamma\,|\,(v,\Gamma)\subset\mathbb{Z}\}$ is the dual lattice of $\Gamma$. We denote the modules by $V_{\nu+\Gamma}$ for any $\nu\in \Gamma^*/\Gamma$. The character\index{lattice construction!character} formula is given by 
\begin{equation}\label{eq:latch}
    \mathrm{ch}\left[V_{\nu+\Gamma}\right](\tau)=\dfrac{1}{\eta(\tau)^{d}}\sum_{\gamma\in\nu+\Gamma}q^{\frac{1}{2}|\gamma|^2}.
\end{equation}

Even lattices arise naturally from Lie algebras since the root lattice of any simply-laced algebra is even. In this context, we usually borrow terminologies from physics and think of bosons which play the role of $X^j$ above \parencite{francesco1997conformal}. For example, let $A_n$ denote the root lattice of $\mathfrak{sl}(n+1)$, then one can build a CFT as follows. Assign a free boson $\phi^i(z)$, $1\le i\le n$ to each simple root $\alpha_i$ of $\mathfrak{sl}(n+1)$ and let $\Phi$ be the vector $(\phi^1,\phi^2,\dots,\phi^n)^T$. Write $\alpha\in A_n$ as a vector under the basis formed by simple roots $\{\alpha_i|1\le i\le n\}$. Then define, for each $\alpha\in A_n$, a field 
\begin{equation}\label{eq:latfield}
    \psi^{\alpha}(z)=:e^{i\alpha\cdot\Phi(z)}:,
\end{equation}
with
\[\alpha\cdot\Phi=\sum_{i}\alpha^i\phi^i, \]
where $\alpha^i$ is the corresponding eigenvalue of $\alpha$ in the Cartan-Weyl basis. The conformal weight of $\psi^{\alpha}$ is $\frac{1}{2}|\alpha|^2$ by construction. The OPEs are given by
\begin{equation}\label{eq:latOPE}
    \psi^{\alpha}(z)\psi^{\beta}(w)\sim \dfrac{c_{\alpha,\beta}\psi^{\alpha+\beta}(w)}{(z-w)^{-(\alpha,\beta)}}+\dfrac{c_{\alpha,\beta}(\alpha\cdot i\partial\Phi)\psi^{\alpha+\beta}(w)}{(z-w)^{-(\alpha,\beta)-1}}+\cdots,
\end{equation}
where $c_{\alpha,\beta}$ are some constants that absorb the information of cocycle operators. The energy-momentum tensor\index{lattice construction!energy-momentum tensor}, corresponding to the state $\ket{\psi_L}$, is given by
\[L(z)=-\frac{1}{2}\sum_{i=1}^n:\partial\phi^i(z)\partial\phi^i(z):.\]
This process can be generalised for the root lattice scaled by $\sqrt{k}$ or $1/\sqrt{k}$ (with relaxed locality axiom) and we will discuss the cases when $k=2$ in Chapter \ref{Chapter4}.

\subsubsection*{WZW models and the Sugawara construction}
Wess-Zumino-Witten (WZW) models\index{Wess-Zumino-Witten (WZW) model} are conformal field theories with additional conserved currents generating affine Lie algebras. This construction is based on Wess-Zumino chiral effective action and mathematically formalised in \parencite{KZWZW,Witten1984WZW,GepnerWitten}. Here we include a brief review following \parencite{francesco1997conformal}.

Given a semi-simple Lie algebra $\mathfrak{g}$, let $g(z)$ be a matrix bosonic field, living on the group manifold $G$ associated with $\mathfrak{g}$, valued in a non-trivial unitary representation of $\mathfrak{g}$ with generators $\{t^a\}$ which are chosen such that the Killing form is $\delta_{a,b}$. Define the (holomorphic) currents as\index{Wess-Zumino-Witten (WZW) model!current}
\[J(z)=-k(\partial_zg)g^{-1}(z)\]
with $J=\sum_a J^a t^a$, then the fields $J^a(z)$ satisfy the OPE
\begin{equation}\label{eq:WZWOPE}
    J^{a}(z)J^b(w)\sim\dfrac{k\delta_{a,b}}{(z-w)^2}+\dfrac{f_{abc}J^c(w)}{z-w},
\end{equation}
where $f_{abc}$ are structure constants\index{structure constants} of $\mathfrak{g}$ under the basis $\{t^a\}$ and we use the Einstein summation convention. This is known as the current algebra associated with the affine Lie algebra $\hat{\mathfrak{g}}_k$ (see Appendix \ref{AppendixA} for a brief review of affine Lie algebras). The commutation relations of the modes $J^a_n$ from the expansion\index{Wess-Zumino-Witten (WZW) model!commutation relation}
\[J^a(z)=\sum_{n\in\mathbb{Z}}J^a_n z^{-n-1}\]
are equivalent to those of the affine Lie algebra $\hat{\mathfrak{g}}_k$ \eqref{eq:KMcomm} which now reads:
\begin{equation}\label{eq:currentcomm}
    [J^a_m,J^b_n]= f_{abc}J^c_{m+n}+km\delta_{a,b}\delta_{m+n,0}.
\end{equation}
The energy-momentum tensor\index{Wess-Zumino-Witten (WZW) model!energy-momentum tensor} is given by
\begin{equation}\label{eq:WZWEnerMon}
    L(z)=\frac{1}{2(k+h^{\vee})}\sum_{a}:J^a(z)J^a(z):
\end{equation}
which gives the central charge\index{Wess-Zumino-Witten (WZW) model!central charge}
\begin{equation}\label{eq:WZWcc}
    c\left(\hat{\mathfrak{g}}_k\right)=\dfrac{k\dim\mathfrak{g}}{k+h^{\vee}},
\end{equation}
where $h^{\vee}$ is the dual Coxeter number of $\mathfrak{g}$.

The Virasoro algebra belongs to a completion of the enveloping algebra of the affine Lie algebra $\hat{\mathfrak{g}}_k$. This result is known as the Sugawara construction\index{Sugawara construction} in the physics literature \parencite{Sugawara}. 

WZW primary fields \index{Wess-Zumino-Witten (WZW) model!primary field, representation}with underlying $\hat{\mathfrak{g}}_k$ structure are in correspondence with the dominant integral weights of $\hat{\mathfrak{g}}$ at level k, that is, for $\Lambda\in\hat{P}^k_+$, there exists a unique field $\phi_{\Lambda}$ whose associated state $\ket{\Lambda}:=\lim_{z\mapsto 0}\phi_{\Lambda}(z)\ket{0}$ satisfies 
\begin{align*}
    &J^a_0\ket{\Lambda}=t^a_{\Lambda}\ket{\Lambda},&J^a_n\ket{\Lambda}=0,\quad\quad\mathrm{for}\;n>0,\;\forall a,
\end{align*}
where $t^a_{\Lambda}$ is the image of $t^a$ under the $\Lambda$-representation restricted to $\mathfrak{g}$. It can be verified that WZW primary fields are also Virasoro primary fields \eqref{eq:Virprifield} with conformal dimension $h_{\Lambda}$ given by
\begin{equation}\label{eq:WZWcondim}
    h_{\Lambda}=\dfrac{(\Lambda,\Lambda+2\rho)}{2(k+h^{\vee})},
\end{equation} 
where $\rho$ is the Weyl vector.

\index{Wess-Zumino-Witten (WZW) model!conformal dimension}The character\index{Wess-Zumino-Witten (WZW) model!character} associated with the highest weight state $\ket{\Lambda}$ is given by
\begin{equation*}
    \mathrm{ch}_{\Lambda}(\tau):=\Tr_{\Lambda}q^{L_0-\frac{c}{24}}=q^{h_{\Lambda}-\frac{c}{24}}\sum_nd(n)q^n,
\end{equation*}
where $d(n)$ is the number of states at level $n$, which leads us back to \eqref{eq:affch}. The characters of WZW models can be computed using the Weyl-Kac character formula \eqref{eq:weylkac}.

\subsubsection*{GKO coset construction}

The Goddard-Kent-Olive (GKO) coset construction\index{Goddard-Kent-Olive (GKO)} \index{coset construction} was presented in \parencite{GKOcoset}. Given an affine Lie algebra $\hat{\mathfrak{g}}_k$ with $k\in\ZZ_{> 0}$. Let $\hat{\mathfrak{p}}_{k'}$ be an affine Lie subalgebra of $\hat{\mathfrak{g}}_k$ at level $k':=x_ek$ , where $x_e$ is the Dynkin index of the embedding $\mathfrak{p}\subset\mathfrak{g}$, then the Sugawara construction gives conformal fields $L^{\mathfrak{g}}(z)$ and $L^{\mathfrak{p}}(z)$ associated with $\hat{\mathfrak{g}}_k$ and $\hat{\mathfrak{p}}_{k'}$ respectively as in \eqref{eq:WZWEnerMon}, with central charges given by \eqref{eq:WZWcc}. One can show that $L^{(\mathfrak{g}/\mathfrak{p})}(z):=L^{\mathfrak{g}}(z)-L^{\mathfrak{p}}(z)$ satisfies
\[[L^{(\mathfrak{g}/\mathfrak{p})}(z),L^{\mathfrak{p}}(z)]=0.\]
Therefore the modes of $L^{(\mathfrak{g}/\mathfrak{p})}$ also form a copy of the Virasoro algebra with central charge\index{coset construction!central charge} given by
\begin{equation}\label{eq:cosetcc}    c\left(\hat{\mathfrak{g}}_k/\hat{\mathfrak{p}}_{k'}\right)= c\left(\hat{\mathfrak{g}}_k\right)-c\left(\hat{\mathfrak{p}}_{k'}\right),
\end{equation}
where $\hat{\mathfrak{g}}_k/\hat{\mathfrak{p}}_{k'}$ denotes the coset theory\footnote{We should mention that we adopt the terminology ``coset" from physics although it is not a quotient but a commutant mathematically.} with energy-momentum tensor $L^{(\mathfrak{g}/\mathfrak{p})}(z)$.

Let $\mathcal{L}^{\mathfrak{g}}_{\Lambda}$ be the representation\index{coset construction!representation} of $\hat{\mathfrak{g}}$ with highest weight $\Lambda$. Assume $x_e=1$, for simplicity. Then $\mathcal{L}^{\mathfrak{g}}_{\Lambda}$ can be decomposed into representations of $\hat{\mathfrak{p}}$ as follows:
\[\mathcal{L}^{\mathfrak{g}}_{\Lambda}=\bigoplus_{\substack{\lambda\in\hat{P}^k_+\\ \mathcal{P}\Lambda-\lambda\in \mathcal{P}Q}}\mathcal{L}^{\mathfrak{g}/\mathfrak{p}}_{\Lambda,\lambda}\otimes \mathcal{L}^{\mathfrak{p}}_{\lambda},\]
where $\hat{P}_+^k$ is the set of dominant weights at level $k$ of $\hat{\mathfrak{p}}$, $\mathcal{P}$ is the projection matrix for weights of $\mathfrak{g}$ projected onto $\mathfrak{p}$, and $Q$ is the root lattice of $\mathfrak{g}$. Then $\mathcal{L}^{\mathfrak{g}/\mathfrak{p}}_{\Lambda,\lambda}$ are the modules of $\hat{\mathfrak{g}}_k/\hat{\mathfrak{p}}_{k}$. Hence, the characters\index{coset construction!character} of $\mathcal{L}^{\mathfrak{g}/\mathfrak{p}}_{\Lambda,\lambda}$, denoted by $\mathrm{ch}_{[\Lambda,\lambda]}$, are determined by
\[\mathrm{ch}_{\Lambda}=\sum_{\substack{\lambda\in\hat{P}^k_+\\ \mathcal{P}\Lambda-\lambda\in \mathcal{P}Q}}\mathrm{ch}_{[\Lambda,\lambda]}\mathrm{ch}_{\lambda}.\]
Normalised characters of coset models evaluated at a certain weight are known as branching functions. 

Let $A$ (resp.$\hat{A}$) be an outer automorphism of $\hat{\mathfrak{g}}$ (resp.$\hat{\mathfrak{p}}$). If $\mathcal{P}A=\hat{A}\mathcal{P}$, then the fields corresponding to $\mathcal{L}^{\mathfrak{g}/\mathfrak{p}}_{\Lambda,\lambda}$ and $\mathcal{L}^{\mathfrak{g}/\mathfrak{p}}_{A\Lambda,\hat{A}\lambda}$ are identified.

The $\widehat{\mathfrak{sl}(2)}$ diagonal cosets $\widehat{\mathfrak{sl}(2)}_k\oplus\widehat{\mathfrak{sl}(2)}_1/\widehat{\mathfrak{sl}(2)}_{k+1}$ are particularly interesting since they describe unitary Virosoro minimal models \parencite{GKOcoset}. It can be readily checked using \eqref{eq:cosetcc} that their central charges fit the values given in \eqref{eq:mincc} with $m=k$.

Another well-known family of cosets is the family of parafermion models in the form of $\widehat{\mathfrak{sl}(2)}_k/\widehat{\mathfrak{u}(1)}$ constructed in \parencite{ZF1985}. A more general version of coset parafermion models introduced in \parencite{GEPNER87} will play an important role in our discussion of coupled free fermions in Chapter \ref{Chapter3} and will be reviewed there.

\subsubsection*{Orbifold construction}

In broad sense, roughly speaking, the ``orbifold construction"\index{orbifold construction} can be used for a process that modifies any conformal field theory by adding some new fields, while removing some others, which has an interpretation in analogue to the procedure of defining ``orbifolds" from manifolds in some cases \parencite{Dixon1985strorb,Dixon1988strorb,dixon1987orb}. For the purpose of later chapters, we shall only focus on the $\mathbb{Z}_2$-orbifold construction from a lattice construction, as described in \parencite{DGM1990orb,Dolan1996}.  

Given a lattice $\Gamma$ of dimension $d$, let $\mathcal{H}(\Gamma)$ be the fock space of its lattice construction. We define in addition the space $\mathcal{H}^T(\Gamma)$ that is built up similarly from a representation $\Gamma_0$ of the gamma matrix algebra associated with $\Gamma$ by the action of operators $\{c^j_r|r\in\mathbb{Z}+\frac{1}{2},1\le j\le d\}$, where $\Gamma_0=\{\gamma_{\mu}|\mu\in\Gamma\}$ is defined by 
\begin{align*}
    &\gamma_{\mu}\gamma_{\mu'}=\epsilon(\mu,\mu')\gamma_{\mu+\mu'}=(-1)^{\mu\cdot\mu'}\gamma_{\mu'}\gamma_{\mu}, &\gamma_{\mu}^2=(-1)^{\frac{|\mu|^2}{2}},
\end{align*}
where $\epsilon(\mu,\mu')=\pm 1$ are determined in \parencite[App.B,C]{DGM1990orb}, and the operators $c^j_r$ satisfy
\[\left[c^i_r,c^j_s\right]=r\delta_{r+s,0}\delta^{ij},\]
$c^{j^{\dagger}}_r=c^j_{-r}$,  and $c^j_r\ket{\gamma_{\mu}}=0$ if $r>0$ and $\gamma_{\mu}\in\Gamma_0$. The operators 
\begin{equation*}
    L_n^c=\frac{1}{2}\sum_{r\in\mathbb{Z}+\frac{1}{2}}:c_rc_{n-r}:+\dfrac{d}{16}\delta_{n,0}
\end{equation*}
satisfy the Virasoro algebra with $c=d$ and the highest weight state of $\mathcal{H}^T(\Gamma)$ has conformal weight $\frac{d}{16}$. Now we can define for each $\mu\in\Gamma$ a new field $\mathcal{V}_T:\mathcal{H}^T(\Gamma)\mapsto\mathcal{H}^T(\Gamma)$ by
\[\mathcal{V}_T\left(\ket{\mu},z\right)=(4z)^{-\frac{|\mu|^2}{2}}:e^{i\mu\cdot R(z)}:\gamma_{\mu},\]
where the components of $R(z)$ are 
\[R^j(z)=i\sum_{r\in\mathbb{Z}+\frac{1}{2}}\frac{c^j_r}{r}z^{-r}.\]
The definition of $\mathcal{V}_T$ can be generalised to any $\ket{\psi}\in\mathcal{H}(\Gamma)$ as given in \parencite{DGM1990orb}. Furthermore, we can define the fields $\mathcal{W}:\mathcal{H}(\Gamma)\mapsto\mathcal{H}^T(\Gamma)$ by
\[\mathcal{W}(\ket{\chi},z)\ket{\psi}=e^{zL^c_{-1}}\mathcal{V}_T(\ket{\psi},z)\ket{\chi}\]
and define its conjugate $\overline{\mathcal{W}}:\mathcal{H}^T(\Gamma)\mapsto\mathcal{H}(\Gamma)$ in a proper sense.

We introduce the reflection map\index{orbifold construction!reflection map} $\theta$ defined on the space $\mathcal{H}(\Gamma)$ by 
\begin{align*}
    &\theta\ket{\gamma}=\ket{-\gamma}, &\theta a^i_n=-a^i_n\theta,
\end{align*}
where $\gamma\in\Gamma$ and on the space $\mathcal{H}^T(\Gamma)$ by\footnote{It was assumed that $d$ is a multiple of 8 in \parencite{DGM1990orb}, but it seems that no reason forbids us to force that $\theta\ket{\gamma_{\mu}}=\ket{-\gamma_{\mu}}$ for any $d$, although a corresponding modification in the definition of $\overline{\mathcal{W}}$ might be needed.}
\begin{align*}
    &\theta\ket{\gamma_{\mu}}=\ket{(-1)^{d/8}\gamma_{\mu}}, &\theta c^i_r=-c^i_r\theta,
\end{align*}
where $\gamma_{\mu}\in\Gamma_0$. Let $\mathcal{H}_0(\Gamma)$ (resp. $\mathcal{H}^T_0(\Gamma)$) be the subspace of $\mathcal{H}(\Gamma)$ (resp. $\mathcal{H}^T(\Gamma)$) on which $\theta$ is the identity map. We note that 
\begin{align*}
    &\theta \mathcal{V}(\ket{\psi},z)=\mathcal{V}(\theta\ket{\psi},z)\theta, & \theta \mathcal{V}_T(\ket{\psi},z)=\mathcal{V}_T(\theta\ket{\psi},z)\theta.
\end{align*}
Now we can consider the space $\Tilde{\mathcal{H}}(\Gamma)=\mathcal{H}_0(\Gamma)\oplus\mathcal{H}_0^T(\Gamma)$ and the fields $\Tilde{\mathcal{V}}:\Tilde{\mathcal{H}}(\Gamma)\mapsto\Tilde{\mathcal{H}}(\Gamma)$ (in the matrix form)
 \begin{align*}
     &\Tilde{\mathcal{V}}(\ket{\psi},z)=\begin{pmatrix}
    \mathcal{V}(\ket{\psi},z)&0\\
    0&\mathcal{V}_T(\ket{\psi},z)
\end{pmatrix}, &\Tilde{\mathcal{V}}(\ket{\chi},z)=\begin{pmatrix}
     0&\overline{\mathcal{W}}(\ket{\chi},z)\\
     \mathcal{W}(\ket{\chi},z)&0
\end{pmatrix},
 \end{align*}
where $\ket{\psi}\in \mathcal{H}_0(\Gamma)$ and $\ket{\chi}\in \mathcal{H}^T_0(\Gamma)$, and it is shown in \parencite{DGM1990orb} that this defines a CFT, which we shall denote by $V^{\mathbb{Z}_2}_{\Gamma}$. 

We note that the states in $\mathcal{H}_0(\Gamma)$ are obtained by acting with an even number of operators $a_n^j$ on $\ket{\gamma}+\ket{-\gamma}$ or by acting with an odd number of operators $a_n^j$ on $\ket{\gamma}-\ket{-\gamma}$. The vacuum in $\mathcal{H}_0(\Gamma)$ is doubly degenerate and $\theta\ket{0}_{\pm}=\pm\ket{0}_{\pm}$. The states in $\mathcal{H}^T_0(\Gamma)$ are similar. The orbifold characters\index{orbifold construction!character} are given by
\begin{align}
    \text{ch}_{++}(q)=&\Tr_{\mathcal{H}_0}q^{L_0-\frac{d}{24}} ,&\text{ch}_{+-}(q)= \Tr_{\mathcal{H}_0}\theta q^{L_0-\frac{d}{24}},\label{eq:orbch-1}\\
    \text{ch}_{-+}(q)=&\Tr_{\mathcal{H}^T_0} q^{L_0+\frac{d}{48}},&\text{ch}_{--}(q)=\Tr_{\mathcal{H}^T_0} \theta q^{L_0+\frac{d}{48}} .\label{eq:orbch-2}
\end{align}

One of the simplest examples of orbifold constructions, namely the $\mathbb{Z}_2$-orbifold\index{orbifold construction!Z@$\mathbb{Z}_2$-orbifold} construction from the free boson CFT has been discussed in detail and the reviews can be found in, for example, \parencite[Chp.8]{SchCFT} and \parencite[Sec.10.4.3]{francesco1997conformal}. Consider the anti-podal $\mathbb{Z}_2$ symmetry that acts by taking the boson $\Phi$ to $-\Phi$. This can be done by introducing a field $\sigma$ with conformal dimension $\frac{1}{16}$ about which $\partial\Phi$ is anti-periodic. The characters are given as follows:
\begin{equation}
\begin{aligned}
    \text{ch}_{++}(q)=\,&q^{-\frac{1}{24}}\prod_{n=1}^{\infty}\dfrac{1}{1-q^n}=\,\frac{1}{\eta(\tau)} ,\\
    \text{ch}_{+-}(q)=\,&q^{-\frac{1}{24}}\prod_{n=1}^{\infty}\dfrac{1}{1+q^n}= \frac{\eta(\tau)}{\eta(2\tau)},\\
    \text{ch}_{-+}(q)=\,&q^{\frac{1}{48}}\prod_{n=1}^{\infty}\dfrac{1}{1-q^{n-\frac{1}{2}}}=\,\frac{\eta(\tau)}{\eta(\tau/2)},\\
    \text{ch}_{--}(q)=\,&q^{\frac{1}{48}}\prod_{n=1}^{\infty}\dfrac{1}{1+q^{n-\frac{1}{2}}}=\frac{\eta(\tau/2)\eta(2\tau)}{\eta(\tau)^2} .
\end{aligned}\label{eq:orbch-boson}
\end{equation}


\subsection{Conformal vertex operators and fusion rules}

The chiral vertex operators (CVOs)\index{chiral vertex operator (CVO)}, also known as intertwining operators \parencite{kac1998vertex}, are equivariant linear maps between representations. Suppose we have a 2D conformal field theory with a set of modules $\{\mathcal{L}_i|i\in \mathcal{I}\}$ for some index set $\mathcal{I}$. We consider CVOs such as follows:   
$$\phi\binom{i}{k\;j}(z):\mathcal{L}_{j}\mapsto \mathcal{L}_{k},$$
whose mode expansions are
\begin{equation}\label{eq:CVOmode}
    \phi\binom{i}{k\;j}(z)=\sum_{n\in\mathbb{Z}}\phi\binom{i}{k\;j}_{n-(h_{k}-h_{j})}z^{-n+(h_{k}-h_{j}-h_{i})},
\end{equation}
where $h_i$ denotes the conformal dimensions of $\mathcal{L}_{i}$. The number of CVOs of type $\binom{i}{k\;j}$ is given by the fusion rules $\mathcal{N}_{ij}^k$ \parencite{BouwknegtUCPF}.

The fusion rules\index{fusion rule} $\mathcal{N}_{ij}^k\in\mathbb{Z}_{\ge0}$ are decomposition coefficients of the commutative and associative formal product ``$\times$'' between modules, which we call the fusion product,
\[\mathcal{L}_{i}\times \mathcal{L}_{j}=\sum_{k\in \mathcal{I}}\mathcal{N}_{ij}^k\mathcal{L}_{k}.\]

For the WZW model with underlying algebra $\hat{\mathfrak{g}}_k$ with $\mathfrak{g}$ semi-simple and $k\in \ZZ_{>0}$, the index set $\mathcal{I}$ is $\hat{P}^k_+$ and the fusion rules can be calculated from the Verlinde formula
\begin{equation*}    \mathcal{N}_{ij}^{k}=\sum_{l\in\hat{P}^k_+}\dfrac{S_{il}S_{jl}S_{kl}}{S_{0l}},
\end{equation*}
where $S$ is the modular matrix that transforms the characters of $\hat{\mathfrak{g}}_k$ under $\tau\mapsto-1/\tau$, and $0$ stands for $k\Lambda_0$. A simple algorithm for calculating the fusion rules for the WZW models associated with affine Lie algebra was given in \parencite{Walton:1990qs}, based on the Kac-Walton formula 
\[\mathcal{N}_{ij}^{k}=\sum_{\substack{w\in\hat{W}\\ w\cdot k\in P_+}}\mathcal{N}_{ij}^{w\cdot k}\det(w),\]
where $\hat{W}$ is the weyl group of $\hat{\mathfrak{g}}$.

Investigations of fusion rules from the view of elementary group theory have also been made, for example, in \parencite{feingold2002type}, where they obtained fusion rules of minimal models and included many explicit fusion rules, especially the ones of $\widehat{\mathfrak{sl}(3)}_{2}$ and $\widehat{\mathfrak{sl}(4)}_{2}$ that we will need.

The space spanned by the states in which $\mathcal{L}_{i}$ and $\mathcal{L}_{j}$ fuse to $\mathcal{L}_{k}$ is called a fusion space $\mathcal{V}^k_{ij}$, whose dimension is $\mathcal{N}^k_{ij}$. In this thesis we will only be considering CFTs with $\mathcal{N}^k_{ij}\le 1$, so each non-trivial fusion space $\mathcal{V}^k_{ij}$ is one-dimensional, and therefore, we do not distinguish the notations between the fusion space $\mathcal{V}^k_{ij}$ and its basis state. 

Similarly, the space spanned by the states in which $\mathcal{L}_{i}$, $\mathcal{L}_{j}$ and $\mathcal{L}_{k}$ fuse to $\mathcal{L}_{l}$ is denoted by $\mathcal{V}^l_{ijk}$. The associativity of fusion rules imply that the fusion space $\mathcal{V}^l_{ijk}$ can be composed in two equivalent ways: either first fusing $\mathcal{L}_{i}$ and $\mathcal{L}_{j}$ to $\mathcal{L}_{p}$, which is then fused
with $\mathcal{L}_{k}$ to give $\mathcal{L}_{l}$ , or alternatively fusing $\mathcal{L}_{j}$ and $\mathcal{L}_{k}$ first, to give $\mathcal{L}_{p'}$, and then fusing with
$\mathcal{L}_{i}$ resulting in $\mathcal{L}_{l}$:
\[\mathcal{V}^l_{ijk}=\sum_p\mathcal{V}^p_{ij}\mathcal{V}^l_{pk}=\sum_{p'}\mathcal{V}^l_{ip'}\mathcal{V}^{p'}_{jk},\]
where the summations are over possible degenerate multi-particle states, also known as conformal blocks, in non-abelian theories where $\mathcal{N}^k_{ij}>1$ for some $i,j,k$. Hence there are two natural bases for $\mathcal{V}^l_{ijk}$ and the two bases are related by a transformation, represented by the $F$-matrix\index{F@$F$-matrix} $\left(F^l_{jki}\right)_{p'p}$.

The operation of exchanging two fields can be described by the $R$-matrix \index{R@$R$-matrix} $R^k_{ij}:\mathcal{V}^k_{ij}\mapsto\mathcal{V}^k_{ji}$. Note that $\left(R^k_{ij}\right)^2: \mathcal{V}^k_{ij}\mapsto\mathcal{V}^k_{ij}$ should be the operator returning each field to its initial position, so we have
\begin{equation}\label{eq:Rmat}
    \left(R^k_{ij}\right)^2=e^{2\pi i (h_k-h_i-h_j)},
\end{equation}
where $h_i$ is the conformal dimension of $\mathcal{L}_{i}$.

A complete set of consistency conditions to determine $R-$ and $F-$matrices given the fusion rules for a CFT has been derived in \parencite{MS1989CFT,MOORE1989422}, which include the pentagon equation and hexagon equation\index{pentagon equation}\index{hexagon equation}.  The pentagon equation reads
\begin{equation}\label{eq:pent}
  \left(F_{qcp}^t\right)_{ad}\left(F_{rsa}^t\right)_{bc}=\sum_e \left(F_{rsq}^d\right)_{ec}\left(F_{esp}^t\right)_{bd}\left(F_{qrp}^b\right)_{ae}.
\end{equation}
\begin{figure}[h!]
\centering
\includegraphics[width=10cm]{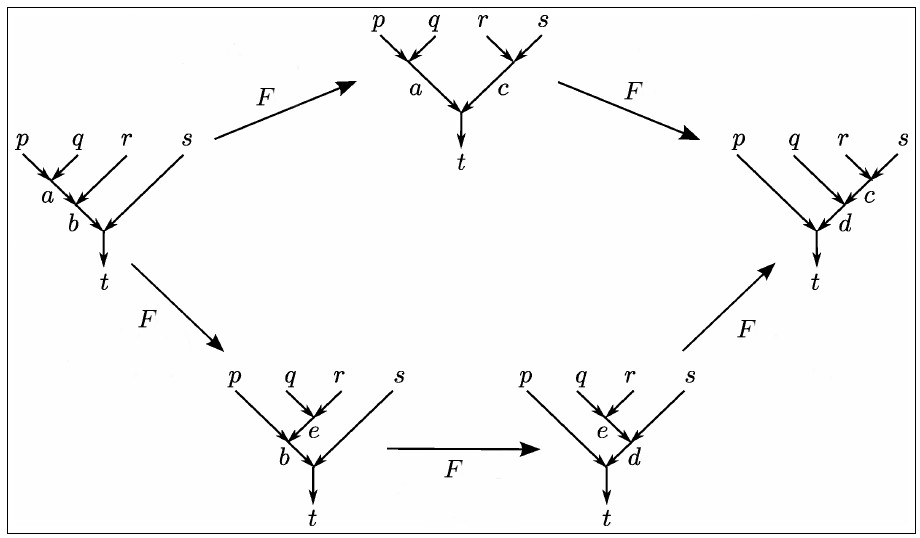}
\caption{A graphical form of the pentagon equation\protect\footnotemark}
\label{fig:pentagon}
\end{figure}
\footnotetext{The picture is edited based on \parencite[Figure 4.9]{Heenan2008}.}

The (fundamental) hexagon equation is given by
\begin{equation}\label{eq:hex}
     R^c_{pr}\left(F_{prq}^s\right)_{ac}R^a_{pq}=\sum_b \left(F_{rpq}^s\right)_{bc}R^s_{pb}\left(F_{qrp}^s\right)_{ab}.
\end{equation}
\begin{figure}[h!]
\centering
\includegraphics[width=10cm]{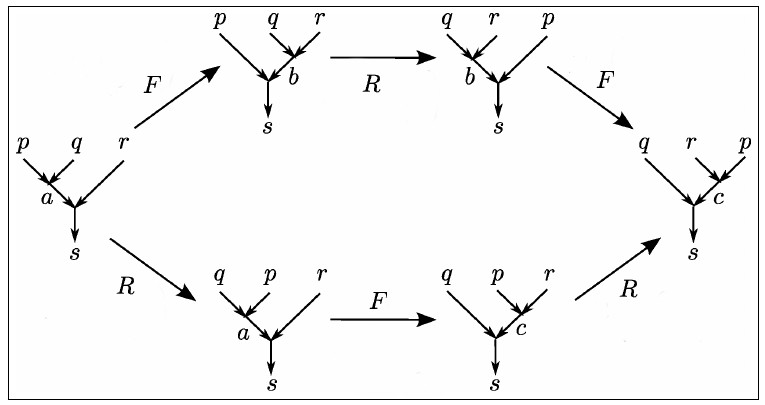}
\caption{A graphical form of the hexagon equation\protect\footnotemark}
\label{fig:hexagon}
\end{figure}

\footnotetext{The picture is edited based on \parencite[Figure 4.10]{Heenan2008}.}

Not all fusion rules have a solution to the polynomial equations and the ones that do have solutions specify physical theories of anyons \parencite{Kitaev_2006}. For theories with fusion rules larger than 1 or degenerate states, solving these equations is thought of as an art. However, as we will be focusing on theories that do not have those obstacles, we will calculate $R-$ and $F-$matrices for our coupled free fermion CFTs by solving the pentagon and hexagon equations.


\subsection{Universal chiral partition functions for exclusion statistics}

 Fractional exclusion statistics\index{exclusion statistics}, as introduced in \parencite{hal1991}, originally in the context of fractional quantum Hall effect, is based on the idea that the number of accessible states for a certain particle depends on the occupation number of all other particles through a statistical interaction matrix\index{statistical interaction matrix} $\mathbf{G}$. The matrix $\mathbf{G}$ is defined by the differential relation
 \begin{equation}\label{eq:exstat}
     \Delta d_{a}=-\sum_{b}G_{ab}\Delta N_b,
 \end{equation}
where $d_{a}$ is the dimension of the Fock space of states of a single particle of species $a$ and  $\{\Delta N_a\}$ is a set of allowed changes of the particle numbers at fixed size and boundary conditions. When $G_{ab} = 0$, there is no reduction in the number of states and particles are bosonic. When $G_{ab} = \delta_{ab}$, the particles obey the Pauli exclusion principle. Therefore fractional exclusion statistics is considered a generalisation of the
Pauli principle. A detailed treatment of exclusion statistics for the case of one quasi-particle was made in \parencite{Wu1994}. There are also other applications such as \parencite{DasnieresdeVeigy:1995rx, ES1998}.

Fractional exclusion statistics arise naturally in quasi-particle descriptions of two-dimensional CFTs. Here, quasi-particles correspond to CVOs between representations of the
chiral algebra and the chiral spectrum (Fock space) is constructed by repeated application of the modes of a preferred set of CVOs on the vacuum. Such a basis was constructed for the $\widehat{\mathfrak{sl}(2)}_{k\ge1}$ WZW models in \parencite{BLS1994,BLS1995}. In Chapter \ref{Chapter3}, we will construct such bases for some coupled free fermion CFTs.

The so-called universal chiral partition function (UCPF)\index{universal chiral partition function (UCPF)}, also known as the fermionic sum representation of characters, 
was introduced in \parencite{KKMM1992,KKMM1993}, in the context of $(\hat{\mathfrak{g}}_1\times \hat{\mathfrak{g}}_1)/\hat{\mathfrak{g}}_2$ coset CFTs and the corresponding integrable lattice models. The general description of UCPF is as follows:
\begin{equation}\label{eq:UCPFgen}
    S_{\mathbf{B}}[\mathbf{a}](\mathbf{u}|q)=\sum_{\substack{N_i\ge 0\\ \text{restrictions}\,\mathbf{Q}}} q^{\frac{1}{2}\mathbf{N}^{\mathbf{T}}\mathbf{B}\mathbf{N}-\frac{1}{2}\mathbf{a}\cdot\mathbf{N}}\prod_{i=1}^n \begin{bmatrix}\big((1-\mathbf{B})\mathbf{N}+\frac{\mathbf{u}}{2}\big)_{i}\\ \mathbf{N}_{i}\end{bmatrix},
\end{equation}
where $\mathbf{N},\mathbf{a},\mathbf{u}$ are $n$ component vectors, $\mathbf{B}$ is an $n\times n$ matrix and for $m$ and $l$ integers, the Gaussian polynomials are defined by
\[    \begin{bmatrix}
    l\\m
\end{bmatrix}=\left\{ 
\begin{aligned}
    &\frac{(q)_l}{(q)_m(q)_{l-m}}&\text{for}\; 0\le m\le l\\
    &0&\text{otherwise},
\end{aligned} \right. \]
where $(q)_m:=\prod_{j=0}^{m} (1-q^{j})$ is the $q$-Pochhammer\index{q@$q$-Pochhammer} symbol. Here the restrictions $\mathbf{Q}$ on the sum are such that the arguments of the Gaussian polynomials are integers.

It has been demonstrated and verified in \parencite{berkovich1999universal,BouwknegtUCPF} that if we set $\mathbf{B}=\mathbf{G}$, then the exclusion statistics \eqref{eq:exstat} will lead to the UCPF \eqref{eq:UCPFgen} with $\mathbf{u}=\infty$, which reduces to the form of
\begin{equation}\label{eq:UCPF}
    S_{\mathbf{G}}[\mathbf{a}](\infty|q)=q^{\delta}\sum_{\substack{N_i\ge 0\\i=1,\dots,m}}\dfrac{q^{\frac{1}{2}\mathbf{N}^{\mathsf{T}}\mathbf{G}\mathbf{N}-\mathbf{a}\cdot \mathbf{N}}}{\prod_i (q)_{N_i}}
\end{equation}
where $q^{\delta}$ for some $\delta\in\mathbb{Q}$ takes account of the restrictions $\mathbf{Q}$, by noting the property of the Gaussian polynomials:
\[\lim_{l\rightarrow\infty}\begin{bmatrix}
    l\\m
\end{bmatrix}=\dfrac{1}{(q)_m}.\]
In this thesis, by a slight abuse of terminology, when we refer to UCPFs, we mostly mean those in the form of \eqref{eq:UCPF} and we will omit the parameter $\mathbf{u}$.

An equation that relates the central charge of a CFT to its corresponding statistical interaction matrix $\bold{G}$ has been derived in \parencite{BouwknegtUCPF}, which reads:
\begin{equation}\label{eq:dilog}
    \dfrac{\pi^2}{6}c=\sum_{a}Li(\xi_a),
\end{equation}
where $\xi_a$ are solutions to the equations
\[\xi_a=\prod_b(1-\xi_b)^{G_{ab}},\]
and $Li(x)$ is Roger's dilogarithm\index{dilogarithm} defined by
\[Li(x)=\dfrac{6}{\pi^2}\left[\sum_{n=1}^{\infty}\dfrac{x^n}{n^2}+\dfrac{1}{2} \text{ln} x\,\text{ln} (1-x)\right].\]
We note that \eqref{eq:dilog} provides the first step of verifying if a statistical interaction matrix $\bold{G}$ is the correct one for a CFT.

In \parencite{berkovich1999universal}, it was conjectured that all CFT characters can be written in the UCPF form \eqref{eq:UCPFgen} based on many known examples (see references therein). For instance, the character formulae of standard modules for affine Lie algebras and their coset spaces were studied in \parencite{georgiev1996}. In particular, they found identities relating string functions to UCPF forms, which was also conjectured in \parencite{KUNIBA_1993} from a Bethe ansatz analysis. We prove similar identities in Chapter \ref{Chapter4} for $\widehat{\mathfrak{sl}(4)}_{2}$ in the context of coupled free fermions. Furthermore, an algorithm for obtaining affine Lie algebra characters in UCPF form, in principle, for any affine Lie algebra $\hat{\mathfrak{g}}$ at an arbitrary level, was given in \parencite{Bou2000,BH2000}.


\section{Outline}

This thesis is organized as follows. 
In Chapter \ref{Chapter2}, we give a review of the Sugawara constructions in terms of free fermions and give a brief investigation of the relation between their characters and string functions. 

The notion of coupled free fermions will be introduced in Chapter \ref{Chapter3}, and we will analyse two families of coupled free fermion CFTs in Chapters \ref{Chapter3} and \ref{Chapter4}. The family discussed in Chapter \ref{Chapter3} is a special case of the coset parafermions, for which a self-contained introduction will be included there. We study the fusion rules of their modules, provide convenient bases of their representation spaces, calculate their characters and UCPFs, and present their decomposition into minimal model modules. 

In Chapter \ref{Chapter4}, we will focus on another family of coupled free fermion CFTs constructed from scaled root lattices and we will have a look at their $\mathbb{Z}_2$-orbifolds. Towards the end of Chapter \ref{Chapter4}, we will reveal a connection among the coset construction, the lattice construction, and the orbifold construction in terms of coupled free fermions by proving equivalences between their characters and UCPFs. We also find simple explicit forms of some string functions.

The last chapter contains a summary and remarks on our results. We also propose future directions to extend the topics discussed in this thesis.

\chapter{Free Fermions} 

\label{Chapter2} 

\section{Free fermion CFT}

A (real) free fermion\index{free fermion} field $\psi(z)$ satisfies the OPE 
\[\psi(z)\psi(w)\sim \frac{1}{z-w},\]
and has conformal dimension $\frac{1}{2}$ with respect to the\index{free fermion! energy-momentum tensor } energy-momentum tensor 
\[L (z) = -\frac{1}{2} :\psi(z)\partial \psi(z):\;.\]
It is also simple to verify that the central charge is $\frac{1}{2}$. 

The free fermion CFT with a single fermion is a simple current extension of the minimal model with central charge $\frac{1}{2}$ (also know as the Ising model). It admits representations with conformal dimension $h=0,\frac{1}{2},\frac{1}{16}$, whose highest weight states are the true vacuum $\ket{0}$, $\ket{\frac{1}{2}}:=\psi(0)\ket{0}$, and $\ket{\frac{1}{16}}:=\sigma(0)\ket{0}$ respectively, where $\sigma(z)$ is a spin field of conformal dimension $\frac{1}{16}$ such that
\[\psi(z)\sigma(w)\sim \dfrac{\sigma(w)}{(z-w)^{1/2}}.\]
When acting on $\ket{0}$ or $\ket{\frac{1}{2}}$, $\psi(z)$ is periodic and said to be in the Neveu-Schwarz(NS)-sector\index{free fermion!Neveu-Schwarz(NS)-sector}, while when acting on $\ket{\frac{1}{16}}$, it is anti-periodic and said to be in the Ramond(R)-sector\index{free fermion! Ramond(R)-sector}. The mode expansion of $\psi(z)$ is given by
\[\psi(z)=\sum_{n}\psi_{n} z^{-n-\frac{1}{2}},\]
where the sum is over half-integers (resp. integers) in the NS- (resp. R-) sector, and the modes satisfy the anti-commutation relations
\[\{\psi_n,\psi_m\}=\delta_{n+m,0}.\]
Consequently, the modes $L_n$ of the energy-momentum tensor are given by
\begin{align*}
    L_n=	\left\{
	\begin{aligned}
	&\frac{1}{2}\sum_{r\in\mathbb{Z}+\frac{1}{2}}\left(r+\frac{1}{2}n\right):\psi_{-r}\psi_{n+r}:&(NS),\\
&\frac{1}{2}\sum_{r\in\mathbb{Z}}\left(r+\frac{1}{2}n\right):\psi_{-r}\psi_{n+r}:+\frac{1}{16}\delta_{n,0}&(R).
	\end{aligned}
	\right.
\end{align*}
The characters\index{free fermion!character} $\chi_{h}$ for the NS-sector are determined straightforwardly by noticing that $\psi_r^2=0$ for any half-integer $r$. On the other hand, in the $R$-sector, we have that $\psi_0^2=\frac{1}{2}$, so we need two highest-weight states, denoted by $\ket{+}$ and $\ket{-}$ such that $\psi_0\ket{+}=\frac{1}{\sqrt{2}}\ket{-}$ and $\psi_0\ket{-}=\frac{1}{\sqrt{2}}\ket{+}$. We record the characters as follows:  
\begin{equation}\label{eq:freeferch}
    \begin{aligned}
        \chi_0(q)+\chi_{\frac{1}{2}}(q)&=\Tr q^{L_0-\frac{c}{24}}=q^{-\frac{1}{48}}\prod_{n=1}^{\infty}(1+q^{n-\frac{1}{2}}),\\
        \chi_0(q)-\chi_{\frac{1}{2}}(q)&=\Tr (-1)^F q^{L_0-\frac{c}{24}}=q^{-\frac{1}{48}}\prod_{n=1}^{\infty}(1-q^{n-\frac{1}{2}}),\\
        \chi_{\frac{1}{16}}(q)&=\Tr q^{L_0-\frac{c}{24}}=2q^{\frac{1}{24}}\prod_{n=1}^{\infty}(1+q^{n}),
    \end{aligned}
\end{equation}
where $F$ is the fermion number operator 
\[F=F_0+\sum_{k> 0} \psi_{-k}\psi_{k},\]
with $F_0$ an operator defined by $F_0\ket{0}=0$ and $F_0\psi_0\ket{0}=1$.

A detailed review of representations of the free fermion CFT has been presented in \parencite[Chp.7]{SchCFT}. 

\section{Free fermion equivalence}

Equivalences between certain WZW models\index{Wess-Zumino-Witten (WZW) model} and free fermion theories have been discussed since the appearance of the WZW theories. In \parencite{KZWZW}, it was argued that a level 1 $\mathfrak{u}(N)$ WZW model could be equivalent to a theory with $2N$ real free fermions. In \parencite{Witten1984WZW}, it was also believed that the level 1 $\mathfrak{so}(N)$ WZW model is equivalent to a theory with $N$ free fermions. These two kinds of equivalence were then proved at the quantum scope in \parencite{FUCHS1987392}, where it was also claimed that these cases exhausted the possibilities of the quantum equivalence between free fermion theories and WZW theories. 

A criterion for constructing free fermion theories from representations of Lie algebras was given in \parencite{GODDARD1985226}. Consider a real highest-weight representation, of dimension $d_{\lambda}$ and highest weight $\lambda$, for the Lie algebra $\mathfrak{g}$, consisting of anti-symmetric matrices $\{t^a|a=1,\dots,\dim\mathfrak{g}\}$:
\[[t^a,t^b]=f_{abc}t^c,\]
where $f_{abc}$ are the structure constants\index{structure constants} of $\mathfrak{g}$.

We introduce $d_{\lambda}$-dimensional real free fermion fields $\{\psi^{a}(z)\mid 1\le a\le \dim\mathfrak{g}\}$, defined on the unit circle in the complex plane. The (uncoupled) free fermions satisfy the OPE:
\[\psi^a(z)\psi^b(z)\sim \dfrac{\delta_{a,b}}{z-w}.\]
The currents $J^a(z)$ are constructed as follows:
\[J^a(z)=-\frac{1}{2}t^a_{bc}:\psi^{b}(z)\psi^{c}(z):,\]
where  $t_{b c}^{a}$ are the matrix entries of $t^{a}$, 
and then the Sugawara energy-momentum tensor $L(z)$ is defined as in \eqref{eq:WZWEnerMon}.
We can also introduce a fermionic energy-momentum tensor 
\[\tilde{L}(z)= - \frac{1}{2} \sum_{a} :\psi^{a}(z)\partial \psi^{a}(z):.\]
It can be shown that $L(z)=\tilde{L}(z)$ if the total antisymmetrisation of $t^a_{bc}t^a_{de}$ vanishes:
\[t^a_{bc}t^a_{de}+t^a_{bd}t^a_{ec}+t^a_{be}t^a_{cd}=0,\]
which is equivalent to the condition
\begin{equation}\label{eq:freefercond}
    \dfrac{2c_{\lambda}}{c_{\theta}+\kappa_{\lambda}}=1,
\end{equation}
where $c_{\lambda}$ is the scalar by which the quadratic Casimir operator acts on the $\lambda$-representation and can be calculated by
\[\sum_{a}t^at^a=-c_{\lambda}I,\]
$\kappa_{\lambda}$ is the Dynkin index of the representation $\lambda$ defined by 
\begin{equation}\label{eq:Dynkindex}
    \Tr(t^at^b)=-\kappa_{\lambda}\delta_{a,b},
\end{equation}
and $\theta$ is the highest root of $\mathfrak{g}$. Note that by putting $a=b$ in \eqref{eq:Dynkindex} and summing over $a$ we obtain
\begin{equation}
    d_{\lambda}c_{\lambda}=\kappa_{\lambda}\dim\mathfrak{g}.
\end{equation}

Since $d_{\theta}=\dim\mathfrak{g}$, we have $c_{\theta}=\kappa_{\theta}$. Hence,  \eqref{eq:freefercond} is manifestly satisfied for the adjoint representation. This is saying that the WZW model constructed from $\hat{\mathfrak{g}}$ at level $h^{\vee}$ can be represented by $\dim \mathfrak{g}$ free fermions \parencite{GODDARD1985111,FUCHS198730}. As explicitly stated in \parencite{KUW1990}, the field $J^a(z)$ in the current algebra of $\hat{\mathfrak{g}}_{h^{\vee}}$ can be written in terms of the fermions as
\[J^a(z)=-\frac{1}{\sqrt{2}}f_{abc}:\psi^b(z)\psi^c(z):.\]
Then using the generalized Wick theorem \big(see, for example, \parencite[App. A]{Bais1988extensions}\big), we have the OPE:
\[J^a(z)J^b(w)\sim \dfrac{h^{\vee}\delta_{a,b}}{(z-w)^2} +\dfrac{f_{abc}J^c(w)}{z-w},\]
which recovers \eqref{eq:WZWOPE}. The central charge is $\frac{\dim\mathfrak{g}}{2}$ given by \eqref{eq:WZWcc}, as desired. 

When the criterium \eqref{eq:freefercond} is not satisfied, there is another method given in \parencite{GODDARD1985226} of constructing free fermion theories, which was been developed to the later widely known GKO coset construction.\index{Goddard-Kent-Olive (GKO)}

\section{Free fermion characters and string functions}

The so-called string functions\index{string function} are defined in \parencite{KACPETERSON} as 
\begin{equation}\label{eq:strfun}
    c^{\Lambda}_{\lambda}(\tau)=q^{s_{\Lambda}(\lambda)}\sum_{n\ge0}\text{mult}_{\Lambda}(\lambda-n\delta)q^n,
\end{equation}
where $q=e^{2\pi i \tau}$ and
\begin{equation}\label{eq:strfunprefac}
s_{\Lambda}(\lambda)=s_{\Lambda}-\dfrac{|\lambda|^2}{2k},
\end{equation}
where
\[s_{\Lambda}=\dfrac{|\Lambda+\hat{\rho}|^2}{2(k+h^{\vee})}-\dfrac{|\hat{\rho}|^2}{2h^{\vee}}.\]
Here $\Lambda\in\hat{P}_+$ is a dominant integral weight, $\lambda\in\hat{P}$ is an integral weight, $\delta$ is the imaginary root, $k\in\mathbb{Z}_{>0}$ is the level of $\Lambda$, and $\hat{\rho}$ is the Weyl vector of $\hat{\mathfrak{g}}$ (see Appendix \ref{AppendixA} for details).

The following theorem shown in \parencite{KACPETERSON} describes modular transformation\index{modular transformation} properties of string functions.
\begin{theorem}\label{thm:strfunmodtrans}
    Let $\mathcal{L}(\Lambda)$ be a $\hat{\mathfrak{g}}$-module with highest weight $\Lambda\in\hat{P}_+$ of level $k\in\mathbb{Z}_{>0}$. Then

    $(1)$ \begin{equation}
        c^{\Lambda}_{\lambda}(-\dfrac{1}{\tau})=(-i\tau)^{-n/2}\sum_{\substack{\Lambda'\in\hat{P}_+\;\text{mod}\;\mathbb{C}\delta\\ \lambda'\in\hat{P}\;\text{mod}\;(kM+\mathbb{C}\delta)\\ \Lambda'(\hat{k})=\lambda'(\hat{k})=k}} \dfrac{i^{|\Delta_+|}}{|P/M|\left(k(k+h^{\vee})\right)^{n/2}}b^{\Lambda,\Lambda'}_{\lambda,\lambda'}c^{\Lambda'}_{\lambda'}(\tau)
    \end{equation}
    where $P$ is the set of integral weights of $\mathfrak{g}$ and
    \[b^{\Lambda,\Lambda'}_{\lambda,\lambda'}=\sum_{w\in W}(\det w)\exp\left(-\dfrac{2\pi i (\Lambda+\rho,w(\Lambda'+\rho))}{k+h^{\vee}}\right).\]

    $(2)$ $\eta(\tau)^{\dim\mathfrak{g}} c^{\Lambda}_{\lambda}(\tau)$ is a holomorphic modular cusp form of weight $|\Delta_+|$ for the group 
    $$\left\{\left(\begin{array}{cc}
         a& b \\
         c& d
    \end{array}\right)\in SL(2,\mathbb{Z})|c\equiv0\;\text{mod}\; (Nk)\bigwedge c\equiv 0\;\text{mod}\; (N(k+h^{\vee}))\right\},$$ where $N$ is the least positive integer such that $N|\mu|^2\in 2\mathbb{Z}$ for all $\mu\in P$ and $\eta(\tau)$ is the Dedekind eta function\index{Dedekind eta function}
    \[\eta(\tau)=e^{\frac{\pi i \tau}{12}}\prod_{n=1}^{\infty}(1-e^{2\pi i\tau n}).\]
\end{theorem}

Explicit expressions of string functions are hard to compute in general even with the algorithm  developed in \parencite{KACPETERSON}, based on the modular transformation properties above. Some explicit examples are included in \parencite{KACPETERSON}. We will obtain new explicit expressions in the following and in Chapter \ref{Chapter3}. 

In this section, we are interested in the free fermion theory arising from the adjoint representation of $\mathfrak{sl}(n+1)$ and we want to explore how we can obtain some string functions\footnote{We will use Dynkin labels to represent weights of $\widehat{\mathfrak{sl}(n+1)}$ wherever suitable, following conventions given in \parencite{francesco1997conformal} and adopt simplified notations for Dynkin labels as indices from \parencite{KACPETERSON}.} of $\widehat{\mathfrak{sl}(n+1)}_{h^{\vee}}$ from the characters calculated out of free fermions.

Observe that by \eqref{eq:WZWcondim} and the Freudenthal-de Vries very strange formula from \parencite{freudenthal}:
\[\dfrac{|\rho|^2}{2h^{\vee}}=\dfrac{\dim\mathfrak{g}}{24},\]
we see that in $\widehat{\mathfrak{sl}(n+1)}_{n+1}$ WZW model, the Weyl vector $\hat{\rho}$ has conformal dimension
\[h(\hat{\rho})=\dfrac{3|\rho|^2}{4h^{\vee}}=\dfrac{\dim\mathfrak{g}}{16},\]
which suggests that the character of the module with the highest weight $\hat{\rho}$ is the same as that of $\dim\mathfrak{g}$ copies of the free fermion in R-sector as given in \eqref{eq:freeferch}, that is,
\[\text{ch}_{\hat{\rho}}(\tau) \propto \prod_{k=1}^{\infty} (1+q^k)^{\dim\mathfrak{g}}.\]
Now, in addition to tracking the multiplicities of weights with $q=e^{2\pi i\tau}$, we shall also keep track of the root $\alpha$ associated with each free fermion $\psi^{\alpha}(z)$ with $e^{\alpha}$. For $\mathfrak{sl}(n+1)$, consider the following expression:
\begin{equation}\label{eq:ferch}
    q^{-\frac{n^2+2n}{48}}e^{\rho}\prod_{\alpha\in\Delta_{+}}(1+e^{-\alpha})\prod_{k=1}^{\infty}(1+q^k)^{n}(1+e^{\alpha}q^k)(1+e^{-\alpha}q^k),
\end{equation}
where the factor $e^{\rho}\prod_{\alpha\in\Delta_{+}}(1+e^{-\alpha})$ is inspired by the Macdonald-Weyl dominator identity from \parencite{macdonald1971affine}:
\[\sum_{w\in\hat{W}}\det(w) e^{w\hat{\rho}}=e^{\hat{\rho}}\prod_{\hat{\alpha}\in\hat{\Delta}_{+}}(1+e^{-\hat{\alpha}})^{\text{mult}(\hat{\alpha})}.\]
Let $\{\alpha_i|1\le i\le n\}$ be the set of simple roots of $\mathfrak{sl}(n+1)$, we set $z_i:=e^{\alpha_i}$. Then \eqref{eq:ferch} can be viewed as a function of variables $\{z_1,\dots,z_n\}$, denoted by $\chi_{R}^{(n)}(z_1,\dots,z_n;q)$. 

We expect that the coefficient of $e^{\rho}=(z_1\cdots z_n)^{\frac{n}{2}}$ of $\chi_{R}^{(n)}(z_1,\dots,z_n;q)$, as a function of $q$, denoted by $\mathcal{O}_{(z_1\cdots z_n)^{\frac{n}{2}}}^{\chi_{R}^{(n)}}$, equals the string function $c^{\hat{\rho}}_{\hat{\rho}}$ of $\widehat{\mathfrak{sl}(n+1)}_{n+1}$:
\begin{conjecture}\label{conj:freeferstrfun-R}
For any $\widehat{\mathfrak{sl}(n+1)}_{n+1}$, we have
\begin{equation}\label{eq:freeferstrfun-R}
    \mathcal{O}_{(z_1\cdots z_n)^{\frac{n}{2}}}^{\chi_{R}^{(n)}}=c^{\hat{\rho}}_{\hat{\rho}}.
\end{equation}
\end{conjecture}

For $\mathfrak{sl}(2)$, \eqref{eq:freeferstrfun-R} is simply a consequence of the Jacobi triple identity:
\begin{equation}\label{eq:JacobitriId}
    \prod_{k=1}^{\infty} (1-q^{k})(1-z^{-1}q^{k-\frac{1}{2}})(1-zq^{k-\frac{1}{2}})=\sum_{k\in\mathbb{Z}}(-1)^kz^kq^{\frac{1}{2}k^2}.
\end{equation}
\begin{corollary}
For $n=1$,
\[\mathcal{O}_{(z_1)^{\frac{1}{2}}}^{\chi_{R}^{(1)}}=c^{11}_{11}.\]
\end{corollary}
\begin{proof}
By \eqref{eq:JacobitriId} with $z\mapsto -zq^{\frac{1}{2}}$, we have
\begin{align}
       &\prod_{k=1}^{\infty} (1-q^{k})(1+z^{-1}q^{k-1})(1+zq^{k})\notag\\
       =&(1-z^{-1})\prod_{k=1}^{\infty} (1-q^{k})(1+z^{-1}q^{k})(1+zq^{k})=\sum_{k\in\mathbb{Z}}z^{k}q^{\frac{1}{2}k^2+\frac{1}{2}k},\label{eq:JacobitriId-1}
\end{align}
which gives
\begin{align*}
   \chi_{R}^{(1)}(z_1;q)=&q^{-\frac{1}{16}}(z_1^{\frac{1}{2}}+z_1^{-\frac{1}{2}})\prod_{k=1}^{\infty}(1+q^k)(1+z_1q^k)(1+z_1^{-1}q^k)\\
   =& q^{-\frac{1}{16}}\dfrac{\prod_{k=1}^{\infty}(1+q^{k})}{\prod_{k=1}^{\infty}(1- q^{k})}\sum_{k\in\mathbb{Z}}z_1^{k+\frac{1}{2}}q^{\frac{1}{2}(k^2+k)}
\end{align*}
Hence, taking $k=0$, we have 
\[\mathcal{O}_{(z_1)^{\frac{1}{2}}}^{\chi_{R}^{(1)}}(q)=q^{-\frac{1}{16}}\dfrac{\prod_{k=1}^{\infty}(1+q^{k})}{\prod_{k=1}^{\infty}(1- q^{k})}=\dfrac{\eta(2\tau)}{\eta(\tau)^2},\]
which is the same as $c^{11}_{11}$ given in \parencite[Example 3]{KACPETERSON}.
\end{proof}
The statement for $\mathfrak{sl}(3)$ is also not hard to prove with further manipulations of the Jacobi triple identity. 
\begin{corollary}\label{cor:sl3ferstr}
For $n=2$,
\[\mathcal{O}_{(z_1z_2)^1}^{\chi_{R}^{(2)}}=c^{111}_{111}.\]
\end{corollary}
\begin{proof}
By \eqref{eq:JacobitriId-1}, we have
\begin{align*}
    \chi_{R}^{(2)}(z_1,z_2;q)=&q^{-\frac{1}{6}}\dfrac{\prod_{k=1}^{\infty}(1+q^{k})^2}{\prod_{k=1}^{\infty}(1- q^{k})^3}\\
    \times&\sum_{k_1,k_2,k_{12}\in\mathbb{Z}}z_1^{k_1+\frac{1}{2}}z_2^{k_2+\frac{1}{2}}(z_1z_2)^{k_{12}+\frac{1}{2}}q^{\frac{1}{2}(k_1^2+k_2^2+k_{12}^2+k_1+k_2+k_{12})}.
\end{align*}
Then, taking $k_1=k_2=-k_{12}$, we have
\[\mathcal{O}_{(z_1z_2)^1}^{\chi_{R}^{(2)}}(q)=q^{-\frac{1}{6}}\dfrac{\prod_{k=1}^{\infty}(1+q^{k})^2}{\prod_{k=1}^{\infty}(1- q^{k})^3}\sum_{k_{12}\in\mathbb{Z}}q^{\frac{3}{2}k_{12}^2-\frac{1}{2}k_{12}}.\]
Note that by \eqref{eq:JacobitriId} with $(q,z)\mapsto(q^3,-q^{-\frac{1}{2}})$, we have
\begin{align*}
    \sum_{k\in\mathbb{Z}}q^{\frac{3}{2}k^2-\frac{1}{2}k}=&\prod_{k=1}^{\infty}(1-q^{3k})(1+q^{3k-1})(1+q^{3k-2}).
\end{align*}
Hence,
\begin{align*}
    \mathcal{O}_{(z_1z_2)^1}^{\chi_{R}^{(2)}}(q)=&q^{-\frac{1}{6}}\dfrac{\prod_{k=1}^{\infty}(1+q^{k})^2}{\prod_{k=1}^{\infty}(1- q^{k})^3}\prod_{k=1}^{\infty}(1-q^{3k})(1+q^{3k-1})(1+q^{3k-2})\\
    =&q^{-\frac{1}{6}}\dfrac{\prod_{k=1}^{\infty}(1+q^k)^3(1-q^{3k})}{\prod_{k=1}^{\infty}(1- q^{k})^3(1+q^{3k})}\\
    =&\dfrac{\eta(2\tau)^3\eta(3\tau)^2}{\eta(\tau)^6\eta(6\tau)},
\end{align*}
which is the same as $c^{111}_{111}$ given in \parencite[Example 4]{KACPETERSON}.
\end{proof}

As for the case of  $\mathfrak{sl}(4)$, the explicit expression of $c^{1111}_{1111}$ is not known but can be conjectured based on its $q-$expansion. We aim to prove that
\begin{corollary}\label{cor:sl4ferstr}
   \[\mathcal{O}_{(z_1z_2z_3)^{\frac{3}{2}}}^{\chi_{R}^{(3)}}(q)=\dfrac{\eta(2\tau)^{15}}{\eta(\tau)^{14}\eta(4\tau)^4}.\]
\end{corollary}
\begin{proof}
    By \eqref{eq:JacobitriId-1}, we have
\begin{align*}
    \chi_{R}^{(3)}(z_1,z_2,z_3;q)=&q^{-\frac{5}{16}}\dfrac{\prod_{k=1}^{\infty}(1+q^{k})^3}{\prod_{k=1}^{\infty}(1- q^{k})^6}\\
    \times&\sum_{k_1,k_2,k_{12}\in\mathbb{Z}}z_1^{k_1+\frac{1}{2}}z_2^{k_2+\frac{1}{2}}z_3^{k_3+\frac{1}{2}}(z_1z_2)^{k_{12}+\frac{1}{2}}(z_2z_3)^{k_{23}+\frac{1}{2}}(z_1z_2z_3)^{k_{123}+\frac{1}{2}}\\
     \times&q^{\frac{1}{2}(k_1^2+k_2^2+k_3^2+k_{12}^2+k_{23}^2+k_{123}^2+k_1+k_2+k_3+k_{12}+k_{23}+k_{123})}.
\end{align*}
Then, taking $k_1=-k_{12}-k_{123},k_2=-k_{12}-k_{23}-k_{123}$ and $k_3=-k_{23}-k_{123}$, we have
\begin{align*}
    \mathcal{O}_{(z_1z_2z_3)^{\frac{3}{2}}}^{\chi_{R}^{(3)}}(q)=&q^{-\frac{5}{16}}\dfrac{\prod_{k=1}^{\infty}(1+q^{k})^3}{\prod_{k=1}^{\infty}(1- q^{k})^6}\\
\times&\sum_{k_{12},k_{23},k_{123}\in\mathbb{Z}}q^{\frac{1}{2}(3k_{12}^2+3k_{23}^2+4k_{123}^2+2k_{12}k_{23}+4k_{12}k_{123}+4k_{23}k_{123}-k_{12}-k_{23}-2k_{123}}).
\end{align*}
Now it amounts to show that 
\[\sum_{k_{12},k_{23},k_{123}\in\mathbb{Z}}q^{\frac{1}{2}(3k_{12}^2+3k_{23}^2+4k_{123}^2+2k_{12}k_{23}+4k_{12}k_{123}+4k_{23}k_{123}-k_{12}-k_{23}-2k_{123}})=\dfrac{(q^2)_{\infty}^{12}}{(q)_{\infty}^5(q^4)_{\infty}^4},\]
whose left-hand side can be rewritten as
\[\sum_{l,m\in\mathbb{Z}}\sum_{\substack{k\in\mathbb{Z}\\k\,\equiv\,l+m\mod 2}}q^{\binom{k}{2}+2\binom{l}{2}+2\binom{m}{2}+l+m},\]
which, by considering the parities of $l,m$ and $k$, can be further written as
\begin{align*}
    &\sum_{u,v,w\in\mathbb{Z}}q^{\binom{2w}{2}+2\binom{2u}{2}+2\binom{2v}{2}+2u+2v}\\
    +&\sum_{u,v,w\in\mathbb{Z}}q^{\binom{2w}{2}+2\binom{2u+1}{2}+2\binom{2v+1}{2}+(2u+1)+(2v+1)}\\
    +&\sum_{u,v,w\in\mathbb{Z}}q^{\binom{2w+1}{2}+2\binom{2u+1}{2}+2\binom{2v}{2}+(2u+1)+2v}\\
    +&\sum_{u,v,w\in\mathbb{Z}}q^{\binom{2w+1}{2}+2\binom{2u}{2}+2\binom{2v+1}{2}+2u+(2v+1)}\\
    :=&S_1+S_2+S_3+S_4.
\end{align*}
Using \eqref{eq:JacobitriId} and \eqref{eq:JacobitriId-1} with appropriate substitutions of $z$ and $q$, we have the following results:
\begin{align*}
    S_1=\dfrac{(q^2)_{\infty}^2(q^8)_{\infty}^{10}}{(q)_{\infty}(q^4)_{\infty}^4(q^{16})_{\infty}^4},&\quad\quad\quad
    S_2=\dfrac{4q^2(q^2)_{\infty}^2(q^{16})_{\infty}^{4}}{(q)_{\infty}(q^8)_{\infty}^2},\\
    S_3=S_4=&\dfrac{2q(q^2)_{\infty}^2(q^8)_{\infty}^{4}}{(q)_{\infty}(q^4)_{\infty}^2}.
\end{align*}
Therefore,
\begin{align*}
    &\sum_{k_{12},k_{23},k_{123}\in\mathbb{Z}}q^{\frac{1}{2}(3k_{12}^2+3k_{23}^2+4k_{123}^2+2k_{12}k_{23}+4k_{12}k_{123}+4k_{23}k_{123}-k_{12}-k_{23}-2k_{123}})\\
    =&\dfrac{(q^2)_{\infty}^2(q^8)_{\infty}^{10}}{(q)_{\infty}(q^4)_{\infty}^4(q^{16})_{\infty}^4}+\dfrac{4q^2(q^2)_{\infty}^2(q^{16})_{\infty}^{4}}{(q)_{\infty}(q^8)_{\infty}^2}+\dfrac{4q(q^2)_{\infty}^2(q^8)_{\infty}^{4}}{(q)_{\infty}(q^4)_{\infty}^2}\\
    =&\dfrac{(q^2)_{\infty}^2(q^8)_{\infty}^{10}}{(q)_{\infty}(q^4)_{\infty}^4(q^{16})_{\infty}^4}\left(1+\dfrac{2q(q^4)_{\infty}^4(q^{16})_{\infty}^{4}}{(q^8)_{\infty}^6}\right)^2\\
    =&\dfrac{(q^2)_{\infty}^2(q^8)_{\infty}^{10}}{(q)_{\infty}(q^4)_{\infty}^4(q^{16})_{\infty}^4}\left(\dfrac{(q^4)_{\infty}^2(q^{16})_{\infty}^2}{(q^8)_{\infty}^5}\right)^2\left(\dfrac{(q^8)_{\infty}^5}{(q^4)_{\infty}^2(q^{16})_{\infty}^2}+\dfrac{2q(q^{16})_{\infty}^{2}}{(q^8)_{\infty}}\right)^2.
\end{align*}
Recall the 2-dissection formula of Ramanujan's theta function \parencite[Eq. (1.9.4)]{Hir2007}:
\[\dfrac{(q^8)_{\infty}^5}{(q^4)_{\infty}^2(q^{16})_{\infty}^2}+\dfrac{2q(q^{16})_{\infty}^{2}}{(q^8)_{\infty}}=\dfrac{(q^2)_{\infty}^5}{(q)_{\infty}^2(q^{4})_{\infty}^2},\]
then we have
\begin{align*}
    &\sum_{k_{12},k_{23},k_{123}\in\mathbb{Z}}q^{\frac{1}{2}(3k_{12}^2+3k_{23}^2+4k_{123}^2+2k_{12}k_{23}+4k_{12}k_{123}+4k_{23}k_{123}-k_{12}-k_{23}-2k_{123}})\\
    =&\dfrac{(q^2)_{\infty}^2(q^8)_{\infty}^{10}}{(q)_{\infty}(q^4)_{\infty}^4(q^{16})_{\infty}^4}\left(\dfrac{(q^4)_{\infty}^2(q^{16})_{\infty}^2}{(q^8)_{\infty}^5}\right)^2\left(\dfrac{(q^2)_{\infty}^5}{(q)_{\infty}^2(q^{4})_{\infty}^2}\right)^2\\
    =&\dfrac{(q^2)_{\infty}^{12}}{(q)_{\infty}^5(q^4)_{\infty}^4},
\end{align*}
as desired.
\end{proof}

It can be seen that the proof for the $\mathfrak{sl}(4)$ case is more subtle than lower rank cases. We hope that the method used in this proof can work for higher rank cases. For $\mathfrak{sl}(n+1)$ with $n>3$, we expect that $c^{\hat{\rho}}_{\hat{\rho}}$ are linear combinations of products of Dedekind eta-functions. Although explicit expressions for higher rank cases are not clear yet, comparing the $q$-expansions of $\mathcal{O}_{(z_1\cdots z_n)^{\frac{n}{2}}}^{\chi_{R}^{(n)}}$ and $c^{\hat{\rho}}_{\hat{\rho}}$ for larger $n$ still gives evidences for \eqref{eq:freeferstrfun-R}. 

For the next step, it is natural to wonder if we could have an analogous result for the NS-sector. We shall experiment with lower rank $\mathfrak{sl}(n+1)$. 

Consider the following expressions according to the first two equations in \eqref{eq:ferch}:
\begin{equation}\label{eq:ferch-NS}
    q^{-\frac{n^2+2n}{48}}\prod_{k=1}^{\infty}(1\pm q^{k-\frac{1}{2}})^{n}\prod_{\alpha\in\Delta_{+}}(1\pm e^{\alpha}q^{k-\frac{1}{2}})(1\pm e^{-\alpha}q^{k- \frac{1}{2}}),
\end{equation}
which again can be viewed as functions of variables $\{z_1,\dots,z_n\}$ and denoted by $\chi^{(n_{\pm})}_{NS}(z_1,\dots,z_n;q)$. We look at the coefficient of $(z_1\cdots z_n)^{0}$ of $\chi_{NS}^{(n_{\pm})}(z_1,\dots,z_n;q)$, as a function of $q$, denoted by $\mathcal{O}_{(z_1\cdots z_n)^{0}}^{\chi_{NS}^{(n_{\pm})}}$. Here we shall focus on $\chi^{(n_{-})}_{NS}$ and omit the minus sign in the notations.

Again with the Jacobi triple identity \eqref{eq:JacobitriId}, we easily obtain, for $n=1$, that 
\[\mathcal{O}_{(z_1)^{0}}^{\chi_{NS}^{(1)}}(q)=\dfrac{\eta(\frac{\tau}{2})}{\eta(\tau)^2},\]
which is the same as $c^{20}_{20}-c^{20}_{02}$ given in \parencite[Example 3]{KACPETERSON}. With Theorem \ref{thm:strfunmodtrans} we notice that 
\[\dfrac{\eta(\frac{\tau}{2})}{\eta(\tau)^2}=c^{20}_{20}(\tau)-c^{20}_{02}(\tau)=c^{11}_{11}\left(-\frac{1}{\tau}\right).\]

By a similar argument as the proof of Corollary \ref{cor:sl3ferstr}, we find, for $n=2$, that
\begin{equation}\label{eq:ferNSsl3}
    \mathcal{O}_{(z_1z_2)^{0}}^{\chi_{NS}^{(2)}}(q)=\dfrac{\eta(\frac{\tau}{2})^2\eta(\frac{3\tau}{2})^2}{\eta(\tau)^5\eta(3\tau)}.
\end{equation}
It is however not obvious how this expression relates to string functions of $\widehat{\mathfrak{sl}(3)}_{3}$. Motivated by Theorem \ref{thm:strfunmodtrans}, we consider the modular transformation of \eqref{eq:ferNSsl3} under $\tau\mapsto -1/\tau$ using the well-known property of $\eta(\tau)$, that is
\begin{equation}\label{eq:etamodtrans}
    \eta\left(-\frac{1}{\tau}\right)=(-i\tau)^{\frac{1}{2}}\eta(\tau).
\end{equation}
Then we see that
\begin{equation}\label{eq:sl3ferstrNS}
    \mathcal{O}_{(z_1z_2)^{0}}^{\chi_{NS}^{(2)}}(-\frac{1}{\tau})=\dfrac{2\eta(2\tau)^2\eta(\frac{2\tau}{3})^2}{\eta(\tau)^5\eta(\frac{\tau}{3})}=2c^{111}_{111}(\tau)+c^{111}_{300}(\tau).
\end{equation}

For $n=3$, with a similar argument as the proof of Corollary \ref{cor:sl4ferstr}, we can obtain that
\begin{align*}
   &\sum_{k_{12},k_{23},k_{123}\in\mathbb{Z}}(-1)^{k_{12}+k_{23}}q^{\frac{1}{2}(3k_{12}^2+3k_{23}^2+4k_{123}^2+2k_{12}k_{23}+4k_{12}k_{123}+4k_{23}k_{123}})\\
    =&\dfrac{(q^4)_{\infty}(q^8)_{\infty}^{8}}{(q^2)_{\infty}^2(q^{16})_{\infty}^4}+\dfrac{4q^2(q^{16})_{\infty}^4(q^{4})_{\infty}^{5}}{(q^2)_{\infty}^2(q^{8})_{\infty}^4}-\dfrac{8q^{\frac{2}{3}}(q^8)_{\infty}^6}{(q^4)_{\infty}^3}\\
    =&\dfrac{(q^4)_{\infty}^5}{(q^2)_{\infty}^2(q^{8})_{\infty}^2}\left(\dfrac{(q^8)_{\infty}^{10}}{(q^4)_{\infty}^4(q^{16})_{\infty}^4}+\dfrac{4q^2(q^{16})_{\infty}^4}{(q^{8})_{\infty}^2}\right)-\dfrac{8q^{\frac{2}{3}}(q^8)_{\infty}^6}{(q^4)_{\infty}^3}.
\end{align*}
Using the 2-dissection formula of Ramanujan's theta function \parencite[Eq. (1.10.1)]{Hir2007}:
\[\dfrac{(q^4)_{\infty}^{10}}{(q^2)_{\infty}^4(q^{8})_{\infty}^4}+\dfrac{4q(q^{8})_{\infty}^{4}}{(q^4)_{\infty}^2}=\dfrac{(q^2)_{\infty}^{10}}{(q)_{\infty}^4(q^{4})_{\infty}^4},\]
we then have
\begin{align*}
    \mathcal{O}_{(z_1z_2z_3)^{0}}^{\chi_{NS}^{(3)}}(q)
    =&\dfrac{\eta(\frac{\tau}{2})^3}{\eta(\tau)^9}\left(\dfrac{\eta(4\tau)^{15}}{\eta(2\tau)^6\eta(8\tau)^6}-\dfrac{8\eta(8\tau)^6}{\eta(4\tau)^3}\right)
\end{align*}
Once again it is hard to observe any relation between this expression and the string functions of $\widehat{\mathfrak{sl}(4)}_{4}$, but with the modular transformation we see that
\[\dfrac{\eta(2\tau)^3}{4\eta(\tau)^9}\left(\dfrac{\eta(\frac{\tau}{4})^{15}}{\eta(\frac{\tau}{2})^6\eta(\frac{\tau}{8})^6}-\dfrac{\eta(\frac{\tau}{8})^6}{\eta(\frac{\tau}{4})^3}\right)=3c^{1111}_{1111}(\tau)+2c^{1111}_{2200}(\tau)+3c^{1111}_{3010}(\tau).\]

It is not yet clear how to formulate a precise statement about the characters of the NS-sector as that of the R-sector, but at least we can conclude that 
\begin{conjecture}\label{conj:freeferstrfun-NS}
    For any $\widehat{\mathfrak{sl}(n+1)}$, the modular transformation of 
   $\mathcal{O}_{(z_1\cdots z_n)^{0}}^{\chi_{NS}^{(n)}}(q)$ under $\tau\mapsto -1/\tau$ can be written as a linear combination of string functions $c^{\Lambda}_{\lambda}(\tau)$ of $\widehat{\mathfrak{sl}(n+1)}_{n+1}$ with $\Lambda=\hat{\rho}$.
\end{conjecture}

One can integrate the results in Conjecture \ref{conj:freeferstrfun-R} and \ref{conj:freeferstrfun-NS} into the algorithm given in \parencite{KACPETERSON} to help calculate the string functions. For example, combining \eqref{eq:sl3ferstrNS} with Corollary \ref{cor:sl3ferstr}, we have 
\[c^{111}_{300}(\tau)=\dfrac{2\eta(2\tau)^2\eta(\frac{2\tau}{3})^2}{\eta(\tau)^5\eta(\frac{\tau}{3})}-\dfrac{2\eta(2\tau)^3\eta(3\tau)^2}{\eta(\tau)^6\eta(6\tau)}.\]
Now transform $c^{111}_{300}$ under $\tau\mapsto -1/\tau$ using Theorem \ref{thm:strfunmodtrans}, we obtain 
\begin{align*}
    &c^{300}_{300}(\tau)+6c^{300}_{111}(\tau)+2c^{300}_{030}(\tau)-2c^{111}_{111}(\tau)-c^{111}_{300}(\tau)
    =\dfrac{3\eta(\frac{\tau}{2})^2\eta(\frac{3\tau}{2})^2}{\eta(\tau)^5\eta(3\tau)}-\dfrac{2\eta(\frac{\tau}{2})^3\eta(\frac{\tau}{3})^2}{\eta(\tau)^6\eta(\frac{\tau}{6})},
\end{align*}
which then gives an expression of $c^{300}_{300}(\tau)+6c^{300}_{111}(\tau)+2c^{300}_{030}(\tau)$. Together with the following expressions given in \parencite[Example 4]{KACPETERSON}:
\begin{align*}
    c^{300}_{300}(\tau)-c^{300}_{030}(\tau)&=\dfrac{1}{\eta(\tau)\eta(3\tau)},\\
    c^{300}_{300}(\tau)-3c^{300}_{111}(\tau)+2c^{300}_{030}(\tau)+c^{111}_{111}(\tau)-c^{111}_{300}(\tau)&=\dfrac{\eta(\frac{\tau}{2})^3\eta(\frac{\tau}{3})^2}{\eta(\tau)^6\eta(\frac{\tau}{6})},
\end{align*}
one can write down separate explicit expressions of $c^{300}_{300},c^{300}_{111},c^{300}_{030},c^{111}_{111},c^{111}_{300}$ purely in terms of Dedekind eta functions, which were not obtained in \parencite{KACPETERSON}. 

\chapter{Coupled Free Fermions: The Coset Construction} 

\label{Chapter3} 


\section{The coset construction}
Given an affine Lie algebra $\hat{\mathfrak{g}}$ at level $k\in\ZZ_{>0}$, one can construct a parafermionic conformal field theory containing parafemionic fields, which are non-local, fractional spin, analytic fields\index{parafermionic!conformal field theory}, as described in \parencite{GEPNER87}. Algebraically, this theory can be thought of as the GKO coset construction\index{coset construction} of the (generalized) WZW model\index{Wess-Zumino-Witten (WZW) model} based on $\hat{\mathfrak{g}}_k$ constrained by $\widehat{\mathfrak{u}(1)}^n$, denoted by $\hat{\mathfrak{g}}_k/\widehat{\mathfrak{u}(1)}^n$, where $n$ is the rank of $\mathfrak{g}$ \parencite{GEPNER89,Bratchikov_2000}.

We shall review the details of this construction in a slightly different language. For simplicity, we only focus on the holomorphic part. We also assume that $\mathfrak{g}$ is simply-laced.

Associate a parafermionic field $\psi^{(\alpha)}$ to each $\alpha\in Q$, where $Q$ is the root lattice of $\mathfrak{g}$, and identify $\psi^{(\alpha)}$ and $\psi^{(\beta)}$ if $\alpha\equiv\beta\mod kQ$. 
Among these parafermionic fields, we are particularly interested in the so-called generating parafermions\index{parafermionic!generating parafermion} $\{\psi^{(\alpha)}\mid \alpha\in\Delta\}$, where $\Delta$ is the set of roots of $\mathfrak{g}$. The OPEs of generating parafermions are to be determined so that the following fields
\begin{align*}
    J^{\alpha}(z)&=\sqrt{\dfrac{2k}{\alpha^2}}c_\alpha\psi_{\alpha}e^{i\alpha\cdot\phi(z)/\sqrt{k}},\;\;\alpha\in \Delta,\\
    J^{i}(z)&=\dfrac{2i\sqrt{k}}{\alpha_i^2}\alpha_i\cdot\partial_z\phi(z)\;,\;\;\alpha_i\in\Delta_0,
\end{align*}
where $c_\alpha$ are cocycles, $\phi(z)$ is the vector of $n$ free bosons and $\Delta_0$ is the set of simple roots of $\mathfrak{g}$, obey the current \index{Wess-Zumino-Witten (WZW) model!current} algebra commutation relations 
\[J^{a}(z)J^b(w)\sim\dfrac{k\kappa(a,b)}{(z-w)^2}+\dfrac{f_{abc}J^c(w)}{z-w},\]
where $J^a(z)$ stands for either $J^{\alpha}(z)$ or $J^{i}(z)$ defined above, $\kappa(a,b)$ is the Killing form and $f_{abc}$ are structure constants of $\mathfrak{g}$ in the Chevalley basis.

It turns out that the cocycles and OPEs are determined to be
\begin{align}
    c_{\alpha}c_{-\alpha}=1,\;\;\;\;c_{\alpha}c_{\beta}=S_{\alpha,\beta}c_{\alpha+\beta},\notag\\
    \psi^{(\alpha)}(z)\psi^{(-\alpha)}(w)\sim\dfrac{1}{(z-w)^{2-|\alpha|^2/k}},\label{eq:paraOPE-1}\\
     \psi^{(\alpha)}(z)\psi^{(\beta)}(w)\sim\dfrac{c_{\alpha,\beta}\psi^{(\alpha+\beta)}(w)}{(z-w)^{1+(\alpha,\beta)/k}},\label{eq:paraOPE-2}
\end{align}
where $S_{\alpha,\beta}$ are some roots of unity and $c_{\alpha,\beta}$ are some constants to be fixed by the Borcherds identity \eqref{eq:paraBorId}.

Denote the energy-momentum tensor of the current algebra in the Sugawara form \eqref{eq:WZWEnerMon} by $L^{(c)}(z)$ and that of the (uncoupled free) bosonic systems by $L^{(b)}(z)$. The energy-momentum tensor $L(z)$ of the parafermionic system is then $$L(z)=L^{(c)}(z)-L^{(b)}(z)$$
and the central charge\index{coset construction!central charge}\index{parafermionic!central charge} of the parafermionic system is
\begin{equation}\label{eq:PFcc}
    c\left(\dfrac{\hat{\mathfrak{g}}_k}{\widehat{\mathfrak{u}(1)}^n}\right)= c\left(\hat{\mathfrak{g}}_k\right)-c\left(\widehat{\mathfrak{u}(1)}^n\right)=\dfrac{k\dim\mathfrak{g}}{k+h^{\vee}}-n,
\end{equation}
where we have used \eqref{eq:WZWcc} and \eqref{eq:cosetcc}.

The conformal dimension of a parafermionic field $\psi^{(\alpha)}$ with respect to $L(z)$ is given by
\begin{equation}\label{eq:genPFcfwt}
    h(\alpha)=-\dfrac{|\alpha|^2}{2k}+n(\alpha),
\end{equation}
where $n(\alpha)$ is the minimal number of roots of $\mathfrak{g}$ from which $\alpha$ is composed. The mode expansions of parafermionic fields $\psi^{(\alpha)}$ are in the form of 
\begin{equation}\label{eq:paramode}
    \psi^{(\alpha)}(z)=\sum_{n\in \epsilon(\alpha)+\mathbb{Z}} \psi^{(\alpha)}_n z^{-n-h(\alpha)},
\end{equation}
where the twisting $\epsilon\in\mathbb{R}/\mathbb{Z}$ depends on relevant OPEs. We may introduce the unitarity by the standard conjugation relation:
\[\left(\psi^{(\alpha)}_n\right)^{\dag}=\psi^{(\alpha)}_{-n},\]
which is consistent with the generalised commutation relations \eqref{eq:gcommrel}. 

One can derive generalised commutation relations between the modes of parafermionic fields from the Borcherds identity\index{Borcherds identity} for parafermionic fields:\index{parafermionic!generalised commutation relation}
\begin{align}\label{eq:paraBorId}
  &\sum_{j\ge0}\binom{m}{j}[[AB]_{n+1+j}C]_{m+k+1-j} \notag\\
   =&(-1)^j\binom{n}{j}\sum_{j\ge0}[A[BC]_{k+1+j}]_{m+n+1-j}-\mu_{AB}(-1)^{\alpha_{AB}-n}[B[AC]_{m+1+j}]_{n+k+1-j},
\end{align}
where $n\equiv\alpha_{AB}\mod\mathbb{Z}$, $ m\equiv\alpha_{AC}\mod\mathbb{Z}$, $k\equiv\alpha_{BC}\mod\mathbb{Z}$, $[AB]_n$ is the $n-th$ product of fields $A$ and $B$ as in the OPE
\[A(z)B(w)=\dfrac{1}{(z-w)^{\alpha_{AB}}}\left([AB]_{\alpha_{AB}}(w)+[AB]_{\alpha_{AB}-1}(w)+[AB]_{\alpha_{AB}-2}(w)+\cdots\right),\]
the singularity $\alpha_{AB}$ can be calculated fro mconfomal dimensions of $A$ and $B$, and the commutation factors $\mu_{AB}$ is introduced by the parafermionic mutual locality axiom:
\begin{equation}\label{eq:OPEcomm}
A(z)B(w)(z-w)^{\alpha_{AB}}=\mu_{AB}B(w)A(z)(w-z)^{\alpha_{AB}}.
\end{equation}

The parafermionic primary fields\index{parafermionic!primary field} of $\hat{\mathfrak{g}}_k/\widehat{\mathfrak{u}(1)}^n$, denoted by $\Phi^{\Lambda}_{\lambda}$, are characterised by $\Lambda\in \hat{P}^{(k)}_+$, a dominant integral weight at level $k$, and an integral weight $\lambda\in \hat{P}$ such that $\Lambda-\lambda\in\hat{Q}$, where $\hat{Q}$ is the root lattice of $\hat{\mathfrak{g}}$. We notice that $\Phi^{\Lambda}_{\lambda}$ and $\Phi^{\Lambda'}_{\lambda'}$ are identified if $$\Lambda'=\Lambda\;\;\text{and}\;\;\lambda'\equiv \lambda\mod k\hat{Q}$$ or $$\Lambda'=\sigma\Lambda\;\;\text{and}\;\;\lambda'=\sigma\lambda,$$
where $\sigma\in\text{Aut}(Dyn(\mathfrak{g}))$. The modules of $\hat{\mathfrak{g}}_k/\widehat{\mathfrak{u}(1)}^n$, denoted by $\mathcal{L}^{\Lambda}_{\lambda}$, are therefore characterised and identified in the same manner.

The generating parafermions $\psi^{(\alpha)}$ act on modules $\mathcal{L}^{\Lambda}_{\lambda}$ as chiral vertex operators (CVOs). Consider a generic CVO \index{chiral vertex operator (CVO)} $$\phi^{\Lambda^{(i)}}_{\lambda^{(i)}}\binom{i}{k\;j}(z):\mathcal{L}^{\Lambda^{(j)}}_{\lambda^{(j)}}\mapsto \mathcal{L}^{\Lambda^{(k)}}_{\lambda^{(k)}},$$
whose mode expansion is, according to \eqref{eq:CVOmode},
\begin{equation}\label{eq:CVOexp}
    \phi^{\Lambda^{(i)}}_{\lambda^{(i)}}\binom{i}{k\;j}(z)=\sum_{n\in\mathbb{Z}}\phi^{\Lambda^{(i)}}_{\lambda^{(i)}}\binom{i}{k\;j}_{n-(h^{\Lambda^{(k)}}_{\lambda^{(k)}}-h^{\Lambda^{(j)}}_{\lambda^{(j)}})}z^{-n+(h^{\Lambda^{(k)}}_{\lambda^{(k)}}-h^{\Lambda^{(j)}}_{\lambda^{(j)}}-h^{\Lambda^{(i)}}_{\lambda^{(i)}})},
\end{equation}
where $h^{\Lambda}_{\lambda}$ is the conformal dimension\index{coset construction!conformal dimension}\index{parafermionic!conformal dimension} of the field $\Phi^{\Lambda}_{\lambda}$, given by
\begin{equation}\label{eq:PFcfwt}
    h^{\Lambda}_{\lambda}=\dfrac{(\Lambda,\Lambda+2\rho)}{2(k+h^\vee)}-\dfrac{|\lambda|^2}{2k}+n^{\Lambda}_{\lambda},
\end{equation}
where $n^{\Lambda}_{\lambda}$ is some integer.
\begin{remark}
If $\Lambda=k\Lambda_0$, then $n^{\Lambda}_{\lambda}$ may be described as the minimal number of finite roots from which the finite part of $\lambda$ is composed. If $\lambda$ is a weight in the representation $\Lambda$, then $n^{\Lambda}_{\lambda}=0$. 
\end{remark}

The number of such CVOs are determined by the fusion rules\index{coset construction! fusion rule} of $\hat{\mathfrak{g}}_k/\widehat{\mathfrak{u}(1)}^n$, which are given  by 
\begin{equation}\label{eq:parafus}
    \Phi^{\Lambda^{(i)}}_{\lambda^{(i)}}\times \Phi^{\Lambda^{(j)}}_{\lambda^{(j)}}=\sum_k \mathcal{N}_{ij}^k\, \Phi^{\Lambda^{(k)}}_{\lambda^{(i)}+\lambda^{(j)}\mod k\hat{Q}}
\end{equation}
where $\mathcal{N}_{ij}^k$ are fusion rules of $\hat{\mathfrak{g}}_k$. 

When a generating parafermion $\psi^{(\alpha)}$ acts on a certain module, it plays the role of some $\phi^{k\Lambda_0}_{\lambda}\binom{i}{k\;j}$ with the finite part of $k\Lambda_0-\lambda$ being $\alpha$ and $\binom{i}{k\;j}$ determined by fusion rules \eqref{eq:parafus}. For this reason, we may also determine the twisting $\epsilon$ in \eqref{eq:paramode} by \eqref{eq:CVOexp}.

The characters\index{coset construction!character}\index{parafermionic!character} of $\mathcal{L}^{\Lambda}_{\lambda}$, denoted by $b^{\Lambda}_{\lambda}$, are defined in the standard way, that is,
\[b^{\Lambda}_{\lambda}(\tau):=\text{Tr}|_{\mathcal{L}^{\Lambda}_{\lambda}}\;e^{\left(L_0-c/24\right)2\pi i\tau}\]
where $L_0$ is the zeroth mode of the energy-momentum tensor $L(z)$ of the parafermionic system. Recall that $L(z)=L^{(c)}(z)-L^{(b)}(z)$, we have, as shown in \parencite{GEPNER87},
\begin{equation}\label{eq:cosetstrfun}
    b^{\Lambda}_{\lambda}(\tau)=\dfrac{\text{Tr}|_{\mathcal{L}^{\Lambda}_{\lambda}}\;e^{\left(L_0^{(c)}-c(\hat{\mathfrak{g}}_k)/24\right)2\pi i\tau}}{\text{Tr}|_{\mathcal{L}^{\Lambda}_{\lambda}}\;e^{\left(L_0^{(b)}-n/24\right)2\pi i\tau}}=\eta(\tau)^{n} c^{\Lambda}_{\lambda}(\tau),
\end{equation}
where $\eta(\tau)$ is the Dedekind eta function and $c^{\Lambda}_{\lambda}(\tau)$ are the string functions \eqref{eq:strfun} of $\hat{\mathfrak{g}}_k$.



\section{Coupled free fermions}
We notice that when $\mathfrak{g}$ is simply-laced and $k=2$, this construction is of particular interest because then the OPEs \eqref{eq:paraOPE-1} and \eqref{eq:paraOPE-2} for generating parafermions reduce to
\begin{align}
\notag\\
    &\psi^{(\alpha)}(z)\psi^{(\alpha)}(w)\sim\dfrac{1}{(z-w)},\label{eq:freefer}\\
     &\psi^{(\alpha)}(z)\psi^{(\beta)}(w)\sim\dfrac{c_{\alpha,\beta}\psi^{(\alpha+\beta)}(w)}{(z-w)^{1/2}},\;\;\text{if}\;\alpha+\beta\in\Delta.\label{eq:couplefer}
\end{align}
Keeping in mind that $\alpha\equiv -\alpha\mod 2Q$ and the conformal dimension of any generating parafermion $\psi^{(\alpha)}$ is $\frac{1}{2}$ according to \eqref{eq:genPFcfwt}, we see that \eqref{eq:freefer} tells us that we have a set of real free fermions $\{\psi^{(\alpha)}\mid \alpha\in \Delta_+\}$ and \eqref{eq:couplefer} tells us that these fermions are coupled to each other according to the root structure of $\mathfrak{g}$. Therefore, we will call such a system a "\textbf{coupled free fermion} CFT" hereafter. \index{coupled free fermion}

It is not hard to see that when $\mathfrak{g}=\mathfrak{sl}_2$, the coupled free fermion CFT simply recovers the free fermion CFT, in which sense coupled free fermion CFTs are natural generalisations of the free fermion CFT. 

For a coupled free fermion CFT $\widehat{\mathfrak{sl}(n+1)}_2/\widehat{\mathfrak{u}(1)}^{n}$, the twisting $\epsilon$ in the mode expansion \eqref{eq:paramode} can only be $0$ or $\frac{1}{2}$, corresponding to NS- or R-sector respectively. The generalised commutation relations\footnote{Also known as $Z$-algebra relations \parencite{Lepowsky1984}.}  between the modes of coupled free fermions derived from \eqref{eq:paraBorId} are given in \parencite{DING1994,BorisPF} as follows:
\begin{equation}\label{eq:gcommrel}
    \begin{aligned}
    \psi^{(\alpha)}_n\psi^{(\alpha)}_m+\psi^{(\alpha)}_m\psi^{(\alpha)}_n&=\delta_{m+n,0},&\\[0.7ex]
    \psi^{(\alpha)}_n\psi^{(\beta)}_m+\mu_{\alpha,\beta}\psi^{(\beta)}_m\psi^{(\alpha)}_n&=0,&\text{if}\;\alpha+\beta\not\in\Delta,\\
    \sum_{l\ge0}\binom{l-\frac{1}{2}}{l}\left(\psi^{(\alpha)}_{m-\frac{1}{2}-l}\psi^{(\beta)}_{n+\frac{1}{2}+l}+\mu_{\alpha,\beta}\psi^{(\beta)}_{n-l}\psi^{(\alpha)}_{m+l}\right)&=c_{\alpha,\beta}\psi^{(\alpha+\beta)}_{m+n},&\text{if}\;\alpha+\beta\in\Delta,
    \end{aligned}
\end{equation}
where the commutator factors $\mu_{\alpha,\beta}$ are some roots of unity that satisfy $$\mu_{\alpha,\beta}\mu_{\beta,\alpha}=1.$$

Looking at \eqref{eq:couplefer}, we see that if we interchange $\psi^{(\alpha)}$ and $\psi^{(\beta)}$, then by the axiom \eqref{eq:OPEcomm}, we expect that
\[\psi^{(\beta)}(z)\psi^{(\alpha)}(w)=(-1)^{-\frac{1}{2}}\mu_{\alpha,\beta}\psi^{(\alpha)}(w)\psi^{(\beta)}(z)\sim \dfrac{\mu_{\alpha,\beta}c_{\alpha,\beta}\psi^{(\alpha+\beta)}(z)}{(-1)^{1/2}(w-z)^{1/2}}\sim\dfrac{\mu_{\alpha,\beta}c_{\alpha,\beta}\psi^{(\alpha+\beta)}(w)}{(z-w)^{1/2}},\]
while by direct calculation we have
\[\psi^{(\beta)}(z)\psi^{(\alpha)}(w)\sim \dfrac{c_{\beta,\alpha}\psi^{(\alpha+\beta)}(w)}{(z-w)^{1/2}},\]
so we have the condition 
$$\mu_{\alpha,\beta}c_{\alpha,\beta}=c_{\beta,\alpha}.$$

It is also calculated in \parencite{DING1994,BorisPF} using \eqref{eq:paraBorId} that when $\mathfrak{g}=\mathfrak{sl}(n+1)$, the energy-momentum tensor of any free fermion subalgebra $$L^{(\alpha)}(z):=-\frac{1}{2}:\psi^{(\alpha)}(z)\partial\psi^{(\alpha)}(z):$$ acting on another fermion $\psi^{(\beta)}(w)$ gives 
\[L^{(\alpha)}(z)\psi^{(\beta)}(w)\sim\dfrac{\frac{1}{16}\psi^{(\beta)}(w)}{(z-w)^2}+\dfrac{\partial\psi^{(\beta)}(w)}{z-w}.\]
Hence the total energy-momentum tensor\index{coupled free fermion!energy-momentum tensor} with the desired central charge \eqref{eq:CFFcc} can be written in terms of coupled free fermions with a carefully chosen overall factor as
\begin{equation}\label{eq:CFFEM}
    L(z)=\frac{4}{n+3}\sum_{\alpha\in \Delta_+}L^{(\alpha)}(z)=-\frac{2}{n+3}\sum_{\alpha\in \Delta_+}:\psi^{(\alpha)}(z)\partial\psi^{(\alpha)}(z):
\end{equation}
whose mode expansion is then given by
\[L(z)=\sum_{n}L_nz^{-n-2},\]
with
\[L_n=\frac{4}{n+3}\sum_{\alpha\in \Delta_+}L_n^{(\alpha)},\]
where
\[L_n^{(\alpha)}=	\left\{
	\begin{aligned}
	&\frac{1}{2}\sum_{r\in\mathbb{Z}+\frac{1}{2}}\left(r+\frac{1}{2}n\right):\psi^{(\alpha)}_{-r}\psi^{(\alpha)}_{n+r}:&(NS),\\
&\frac{1}{2}\sum_{r\in\mathbb{Z}}\left(r+\frac{1}{2}n\right):\psi^{(\alpha)}_{-r}\psi^{(\alpha)}_{n+r}:+\frac{1}{16}\delta_{n,0}&(R).
	\end{aligned}
	\right.\]
Here, whether the mode numbers are integers or half-integers is determined by \eqref{eq:CVOexp} depending on which sector the fermion $\psi^{(\alpha)}(z)$ lies in for a particular module. 

It is easily checked that the modes of $L(z)$ form the Virasoro algebra with central charge
\begin{equation}\label{eq:CFFcc}
    c\left(\dfrac{\widehat{\mathfrak{sl}(n+1)}_2}{\widehat{\mathfrak{u}(1)}^{n}}\right)=\dfrac{n(n+1)}{n+3},
\end{equation}
which agrees with \eqref{eq:PFcc}. We notice that this number is the same as the sum of the first $n$ central charges of the discrete series of unitary minimal models\index{minimal model}, that is,
\begin{equation}\label{eq:CFFMINcc}
    c\left(\dfrac{\widehat{\mathfrak{sl}(n+1)}_2}{\widehat{\mathfrak{u}(1)}^{n}}\right)=\sum_{m=1}^{n} \left(1-\dfrac{6}{(m+2)(m+3)}\right).
\end{equation}
Indeed, it has been verified in \cite{BelGep} the coset parafermionic CFTs $\widehat{\mathfrak{sl}(n+1)}_2/\widehat{\mathfrak{u}(1)}^{n}$ can be decomposed into minimal models, taking these cosets as special cases of the AGT correspondence. We will also investigate this kind of decomposition in the examples below.



\section{Examples}\label{sec:cosetexp}
The example of $\widehat{\mathfrak{sl}(2)}_2/\widehat{\mathfrak{u}(1)}$ has been studied in \parencite{francesco1997conformal}. We shall study the structure of $\widehat{\mathfrak{sl}(n+1)}_2/\widehat{\mathfrak{u}(1)}^n$ with $n=2,3$ in detail, explore how their modules decompose to those of minimal models, and give UCPF forms of their characters.

\subsection{$\widehat{\mathfrak{sl}(3)}_2/\widehat{\mathfrak{u}(1)}^2$}
Let $\alpha_1,\alpha_2$ denote the simple roots of $\mathfrak{sl}(3)$. Then the positive roots of $\mathfrak{sl}(3)$ are $\{\alpha_1,\alpha_2,\alpha_3:=\alpha_1+\alpha_2\}$ and we have 3 coupled free fermions $\psi^{(1)}:=\psi^{(\alpha_1)}, \psi^{(2)}:=\psi^{(\alpha_2)}$ and $\psi^{(3)}:=\psi^{(\alpha_3)}$.

For $\widehat{\mathfrak{sl}(3)}_2/\widehat{\mathfrak{u}(1)}^2$, due to the simple structure of $Q/2Q\cong\mathbb{Z}_2\times\mathbb{Z}_2$, the OPEs \eqref{eq:freefer} and \eqref{eq:couplefer}, and the generalised commutation relations \eqref{eq:gcommrel} can be summarised as
\begin{equation}\label{eq:sl3OPE}
    \begin{aligned}
    &\psi^{(i)}(z)\psi^{(i)}(w)\sim\dfrac{1}{(z-w)},\\
     &\psi^{(i)}(z)\psi^{(j)}(w)\sim\dfrac{c_{ij}\psi^{(k)}(w)}{(z-w)^{1/2}},\;\;\text{if}\;k\equiv i+j\bmod 2Q,
    \end{aligned}
\end{equation}
and
\begin{align}
    \psi^{i}_n\psi^{i}_m+\psi^{i}_m\psi^{i}_n&=\delta_{m+n,0},&\label{eq:sl3commrel-1}\\
    \sum_{l\ge0}\binom{l-\frac{1}{2}}{l}\left(\psi^{(i)}_{m-\frac{1}{2}-l}\psi^{(j)}_{n+\frac{1}{2}+l}+\mu_{ij}\psi^{(j)}_{n-l}\psi^{(i)}_{m+l}\right)&=c_{ij}\psi^{(k)}_{m+n},&k\equiv i+j\bmod 2Q, \label{eq:sl3commrel-2}
\end{align}
respectively, with constants $c_{ij}:=c_{\alpha_i,\alpha_j}$ and $\mu_{ij}:=\mu_{\alpha_i,\alpha_j}$ given in \parencite{BorisPF} as follows:
\begin{equation}\label{eq:sl3parameter}
    \begin{aligned}
    &c_{ij}=\mu_{ij}c_{ji}, \; \mu_{ij}\mu_{ji}=1\\
    &\mu_{12}=\mu_{23}=\mu_{31}=x^2,\\
    &c_{12}=c_{23}=c_{31}=\frac{x}{\sqrt{2}},
    \end{aligned}
\end{equation}
where $x$ is an 8th root of unity, which we may choose to be $e^{-\frac{i\pi}{4}}$ here.

The total energy-momentum tensor $L(z)$, in this case, can be written as
\[L(z)=-\frac{2}{5}\sum_{i=1}^3:\psi^{(i)}(z)\partial\psi^{(i)}(z):\]
and by \eqref{eq:CFFcc}, the central charge is
\[ c\left(\dfrac{\widehat{\mathfrak{sl}(3)}_2}{\widehat{\mathfrak{u}(1)}^2}\right)=\dfrac{2(2+1)}{2+3}=\dfrac{6}{5}.\]

Now we search for all inequivalent primary fields \footnote{Note that here (and in the next subsection) we follow physicist's convention and only classify modules obtained from the branching rules but not all coset modules in more precise mathematical sense.} $\Phi^{\Lambda}_{\lambda}$ (hence also all inequivalent modules $\mathcal{L}^{\Lambda}_{\lambda}$). Given the criteria that $\Phi^{\Lambda}_{\lambda}=\Phi^{\sigma\Lambda}_{\sigma\lambda}$, where $\sigma$ is a permutation in $\mathbb{Z}_3$ (as a cyclic subgroup of $S_3$) in this case, we see that for candidates of highest weights $\Lambda$, it is sufficient to consider the set $\{[2,0,0],[1,1,0]\}$. Then by noting that $Q/2Q$ has 4 cosets which can be represented by $\{[0,0,0],[-1,2,-1],[-1,-1,2],$ $[-2,1,1]\}$, we know that there are 8 inequivalent primary fields in total as shown in Table \ref{tab:sl3prifield} with their conformal dimensions $h^{\Lambda}_{\lambda}$ calculated using \eqref{eq:PFcfwt} and some ad hoc fusion channel labels for convenience.
\begingroup
\renewcommand*{\arraystretch}{1.5}
\begin{table}[h!]
    \centering
      \begin{tabular}[c]{c|cccccccc}
        $\Phi^{\Lambda}_{\lambda}$ & $\Phi^{200}_{200}$ & $\Phi^{200}_{12-1}$&$\Phi^{200}_{1-12}$ &$\Phi^{200}_{011}$ &$\Phi^{110}_{002}$& $\Phi^{110}_{-121}$ &$\Phi^{110}_{110}$ & $\Phi^{110}_{2-11}$   \\
    \hline 
        $h^{\Lambda}_{\lambda}$ & 0&$\frac{1}{2}$&$\frac{1}{2}$&$\frac{1}{2}$&$\frac{3}{5}$&$\frac{1}{10}$&$\frac{1}{10}$&$\frac{1}{10}$\\
    \hline
    Label & [0] & [1] & [2] & [3] & [$\tau$] & [$1'$] & [$2'$]  & [$3'$]  
    \end{tabular}
    \caption{Inequivalent primary fields in $\widehat{\mathfrak{sl}(3)}_2/\widehat{\mathfrak{u}(1)}^2$}
    \label{tab:sl3prifield}
\end{table}
\endgroup
\newline We say the modules with $\Lambda=[2,0,0]$ are in the "untwisted sector" and those with $\Lambda=[1,1,0]$ are in the "twisted sector". The relevant fusion rules for our purpose are summarised in Table \ref{tab:sl3PFfus}.

\begingroup
\renewcommand*{\arraystretch}{1.5}
\begin{table}[h!]
    \centering
      \begin{tabular}[c]{c|cccccccc}
        $\times$  & [0] & [1] & [2] & [3] & [$\tau$] & [$1'$] & [$2'$]  & [$3'$]  \\
    \hline
      $[1]$ & [1] & [0] & [3] & [2] &  [$1'$] & [$\tau$] & [$3'$] & [$2'$] \\
      $[2]$ & [2] & [3] & [0] & [1] &  [$2'$] & [$3'$] & [$\tau$] & [$1'$] \\
      $[3]$ & [3] & [2] & [1] & [0] &  [$3'$] & [$2'$]  &  [$1'$] & [$\tau$]
    \end{tabular}
    \caption{Subset of fusion rules for $\widehat{\mathfrak{sl}(3)}_2/\widehat{\mathfrak{u}(1)}^2$}
    \label{tab:sl3PFfus}
\end{table}
\endgroup

\begin{remark}
Note that when $\mathfrak{g}=\mathfrak{sl}(n)$ and $k=2$, for which $N_{ij}^k$ is either 0 or 1, the indices $\binom{i}{k\;j}$ in \eqref{eq:CVOexp} are uniquely determined and can always be understood in context, so we will omit these indices hereafter. 
\end{remark}

The expressions of string functions\index{string function} of $\widehat{\mathfrak{sl}(3)}_2$ are given in \parencite{KACPETERSON}, and therefore by \eqref{eq:cosetstrfun}, the characters $b^{\Lambda}_{\lambda}$ of the modules $\mathcal{L}^{\Lambda}_{\lambda}$ are given by:
\begin{align*}
   b_{200}^{200}(\tau) &=\eta(\tau)^{-2}q^{\frac{1}{30}}\big[(q^\frac{1}{2})_\infty(q,q^{\frac{3}{2}},q^{\frac{5}{2}};q^{\frac{5}{2}})_\infty+ q^{\frac{1}{2}}(q^2;q^2)_\infty(q^2,q^{8},q^{10};q^{10})_\infty\big],\\
	b_{011}^{200}(\tau)&=\eta(\tau)^{-2}q^{\frac{8}{15}}(q^2)_\infty(q^2,q^{8},q^{10};q^{10})_\infty,\\
	b_{110}^{110}(\tau)&=\eta(\tau)^{-2}q^{\frac{2}{15}}(q^2)_\infty(q^4,q^{6},q^{10};q^{10})_\infty,\\
	b_{002}^{110} (\tau)&=\eta(\tau)^{-2}q^{\frac{2}{15}}\big[(q^2)_\infty(q^4,q^{6},q^{10};q^{10})_\infty-(q^\frac{1}{2};q^\frac{1}{2})_\infty(q^{\frac{1}{2}},q^{2},q^{\frac{5}{2}};q^{\frac{5}{2}})_\infty\big],
\end{align*}
where the $q$-Pochhammer\index{q@$q$-Pochhammer} symbols are defined for $n\in\mathbb{N}\cup\{\infty\}$:
\begin{align*}
	(a;q)_n:=\prod_{j=0}^{n-1} (1-aq^{j} ), \quad\quad &(q)_n:=(q;q)_n,\\
	(a_1, a_2, \ldots, a_r;q)_{n} := (a_1;q)_n & (a_2;q)_n \cdots (a_r;q)_n.
\end{align*}

According to the fusion rules, we see that, for example, when $\psi^{(1)}$ acts on $\mathcal{L}^{200}_{200}$, it behaves as the CVO $\phi^{200}_{12-1}:\mathcal{L}^{200}_{200}\mapsto \mathcal{L}^{200}_{12-1}$ and when $\psi^{(2)}$ acts on $L^{110}_{110}$, it behaves as the CVO $\phi^{200}_{1-12}:\mathcal{L}^{110}_{110}\mapsto \mathcal{L}^{110}_{2-11}$. One observes that the group of coupled free fermions acts transitively on both the set $\left\{[0], [1],[2], [3]\right\}$ and the set $\left\{[\tau], [1'],[2'], [3']\right\}$.

Because all fusion number $\mathcal{N}^i_{jk}\le 1$ for $\widehat{\mathfrak{sl}(n+1)}_2/\widehat{\mathfrak{u}(1)}^{n}$, solving the pentagon equations \eqref{eq:pent} and hexagon equations \eqref{eq:hex} for coupled free fermions (or more precisely for the set of fields with $\Lambda=2\Lambda_0$) becomes doable. We focus on the fusion rules of the untwisted sector. Given $i,j,k\in\{0,1,2,3\}$, all other indices of the F-matrix
$\left(F^l_{jki}\right)_{pq}$ are uniquely determined by the rules
$\begin{array}{ccc} [p]=[i]\times [j], &[q]=[j]\times [k], &[l]=[i]\times [q]=[p]\times [k].
\end{array}$
This further implies that given $q,c,p,r\in\{0,1,2,3\}$, all other indices in the pentagon equation are uniquely determined and the sum can be ignored. Similarly, the triple $\{p,r,q\}$ is enough to determine the hexagon equation.

We may choose the phases of F-matrices so that 
\begin{equation}\label{eq:Fmat}
    \left(F_{jki}^l\right)_{pq}=1,\quad\text{if any of $p,q,j,k,i,l$ is $0$},
\end{equation}
and use \eqref{eq:Rmat}. 
 Then we can solve the set of pentagon and hexagon equations for $\widehat{\mathfrak{sl}(3)}_2/\widehat{\mathfrak{u}(1)}^2$ and the solution is simply:
\begin{align}\label{eq:sl3RF}
&\left(F^i_{ijj}\right)_{kk}=\left(F^2_{211}\right)_{33}=\left(R^1_{23}\right)^2=-1, \quad\text{for $i\ne j$ and $k=i\times j$},
\end{align}
as one can expect considering the $\mathbb{Z}_3$ symmetry of the fusion channels $\{[1],[2],[3]\}$. This solution also confirms 
the constants given in \eqref{eq:sl3parameter} for the generalised commutation relations.

To study the modules $\mathcal{L}^{\Lambda}_{\lambda}$, we first need to find a basis for the state space of $\widehat{\mathfrak{sl}(3)}_2/\widehat{\mathfrak{u}(1)}^2$. In \parencite{Ard2002}, the author found a basis that involves the modes of parafermions associated with simple roots, including positive modes.\footnote{It was only done in the paper for the untwisted sector, but it can be done analogously for twisted sector.} In \parencite{BorisPF}, the author claimed that a kind of Poincare-Birkhoff-Witt (PBW) theorem should hold, but he was not able to state it. However, with our understanding of coupled free fermions in terms of chiral vertex operators together with the explicit generalised commutation relations, it is straightforward to state a basis of PBW type in terms of non-positive modes of all coupled free fermions.

Let us focus on the untwisted sector for now and let $\ket{0}$ be the (true) vacuum state with $L_0\ket{0}=0$ and $\psi^{(i)}_{n}\ket{0}=0$ for any $i=1,2,3$ and $n\ge 0$. 

We intend to build it with a special ordering of applying modes to the vacuum that is all $\psi^{(1)}$-modes applying first, followed by all $\psi^{(2)}$-modes and then all $\psi^{(3)}$-modes. Therefore we firstly consider the standard basis of a free (Majorana) fermion in terms of $\psi^{(1)}$-modes consisting of all states in the form of
\[\psi^{(1)}_{-N+\frac{1}{2}-s_N}\dots\psi^{(1)}_{-\frac{3}{2}-s_2}\psi^{(1)}_{-\frac{1}{2}-s_1}\ket{0}\]
with $N\ge0$, $s_N\ge\dots s_2\ge s_1>0$. 


Now for any state involving $\psi^{(2)}$-modes, the mode numbers of $\psi^{(2)}$-modes will be either integer or half-integer depending on the number of $\psi^{(1)}$-modes applied to the vacuum. Explicitly speaking, when the number of $\psi^{(1)}$-modes is odd, $\psi^{(2)}$ behaves as the CVO from sector $[1]$ to $[3]$, whose mode numbers are integers, while when the number of $\psi^{(1)}$-modes is even, $\psi^{(2)}$ behaves as the CVO from sector $[0]$ to $[2]$, whose mode numbers are half-integers. With this observation, it is reasonable to consider the states in the form of
\begin{equation}\label{eq:sl3basis-12}
    \psi^{(2)}_{-N_2-\frac{N_1}{2}+\frac{1}{2}-s^{(2)}_{N_2}}\dots\psi^{(2)}_{-\frac{N_1}{2}-\frac{1}{2}-s^{(2)}_{1}}\psi^{(1)}_{-N_1+\frac{1}{2}-s^{(1)}_{N_1}}\dots\psi^{(1)}_{-\frac{1}{2}-s^{(1)}_{1}}\ket{0}
\end{equation}
with $N_1,N_2\ge0$ and $s^{(i)}_N\ge\dots s^{(i)}_2\ge s^{(i)}_1\ge0$ for $i=1,2$, which automatically takes care of the integer or half-integer nature of the mode numbers of $\psi^{(2)}$-modes. 

We try a similar process to apply any $\psi^{(3)}$-modes, that is, consider the states in the form of
\begin{equation}\label{eq:sl3basis-123}
    \begin{aligned}
      \psi^{(3)}_{-N_3-\frac{N_1+N_2}{2}+\frac{1}{2}-s^{(3)}_{N_3}}\dots\psi^{(3)}_{-\frac{N_1+N_2}{2}-\frac{1}{2}-s^{(3)}_{1}}&\psi^{(2)}_{-N_2-\frac{N_1}{2}+\frac{1}{2}-s^{(2)}_{N_2}}\dots\\
    &\dots\psi^{(2)}_{-\frac{N_1}{2}-\frac{1}{2}-s^{(2)}_{1}}\psi^{(1)}_{-N_1+\frac{1}{2}-s^{(1)}_{N_1}}\dots\psi^{(1)}_{-\frac{1}{2}-s^{(1)}_{1}}\ket{0}
    \end{aligned}
\end{equation}
with $N_1,N_2,N_3\ge0$ and $s^{(i)}_N\ge\dots s^{(i)}_2\ge s^{(i)}_1\ge0$ for $i=1,2,3$. 

The hard part is to make sure that the set of all such states is linearly independent. This is the reason that we do not include any $\psi^{(2)}$-mode whose mode number is larger than $\frac{N_1}{2}-\frac{1}{2}$ in \eqref{eq:sl3basis-12}. For example, if we include $\psi^{(2)}_{0}\psi^{(1)}_{-\frac{1}{2}}\ket{0}$ in our "basis", then the generalised commutation relation \eqref{eq:sl3commrel-2} with $i=2,j=1,m=\frac{1}{2},n=-1$ gives that
\begin{align*}
    &\sum_{l\ge 0}\binom{l-\frac{1}{2}}{l}\left(\psi^{(2)}_{0-l}\psi^{(1)}_{-\frac{1}{2}+l}+\mu_{21}\psi^{(1)}_{-1-l}\psi^{(2)}_{\frac{1}{2}+l}\right)\ket{0}=c_{21}\psi^{(3)}_{-\frac{1}{2}}\ket{0}\\
    \implies&\psi^{(2)}_{0}\psi^{(1)}_{-\frac{1}{2}}\ket{0}=c_{21}\psi^{(3)}_{-\frac{1}{2}}\ket{0},
\end{align*}
using the fact that $\psi^{(i)}_{n}\ket{0}=0$ for any $i=1,2,3$ and $n\ge 0$. But this means our "basis" is not linearly independent, which is a contradiction. 


It is believed that one can inductively prove both the independence and the spanning property of our proposed basis with extensive manipulations of \eqref{eq:sl3commrel-2}. It would be rather tedious to introduce such a proof here, so we may instead claim that this set is a correct basis using the dimension theorem so that we only need to do an inductive proof for the spanning property. The dimensionality can be proven by a combinatorial argument similar to that in \parencite{Ard2002}, but connecting to a different UCPF \index{universal chiral partition function (UCPF)} form, which we claim to be (up to an overall factor) 
\begin{equation}\label{eq:sl3UCPF}
  \sum_{N_i\ge 0}\dfrac{q^{\frac{1}{2}\mathbf{N}^{\mathsf{T}}\cdot\mathbf{G}_3\cdot\mathbf{N}}}{\prod_i (q)_{N_i}},
\end{equation}
where $\bold{N}=(N_1,N_2,N_3)^T$, $(q)_{N_i}=\prod_{j=1}^{N_i}(1-q^j)$ and \[\bold{G}_3=\begin{pmatrix} 
    2& 1& 1\\
    1& 2& 1\\
    1& 1& 2
    \end{pmatrix}.\]

Note that the lowest energy state in the form of \eqref{eq:sl3basis-123} with $N_1$ $\psi^{(1)}$-modes, $N_2$ $\psi^{(2)}$-modes and $N_3$ $\psi^{(3)}$-modes is
\[\psi^{(3)}_{-N_3-\frac{N_1+N_2}{2}+\frac{1}{2}}\dots\psi^{(3)}_{-\frac{N_1+N_2}{2}-\frac{1}{2}}\psi^{(2)}_{-N_2-\frac{N_1}{2}+\frac{1}{2}}\dots\psi^{(2)}_{-\frac{N_1}{2}-\frac{1}{2}}\psi^{(1)}_{-N_1+\frac{1}{2}}\dots\psi^{(1)}_{-\frac{1}{2}}\ket{0}\]
whose energy ($L_0$-eigenvalue) is
\begingroup
\allowdisplaybreaks
\begin{align*}
      &\frac{1}{2}\left(\sum_{k=1}^{N_1}(2k-1)+\sum_{k=1}^{N_2}(N_1+2k-1)+\sum_{k=1}^{N_3}(N_1+N_2+2k-1)\right)\\
    =&\frac{1}{2}\left(N_1^2+N_2^2+N_3^2+N_1N_2+N_1N_3+N_2N_3\right)\\
    =&\frac{1}{2}(N_1,N_2,N_3)\cdot\frac{1}{2}\begin{pmatrix} 
    2& 1& 1\\
    1& 2& 1\\
    1& 1& 2
    \end{pmatrix}\cdot\begin{pmatrix} 
    N_1\\
    N_2\\
    N_3
    \end{pmatrix},
\end{align*}
\endgroup
which gives the term $q^{\frac{1}{2}\bold{N}^{T}\cdot\bold{G_3}\cdot\bold{N}}$ in the UCPF \eqref{eq:sl3UCPF}. 

The set $\{s^{(i)}_j:1\le j\le N_i\}$ can be viewed as a partition of a non-negative integer with $N_i$ parts and then the condition $s^{(i)}_{N_i}\ge\dots s^{(i)}_2\ge s^{(i)}_1\ge0$ is saying that each partition should be counted exactly once. In \eqref{eq:sl3UCPF}, such partitions are taken into account by the generating function $(q)_{N_i}$.

We can do a plausibility check using the dilogarithm\index{dilogarithm} \eqref{eq:dilog} with $\bold{G}=\bold{G}_3$. We verified that the central charge $c$ here is indeed $\frac{6}{5}$, as desired.

The UCPF \eqref{eq:sl3UCPF} should be the sum of characters of all modules in the untwisted sector, if our basis is correct, so it amounts to show that 
\begin{equation}\label{eq:sl3id-untwi}
    q^{-\frac{1}{20}}\sum_{N_i\ge 0}\dfrac{q^{\frac{1}{2}\mathbf{N}^{\mathsf{T}}\cdot\mathbf{G}_3\cdot\mathbf{N}}}{\prod_i (q)_{N_i}}=b_{200}^{200}+3b_{011}^{200},
\end{equation}
where we used the fact that $b_{011}^{200}=b_{12-1}^{200}=b_{1-12}^{200}$ considering \eqref{eq:strfunprop}. Since explicit expressions of string functions of $\widehat{\mathfrak{sl}(3)}_2$ are known \parencite{KACPETERSON}, proving this identity is possible. Indeed this has been done with the help of Shane Chern and the result, together with many other important results relevant to this thesis, are presented in our joint paper \parencite{BCH}. 

As for the spanning property, we prove by strong induction that the set of states in the form of \eqref{eq:sl3basis-123} spans the Fock space of the untwisted sector of $\widehat{\mathfrak{sl}(3)}_2/\widehat{\mathfrak{u}(1)}^2$.

\begin{proof}[Proof of the spanning property of \eqref{eq:sl3basis-123}]
We say a state is of length $n$ if its total number of modes is $n$. We say a state in the Fock space of the untwisted sector of $\widehat{\mathfrak{sl}(3)}_2/\widehat{\mathfrak{u}(1)}^2$ of length $n$ is well-ordered if it is $0$ or a linear combination of the states in the form of \eqref{eq:sl3basis-123} of length less or equal than $n$. Let $P(n)$ be the statement that any state in the Fock space of the untwisted sector of $\widehat{\mathfrak{sl}(3)}_2/\widehat{\mathfrak{u}(1)}^2$ with $n$ modes can be well-ordered.

$P(1)$ is obvious since we have $\psi^{(i)}_{m}\ket{0}=0$ for any $i=1,2,3$ and $m\ge 0$. Now suppose $P(n)$ for all $n<N$, we intend to prove $P(N)$. With this inductive hypothesis, we can ignore the terms on the right-hand side of \eqref{eq:sl3commrel-1} and \eqref{eq:sl3commrel-2} in the following discussion. The exact coefficients in the generalised commutation relations are not crucial to this discussion either. We use the notation $\sim$ to suppress such non-crucial information and view \eqref{eq:sl3commrel-1} and \eqref{eq:sl3commrel-2} as 
\begin{align}
    &\psi^{(i)}_n \psi^{(i)}_m \sim  \psi^{(i)}_m \psi^{(i)}_n \label{eq:sl3com-sim-1},\\
    &\psi^{(i)}_{r}\psi^{(j)}_{s}\sim\sum_{l\ge 1}\psi^{(i)}_{r-l}\psi^{(j)}_{s+l}+\sum_{l\ge 0}\psi^{(j)}_{s-\frac{1}{2}-l}\psi^{(i)}_{r+\frac{1}{2}+l}\label{eq:sl3com-sim-2}.
\end{align}
We observe two facts about \eqref{eq:sl3com-sim-1} and \eqref{eq:sl3com-sim-2}:
\begin{enumerate}[label=\textbf{F.\arabic*}]
    \item They will not change the number of $\psi^{(i)}$-modes for each $i=1,2,3$.\label{fact:sl3-basis-1}
    \item They will not change the total mode number of a state. \label{fact:sl3-basis-2}   
\end{enumerate}

Let $\ket{\psi}:=\psi^{(i_N)}_{m_N}\psi^{(i_{N-1})}_{m_{N-1}}\cdots \psi^{(i_1)}_{m_1}\ket{0}$ be a state with $N$ modes. Because $P(N-1)$, we can assume that $\ket{\psi'}:=\psi^{(i_{N-1})}_{m_{N-1}}\cdots \psi^{(i_1)}_{m_1}\ket{0}$ is in the form of \eqref{eq:sl3basis-123} with $N_1$ $\psi^{(1)}$-modes, $N_2$ $\psi^{(2)}$-modes and $N_3$ $\psi^{(3)}$-modes. We have the following cases:
\begin{enumerate}[label=\textbf{C.\arabic*}]
    \item If $i_N=i_{N-1}=i$ and 
    \begin{enumerate}
        \item $m_N<m_{N-1}$, then we are done.
        \item $m_N= m_{N-s}$ for some $s\le N_i$, then by using \eqref{eq:sl3com-sim-1} repeatedly we have $\ket{\psi}\sim 0$.
        \item $m_{N-s}<m_N< m_{N-s-1}$ for some $0<s< N_i$, or $m_{N-s}<m_N\le -\sum_{j=1}^i \frac{N_j}{2}-\frac{1}{2}$ for $s=N_i$, then by using \eqref{eq:sl3com-sim-1} repeatedly we have
    \[\ket{\psi}\sim \psi^{(i_{N-1})}_{m_{N-1}} \cdots\psi^{(i_{N-s})}_{m_{N-s}}\psi^{(i_N)}_{m_N}\psi^{(i_{N-s-1})}_{m_{N-s-1}}\cdots \psi^{(i_1)}_{m_1}\ket{0}\]
    which is well-ordered.
    \item $m_N> -\sum_{j=1}^i \frac{N_j}{2}-\frac{1}{2}$, then by using \eqref{eq:sl3com-sim-1} repeatedly we have
    \[\ket{\psi}\sim \psi^{(i)}_{m_{N-1}} \cdots\psi^{(i)}_{m_{N-N_i}}\psi^{(i)}_{m_N}\psi^{(i_{N-N_i-1})}_{m_{N-N_i-1}}\cdots \psi^{(i_1)}_{m_1}\ket{0}.\]
    Because of $P(N-N_i)$ and \ref{fact:sl3-basis-1}, we see that $$\psi^{(i)}_{m_N}\psi^{(i_{N-N_i-1})}_{m_{N-N_i-1}}\cdots \psi^{(i_1)}_{m_1}\ket{0}\sim \sum_k\psi^{(i)}_{m^{(k)}_N}\psi^{(i_{N-N_i-1})}_{m^{(k)}_{N-N_i-1}}\cdots \psi^{(i_1)}_{m^{(k)}_1}\ket{0}=:\sum_k\ket{\psi_{(k)}}$$ with $m^{(k)}_N\le -\sum_{j=1}^i \frac{N_j}{2}-\frac{1}{2}$,$\forall k$. Recall that $\ket{\psi'}$ is in the form of \eqref{eq:sl3basis-123}, so we know $m_{N-1}<\cdots<m_{N-N_i}\le -\sum_{j=1}^i \frac{N_j}{2}-\frac{1}{2}$. Therefore we have must have for each $k$ that $m^{(k)}_N\le m_{N-1}$, or $m^{(k)}_N\ge m_{N-N_i}$, or there exists $1< s< N_i$ such that $m_{N-s}\le m^{(k)}_N\le m_{N-s-1}$. Hence by using \eqref{eq:sl3com-sim-1} repeatedly we know $\psi^{(i)}_{m_{N-1}} \cdots\psi^{(i)}_{m_{N-N_i}}\ket{\psi_{(k)}}$ can be well-ordered. \label{case:sl3-1-4} 
    \end{enumerate} \label{case:sl3-1}
    \item If $i=i_N>i_{N-1}=j$ and 
    \begin{enumerate}
        \item $m_N\le-\frac{N}{2}$, then we are done.
        \item $m_N> -\frac{N}{2}$, then by \eqref{eq:sl3com-sim-2} we have     
            \begin{align*}
                 &\psi^{(i)}_{m_N}\psi^{(j)}_{m_{N-1}}\psi^{(i_{N-2})}_{m_{N-2}}\cdots \psi^{(i_1)}_{m_1}\ket{0}\tag{$\medsquare$}\\
           \sim&\sum_{l_1\ge 1}\psi^{(i)}_{m_N-l_1}\psi^{(j)}_{m_{N-1}+l_1}\psi^{(i_{N-2})}_{m_{N-2}}\cdots \psi^{(i_1)}_{m_1}\ket{0}\tag{$\medstar$}\\
           +&\sum_{l_1\ge 0}\psi^{(j)}_{m_{N-1}-\frac{1}{2}-l_1}\psi^{(i)}_{m_N+\frac{1}{2}+l_1}\psi^{(i_{N-2})}_{m_{N-2}}\cdots \psi^{(i_1)}_{m_1}\ket{0}.\tag{$\meddiamond$}
            \end{align*}
We note that because of $P(N-1)$ and \ref{fact:sl3-basis-1}, we know that for the terms in $(\meddiamond)$ we have
\begin{align*}
    &\psi^{(j)}_{m_{N-1}-\frac{1}{2}-l_1}\psi^{(i)}_{m_N+\frac{1}{2}+l_1}\psi^{(i_{N-2})}_{m_{N-2}}\cdots \psi^{(i_1)}_{m_1}\ket{0}\\
    \sim\,&\psi^{(j)}_{m_{N-1}-\frac{1}{2}-l_1}\sum_{k}\psi^{(i)}_{m^{(k)}_{N}}\psi^{(i_{N-2})}_{m^{(k)}_{N-2}}\cdots \psi^{(i_1)}_{m^{(k)}_1}\ket{0}, &\mathrm{with}\;\;m^{(k)}_{N}\le -\frac{N}{2}+\frac{1}{2}.
\end{align*}
Now using \eqref{eq:sl3com-sim-2} again, we have
\begin{align*}
    &\psi^{(j)}_{m_{N-1}-\frac{1}{2}-l_1}\psi^{(i)}_{m^{(k)}_{N}}\psi^{(i_{N-2})}_{m^{(k)}_{N-2}}\cdots \psi^{(i_1)}_{m^{(k)}_1}\ket{0}\tag{$\filledsquare$}\\
    \sim& \sum_{r_1\ge 1}\psi^{(j)}_{m_{N-1}-\frac{1}{2}-l_1-r_1}\psi^{(i)}_{m^{(k)}_{N}+r_1}\psi^{(i_{N-2})}_{m^{(k)}_{N-2}}\cdots \psi^{(i_1)}_{m^{(k)}_1}\ket{0}\tag{$\filledstar$}\\
    +& \sum_{r_1\ge 0} \psi^{(i)}_{m^{(k)}_{N}-\frac{1}{2}-r_1}\psi^{(j)}_{m_{N-1}-l_1+r_1}\psi^{(i_{N-2})}_{m^{(k)}_{N-2}}\cdots \psi^{(i_1)}_{m^{(k)}_1}\ket{0}\tag{$\filleddiamond$}
\end{align*}
Since $m^{(k)}_{N}-\frac{1}{2}-r_1\le -\frac{N}{2}$ for any $r_1$ and $P(N-1)$, we see that ($\filleddiamond$) can be well-ordered. Then by using ($\filledstar$) repeatedly, we have  
\begin{align*}      (\filledsquare)\sim&\sum_{l_1\ge1}\sum_{l_2\ge1}\cdots\sum_{l_s\ge1}\psi^{(j)}_{m_{N-1}-\frac{1}{2}-l_1-(r_1+\cdots+r_s)}\psi^{(i)}_{m^{(k)}_{N}+r_1+\cdots+r_s}\psi^{(i_{N-2})}_{m^{(k)}_{N-2}}\cdots \psi^{(i_1)}_{m^{(k)}_1}\ket{0}\\
        +&\text{well-ordered terms},
 \end{align*}
 where $s$ is a large enough number so that $m^{(k)}_{N}+s+\sum_{j=1}^{N-2}m^{(k)}_j\ge 0$. Hence by $P(N-1)$ and \ref{fact:sl3-basis-2}, we know that $\psi^{(i)}_{m^{(k)}_{N}+r_1+\cdots+r_s}\psi^{(i_{N-2})}_{m^{(k)}_{N-2}}\cdots \psi^{(i_1)}_{m^{(k)}_1}\ket{0}$ is $0$ since it has non-negative total mode number, which gives that ($\filledsquare$) and therefore $(\meddiamond)$ can be well-ordered. Similarly, by using ($\medstar$) repeatedly, we see ($\medsquare$) can be well-ordered. \label{case:sl3-2b}  
    \end{enumerate} \label{case:sl3-2}   
    \item If $i=i_N<i_{N-1}=j$, then by \eqref{eq:sl3com-sim-2} we have     
            \begin{align*}
                 &\psi^{(i)}_{m_N}\psi^{(j)}_{m_{N-1}}\psi^{(i_{N-2})}_{m_{N-2}}\cdots \psi^{(i_1)}_{m_1}\ket{0}\tag{$\heartsuit$}\\
           \sim&\sum_{l_1\ge 1}\psi^{(i)}_{m_N-l_1}\psi^{(j)}_{m_{N-1}+l_1}\psi^{(i_{N-2})}_{m_{N-2}}\cdots \psi^{(i_1)}_{m_1}\ket{0}\tag{$\medcircle$}\\
           +&\sum_{l_1\ge 0}\psi^{(j)}_{m_{N-1}-\frac{1}{2}-l_1}\psi^{(i)}_{m_N+\frac{1}{2}+l_1}\psi^{(i_{N-2})}_{m_{N-2}}\cdots \psi^{(i_1)}_{m_1}\ket{0}.\tag{$\triangle$}
            \end{align*}
    Here ($\triangle$) can be well-ordered by \ref{case:sl3-2} and therefore ($\heartsuit$) can be well-ordered using ($\medcircle$) repeatedly, as argued in \ref{case:sl3-2b}. \label{case:sl3-3}
\end{enumerate}
\end{proof}

Similarly, we can find a basis for twisted sectors. Let $\ket{i'}$ be  the highest weight vector of channel $[i']$. It satisfies $L_0^{(i)}\ket{i'}=0$ and $L_0^{(j)}\ket{i'}=\frac{1}{16}$ for all $j\ne i$. In other words, it is an untwisted true vacuum with respect to one of the free fermion subalgebras but a twisted vacuum with respect to the other two subalgebras. 

Note that we have relations among these vacua $\ket{i'}$ according to the fusion rules, such as $\psi^{(2)}_0\ket{1'}\propto\ket{3'}$ and $\psi^{(3)}_0\ket{2'}\propto\ket{1'}$. For consistency, we fix a normalisation for these vacua relative to $\ket{1'}$. We choose $\ket{2'}=\psi^{(3)}_0\ket{1'}$ and $\ket{3'}=\psi^{(2)}_0\ket{1'}$. The normalisation factors for other relations are then determined by the generalised commutation relations \eqref{eq:sl3commrel-1} and \eqref{eq:sl3commrel-2}, such as $\psi^{(3)}_0\ket{2'}=\psi^{(3)}_0\psi^{(3)}_0\ket{1'}=\frac{1}{2}\ket{1'}$ and $\psi^{(1)}_0\ket{2'}=\psi^{(1)}_0\psi^{(3)}_0\ket{1'}=c_{13}\ket{3'}$.
Then a basis for the twisted sector can be chosen to be the states in the form of 
\begin{equation}\label{eq:sl3basis-twi}
    \begin{aligned}
      \psi^{(3)}_{-N_3-\frac{N_1+N_2}{2}+1-s^{(3)}_{N_3}}\dots\psi^{(3)}_{-\frac{N_1+N_2}{2}-s^{(3)}_{1}}&\psi^{(2)}_{-N_2-\frac{N_1}{2}+1-s^{(2)}_{N_2}}\dots\\
    &\dots\psi^{(2)}_{-\frac{N_1}{2}-s^{(2)}_{1}}\psi^{(1)}_{-N_1+\frac{1}{2}-s^{(1)}_{N_1}}\dots\psi^{(1)}_{-\frac{1}{2}-s^{(1)}_{1}}\ket{1'}
    \end{aligned}
\end{equation}
with $N_1,N_2,N_3\ge0$ and $s^{(i)}_N\ge\dots s^{(i)}_2\ge s^{(i)}_1\ge0$ for $i=1,2,3$. 

The lowest energy state in the form of \eqref{eq:sl3basis-twi} with $N_1$ $\psi^{(1)}$-modes, $N_2$ $\psi^{(2)}$-modes and $N_3$ $\psi^{(3)}$-modes is
\[\psi^{(3)}_{-N_3-\frac{N_1+N_2}{2}+1}\dots\psi^{(3)}_{-\frac{N_1+N_2}{2}}\psi^{(2)}_{-N_2-\frac{N_1}{2}+1}\dots\psi^{(2)}_{-\frac{N_1}{2}}\psi^{(1)}_{-N_1+\frac{1}{2}}\dots\psi^{(1)}_{-\frac{1}{2}}\ket{1'}\]
whose energy ($L_0$-eigenvalue) is $$\frac{1}{2}\bold{N}^{T}\cdot\bold{G_3}\cdot\bold{N}-\frac{1}{2}(N_2+N_3)+\frac{1}{10},$$ so the corresponding UCPF is (up to an overall factor)
\begin{equation}\label{eq:sl3UCPF-twi}
     q^{\frac{1}{10}}\sum_{N_i\ge 0}\dfrac{q^{\frac{1}{2}\mathbf{N}^{\mathsf{T}}\cdot\mathbf{G}_3\cdot\mathbf{N}-\frac{1}{2}(N_2+N_3)}}{\prod_i (q)_{N_i}},
\end{equation}
which is the same as the sum of characters of all twisted sectors, that is,
\begin{equation}\label{eq:sl3id-twi}
q^{\frac{1}{10}-\frac{1}{20}}\sum_{N_i\ge 0}\dfrac{q^{\frac{1}{2}\mathbf{N}^{\mathsf{T}}\cdot\mathbf{G}_3\cdot\mathbf{N}-\frac{1}{2}(N_2+N_3)}}{\prod_i (q)_{N_i}}=b_{002}^{110}+3b_{110}^{110}.
\end{equation}
The proofs of \eqref{eq:sl3id-untwi} and \eqref{eq:sl3id-twi} are included in \parencite{BCH} and rely on the following theorem proved by S. Chern:

\begin{theorem}\label{th:G3-general}
\begin{align*}
&\sum_{N_1,N_2,N_3\ge 0} \frac{z_1^{N_1}z_2^{N_2}z_3^{N_3}q^{\mathbf{N}^{\mathsf{T}}\cdot\mathbf{G}_3\cdot\mathbf{N}}}{(q^2)_{N_1}(q^2)_{N_2}(q^2)_{N_3}}\\
=& (-z_1q;q^2)_\infty \sum_{M\ge 0}\frac{z_2^M z_3^M q^{3M^2} (-z_2 z_3^{-1}q;q^2)_M (-z_2^{-1}z_3q;q^2)_M}{(-z_1q;q^2)_M (q^2)_{2M}}\\
+& (-z_1 q^2;q^2)_\infty \sum_{M\ge 0} \frac{(z_2+z_3) z_2^M z_3^M q^{3M^2+3M+1} (-z_2z_3^{-1}q^2;q^2)_M (-z_2^{-1}z_3q^2;q^2)_M}{(-z_1q^2;q^2)_{M} (q^2)_{2M+1}}.
\end{align*}
\end{theorem}
\begin{proof}
    See \parencite[Theorem 7]{BCH}.
\end{proof}

As mentioned earlier, the central charge of this coupled free fermion model is $\frac{6}{5}$, which is the sum of $\frac{1}{2}$ and $\frac{7}{10}$ which are central charges of minimal models with $i=1,2$, so we expect the decomposition of $\mathcal{L}^{\Lambda}_{\lambda}$ into the tensor products of modules of minimal models. Indeed it is not hard to observe that (up to a sufficiently large order of $q$) the parafermionic characters $b^{\Lambda}_{\lambda}$ can be written as characters of certain combinations of minimal models, as follows:
\begin{equation}\label{eq:sl3mindecomp}
    \begin{aligned}
      b^{200}_{200}&=\text{ch}\left[\mathcal{L}\textstyle\left(\frac{1}{2},0\right)\otimes\mathcal{L}\left(\frac{7}{10},0\right)\oplus\mathcal{L}\left(\frac{1}{2},\frac{1}{2}\right)\otimes\mathcal{L}\left(\frac{7}{10},\frac{3}{2}\right)\right]\\
    b^{200}_{011}&=\text{ch}\textstyle\left[\mathcal{L}\left(\frac{1}{2},\frac{1}{2}\right)\otimes\mathcal{L}\left(\frac{7}{10},0\right)\oplus\mathcal{L}\left(\frac{1}{2},0\right)\otimes\mathcal{L}\left(\frac{7}{10},\frac{3}{2}\right)\right]\\
    &=\text{ch}\textstyle\left[\mathcal{L}\left(\frac{1}{2},\frac{1}{16}\right)\otimes\mathcal{L}\left(\frac{7}{10},\frac{7}{16}\right)\right]\\
    b^{110}_{002}&=\text{ch}\textstyle\left[\mathcal{L}\left(\frac{1}{2},0\right)\otimes\mathcal{L}\left(\frac{7}{10},\frac{3}{5}\right)\oplus\mathcal{L}\left(\frac{1}{2},\frac{1}{2}\right)\otimes\mathcal{L}\left(\frac{7}{10},\frac{1}{10}\right)\right]\\
    b^{110}_{110}&=\text{ch}\textstyle\left[\mathcal{L}\left(\frac{1}{2},0\right)\otimes\mathcal{L}\left(\frac{7}{10},\frac{1}{10}\right)\oplus\mathcal{L}\left(\frac{1}{2},\frac{1}{2}\right)\otimes\mathcal{L}\left(\frac{7}{10},\frac{3}{5}\right)\right]\\&=\text{ch}\textstyle\left[\mathcal{L}\left(\frac{1}{2},\frac{1}{16}\right)\otimes\mathcal{L}\left(\frac{7}{10},\frac{3}{80}\right)\right]
    \end{aligned}
\end{equation}

We shall further investigate and confirm the decomposition by looking for the highest weight vectors of relevant $\mathcal{L}\left(\frac{1}{2},0\right)\otimes \mathcal{L}\left(\frac{7}{10},0\right)$-modules in the Fock space of $\widehat{\mathfrak{sl}(3)}_2/\widehat{\mathfrak{u}(1)}^2$. 

To do that, we first deconstruct the total energy-momentum tensor $L$ into $L^{(\frac{1}{2})}$ and $L^{(\frac{7}{10})}$ such that the modes of $L^{(\frac{1}{2})}$ (resp. $L^{(\frac{7}{10})}$) form the Virosoro algebra with central charge $\frac{1}{2}$ (resp. $\frac{7}{10}$). It turns out that we can choose (up to permutations)
\begin{align*}
   & L^{(\frac{1}{2})}=L^{(1)},&L^{(\frac{7}{10})}=-\frac{1}{5}L^{(1)}+\frac{4}{5}\left(L^{(2)}+L^{(3)}\right).
\end{align*}

Let $\mathscr{E}_n$ denote the $L_0$-eigenspace with eigenvalue $n$. Results in \eqref{eq:sl3mindecomp} suggest to look at the simultaneous eigenvectors of $L_0^{(\frac{1}{2})}$ and $L_0^{(\frac{7}{10})}$ when restricted to the subspaces $\mathscr{E}_0,\mathscr{E}_2,\mathscr{E}_{\frac{1}{2}},\mathscr{E}_{\frac{3}{2}},\mathscr{E}_{\frac{3}{5}},\mathscr{E}_{\frac{1}{10}},\mathscr{E}_{\frac{11}{10}}$. It is sufficient to compute the matrix realisation for one of $L_0^{(\frac{1}{2})}$ and $L_0^{(\frac{7}{10})}$, as the other will be given for free recalling that $L_0=L_0^{(\frac{1}{2})}+L_0^{(\frac{7}{10})}$ and $L_0|_{\mathscr{E}_n}=n\bold{I}$, so we focus on the computation of $L_0^{(\frac{1}{2})}$.

$\mathscr{E}_0$ is trivial and it is obvious that $L_0^{(\frac{1}{2})}|_{\mathscr{E}_0}=L_0^{(\frac{7}{10})}|_{\mathscr{E}_0}=0$, so the highest weight vector of $\mathcal{L}\left(\frac{1}{2},0\right)\otimes\mathcal{L}\left(\frac{7}{10},0\right)$ is $\ket{0}$. 

$\mathscr{E}_2$ has a basis $\left\{\psi^{(1)}_{-\frac{3}{2}}\psi_{-\frac{1}{2}}^{(1)}\ket{0},\psi^{(2)}_{-\frac{3}{2}}\psi_{-\frac{1}{2}}^{(2)}\ket{0},\psi^{(3)}_{-\frac{3}{2}}\psi_{-\frac{1}{2}}^{(3)}\ket{0}\right\}$ under which the operators $L_0^{(\frac{1}{2})}$ is realised as
\begin{align*}
    L_0^{(\frac{1}{2})}|_{\mathscr{E}_2}=\left(\begin{array}{rrr}
2&\frac{1}{4}&\frac{1}{4}\\
0&\frac{1}{4}&-\frac{1}{4}\\
0&-\frac{1}{4}&\frac{1}{4}
\end{array}\right),
\end{align*}
so the highest weight vector of $\mathcal{L}\left(\frac{1}{2},\frac{1}{2}\right)\otimes\mathcal{L}\left(\frac{7}{10},\frac{3}{2}\right)$ or, in other words, the simultaneous eigenvector of $L_0^{(1/2)}|_{\mathscr{E}_2}$ and $L_0^{(7/10)}|_{\mathscr{E}_2}$ with eigenvalue $\frac{1}{2}$ and $\frac{3}{2}$ respectively is $\psi^{(2)}_{-\frac{3}{2}}\psi_{-\frac{1}{2}}^{(2)}\ket{0}-\psi^{(3)}_{-\frac{3}{2}}\psi_{-\frac{1}{2}}^{(3)}\ket{0}$.

Our basis of $\mathscr{E}_{\frac{1}{2}}$ is $\big\{\psi^{(1)}_{-\frac{1}{2}}\ket{0},\psi^{(2)}_{-\frac{1}{2}}\ket{0},\psi^{(3)}_{-\frac{1}{2}}\ket{0}\big\}$ where all the states are eigenvectors of both $L_0^{(\frac{1}{2})}$ and $L_0^{(\frac{7}{10})}$, so the highest weight vector of $\mathcal{L}\left(\frac{1}{2},\frac{1}{2}\right)\otimes\mathcal{L}\left(\frac{7}{10},0\right)$ is $\psi^{(1)}_{-\frac{1}{2}}\ket{0}$ and the two copies of $\mathcal{L}\left(\frac{1}{2},\frac{1}{16}\right)\otimes\mathcal{L}\left(\frac{7}{10},\frac{7}{16}\right)$ are generated by $\psi^{(2)}_{-\frac{1}{2}}\ket{0}$ and $\psi^{(3)}_{-\frac{1}{2}}\ket{0}$ respectively. These results imply that $\left(\mathcal{L}\left(\frac{1}{2},\frac{1}{2}\right)\otimes\mathcal{L}\left(\frac{7}{10},0\right)\right)\oplus\left(\mathcal{L}\left(\frac{1}{2},\frac{1}{16}\right)\otimes\mathcal{L}\left(\frac{7}{10},\frac{7}{16}\right)\right)$ is the decomposition of $\mathcal{L}^{200}_{12-1}$ and hence to find the highest weight vector of $\mathcal{L}\left(\frac{1}{2},0\right)\otimes\mathcal{L}\left(\frac{7}{10},\frac{3}{2}\right)$, we can further restrict the space $\mathscr{E}_{\frac{3}{2}}$ to its subspace $\mathscr{E}_{\frac{3}{2}}^{[1]}:=\mathscr{E}_{\frac{3}{2}}\cap \mathcal{L}^{200}_{12-1}$, so we consider the basis $\big\{\psi^{(1)}_{-\frac{3}{2}}\ket{0},\psi^{(3)}_{-1}\psi^{(2)}_{-\frac{1}{2}}\ket{0}\big\}$ under which we have
\begin{align*}
    &L_0^{(\frac{1}{2})}|_{\mathscr{E}_{\frac{3}{2}}^{[1]}}=\left(\begin{array}{cc}
\frac{3}{2}&\frac{3(1+i)}{8}\\
0&0
\end{array}\right),
\end{align*}
and therefore the desired highest weight vector is $\psi^{(1)}_{-\frac{3}{2}}\ket{0}+(2i-2)\psi^{(3)}_{-1}\psi^{(2)}_{-\frac{1}{2}}\ket{0}$.

We go through the same process for the twisted sector and the results are summarised in Table \ref{tab:sl3hwv}, together with a collection of results for untwisted sectors above.
\begingroup
\renewcommand*{\arraystretch}{1.2}
\setlength{\LTpre}{1pt}
\setlength{\LTpost}{-1pt}
\begin{center}
\begin{longtable}{c|c}
      Module of minimal models & Highest weight vector(s)\\
      \hline
      $\mathcal{L}\left(\frac{1}{2},0\right)\otimes\mathcal{L}\left(\frac{7}{10},0\right)$&$\ket{0}$\\
      $\mathcal{L}\left(\frac{1}{2},\frac{1}{2}\right)\otimes\mathcal{L}\left(\frac{7}{10},\frac{3}{2}\right)$&$\psi^{(2)}_{-\frac{3}{2}}\psi_{-\frac{1}{2}}^{(2)}\ket{0}-\psi^{(3)}_{-\frac{3}{2}}\psi_{-\frac{1}{2}}^{(3)}\ket{0}$\\
      $\mathcal{L}\left(\frac{1}{2},\frac{1}{2}\right)\otimes\mathcal{L}\left(\frac{7}{10},0\right)$&$\psi^{(1)}_{-\frac{1}{2}}\ket{0}$\\
      $\mathcal{L}\left(\frac{1}{2},0\right)\otimes\mathcal{L}\left(\frac{7}{10},\frac{3}{2}\right)$&$\psi^{(1)}_{-\frac{3}{2}}\ket{0}+(2i-2)\psi^{(3)}_{-1}\psi^{(2)}_{-\frac{1}{2}}\ket{0}$\\
      $\mathcal{L}\left(\frac{1}{2},\frac{1}{16}\right)\otimes\mathcal{L}\left(\frac{7}{10},\frac{7}{16}\right)$&$\psi^{(2)}_{-\frac{1}{2}}\ket{0},\psi^{(3)}_{-\frac{1}{2}}\ket{0}$\\
      $\mathcal{L}\left(\frac{1}{2},0\right)\otimes\mathcal{L}\left(\frac{7}{10},\frac{3}{5}\right)$&$\psi^{(1)}_{-\frac{1}{2}}\ket{1'}+(2i-2)\psi^{(3)}_{-\frac{1}{2}}\psi^{(2)}_{0}\ket{1'}$ \\
      $\mathcal{L}\left(\frac{1}{2},\frac{1}{2}\right)\otimes\mathcal{L}\left(\frac{7}{10},\frac{1}{10}\right)$&$\psi^{(1)}_{-\frac{1}{2}}\ket{1'}$\\
      $\mathcal{L}\left(\frac{1}{2},0\right)\otimes\mathcal{L}\left(\frac{7}{10},\frac{1}{10}\right)$&$\ket{1'}$\\
      $\mathcal{L}\left(\frac{1}{2},\frac{1}{2}\right)\otimes\mathcal{L}\left(\frac{7}{10},\frac{3}{5}\right)$&$\psi^{(2)}_{-1}\psi^{(2)}_{0}\ket{1'}-\psi^{(3)}_{-1}\psi^{(3)}_{0}\ket{1'}$\\
      $\mathcal{L}\left(\frac{1}{2},\frac{1}{16}\right)\otimes\mathcal{L}\left(\frac{7}{10},\frac{3}{80}\right)$&$\psi^{(2)}_0\ket{1'},\psi^{(3)}_0\ket{1'}$\\
    \caption{Virasoro highest weight vectors in $\widehat{\mathfrak{sl}(3)}_2/\widehat{\mathfrak{u}(1)}^2$}
    \label{tab:sl3hwv}
\end{longtable}
\end{center}
\endgroup
As shown in \parencite{DING1994}, a $W_3$-algebra can be constructed by $\widehat{\mathfrak{sl}(3)}_2/\widehat{\mathfrak{u}(1)}^2$ coupled free fermions. $W$-algebras\index{W@$W$-algebra} are extensions of conformal algebras that have higher spin fields $\{W^{(s_i)}(z)|i\in I\}$, where $I$ is an index set, such that the entire space of fields is spanned by normal ordered products of the fields $W^{(s_i)}(z)$ and their derivatives. The $W_3$-algebra, introduced by \parencite{zamolodchikov1995infinite}, is a kind of $W$-algebra that has a spin-3 field. It was shown in \parencite{watts1990w} that the $W_3$-algebra can be constructed from the diagonal cosets $\hat{\mathfrak{g}}_{k}\times\hat{\mathfrak{g}}_{1}/\hat{\mathfrak{g}}_{k+1}$, whose modules are characterised by $\Lambda^+\in \hat{P}^k_+$ and $\Lambda^-\in \hat{P}^{k+1}_+$, for which the conformal dimension\index{W@$W$-algebra!conformal dimension} is given by \parencite[eq. (7.53)]{BouwknegtWalg}:
\[h(\Lambda^+,\Lambda^-)=\dfrac{c-n}{24}+\dfrac{|(m+1)(\Lambda^+ +\rho)-m(\Lambda^- +\rho)|^2}{2m(m+1)},\]
where $n=\rank \mathfrak{g}$, $m=k+h^{\vee}$, $\rho$ is the weyl vector, and  $c$ is the central charge given by
\[c(m)=2-\dfrac{24}{m(m+1)}.\]
The character formula is given by \parencite[eq. (7.55)]{BouwknegtWalg}:
\[\mathscr{W}\left(c,h(\Lambda^+,\Lambda^-)\right)(q)=\dfrac{1}{\eta(\tau)^n}\sum_{w\in \hat{W}}\det(w)e^{\frac{1}{2m(m+1)}|(m+1)w(\Lambda^+ +\rho)-m(\Lambda^- +\rho)|^2}.\]
Now comparing the W-characters \index{W@$W$-algebra!character} of  $\widehat{\mathfrak{sl}(3)}_{2}\times\widehat{\mathfrak{sl}(3)}_{1}/\widehat{\mathfrak{sl}(3)}_{3}$ with the  characters of $\widehat{\mathfrak{sl}(3)}_2/\widehat{\mathfrak{u}(1)}^2$, we find that
\begin{align*}
    &b^{200}_{200}=\mathscr{W}\left(\frac{6}{5},0\right)+2\mathscr{W}\left(\frac{6}{5},2\right), & b^{200}_{011}=\mathscr{W}\left(\frac{6}{5},\frac{1}{2}\right),\\
    &b^{110}_{002}=2\mathscr{W}\left(\frac{6}{5},\frac{3}{5}\right)+\mathscr{W}\left(\frac{6}{5},\frac{8}{5}\right), & b^{110}_{110}=\mathscr{W}\left(\frac{6}{5},\frac{1}{10}\right).
\end{align*}

\subsection{$\widehat{\mathfrak{sl}(4)}_2/\widehat{\mathfrak{u}(1)}^3$}

Let $\alpha_1,\alpha_2,\alpha_3$ denote the simple roots of $\mathfrak{sl}(4)$ such that $(\alpha_1,\alpha_3)=0$. Then the positive roots are $\Delta_+=\{\alpha_1,\alpha_2,\alpha_3,\alpha_4:=\alpha_1+\alpha_2,\alpha_5:=\alpha_2+\alpha_3,\alpha_6:=\alpha_1+\alpha_2+\alpha_3\}$ and therefore we have 6 coupled free fermions $\{\psi^{(i)}:=\psi^{(\alpha_i)}\mid \alpha_i\in\Delta_+\}$. 

Here are some ad hoc terminologies for convenience. We call $\{i,j,k\}$ an $\mathfrak{sl}(3)$-triple if $\{\psi^{(i)},\psi^{(j)},\psi^{(k)}\}$ satisfies the OPEs \eqref{eq:sl3OPE} for coupled free fermions in $\widehat{\mathfrak{sl}(3)}_2/\widehat{\mathfrak{u}(1)}^2$ (with compatible choices of constants). We see that there are four $\mathfrak{sl}(3)$-triples which are $\{1,2,4\}, \{1,5,6\},\{2,3,5\},\{3,4,6\}$. We call $\{i,j\}$ an uncoupled pair if there exists no positive root $\alpha\in\Delta_+$ such that $\alpha_i+\alpha_j\equiv \alpha\mod 2Q$. We denote the set of uncoupled pairs $\{\{1,3\},\{2,6\},\{4,5\}\}$ by $\Xi$ and we have $$\alpha_1+\alpha_3\equiv \alpha_2+\alpha_6\equiv\alpha_4+\alpha_5\mod2Q.$$
The total energy-momentum tensor $L$, in this case, can be written as
\[L(z)=-\frac{1}{3}\sum_{i=1}^6:\psi^{(i)}(z)\partial\psi^{(i)}(z):=\dfrac{2}{3}\sum_{i=1}^6L^{(i)}(z)\]
and the central charge is 2 by \eqref{eq:CFFcc}.

To find all inequivalent primary fields, consider for the highest weights $\Lambda$ the set $\{[2,0,0,0],[1,1,0,0],[0,1,0,1]\}$ and note that $Q/2Q$ has 8 cosets which are represented by $\{[0,0,0,0],[-1,2,-1,0],[0,-1,2,-1],[-1,0,-1,2],[-1,1,1,-1],\newline[-1,-1,1,1], [-2,1,0,1],[-2,0,2,0]\}$. The 24 inequivalent primary fields are shown in Table \ref{tab:sl4prifield}.
\begingroup
\renewcommand*{\arraystretch}{1.4}
\begin{table}[h!]
    \centering
      \begin{tabular}[c]{c|cccccccc}
        $\Phi^{\Lambda}_{\lambda}$ & $\Phi^{2000}_{2000}$ & $\Phi^{2000}_{12-10}$&$\Phi^{2000}_{2-12-1}$ &$\Phi^{2000}_{10-12}$ &$\Phi^{2000}_{111-1}$& $\Phi^{2000}_{1-111}$ & $\Phi^{2000}_{0101}$ &$\Phi^{2000}_{0020}$ \\
    \hline 
        $h^{\Lambda}_{\lambda}$ & 0&$\frac{1}{2}$&$\frac{1}{2}$&$\frac{1}{2}$&$\frac{1}{2}$&$\frac{1}{2}$&$\frac{1}{2}$&1\\
    \hline
    Label & [0] & [1] & [2] & [3] & [4] & [5] & [6]  & [$\eta$]  \\
    \hline \hline
     $\Phi^{\Lambda}_{\lambda}$ & $\Phi^{0101}_{2000}$ & $\Phi^{0101}_{12-10}$ & $\Phi^{0101}_{21-21}$&$\Phi^{0101}_{10-12}$ &$\Phi^{0101}_{111-1}$& $\Phi^{0101}_{1-111}$ &$\Phi^{0101}_{0101}$ &$\Phi^{0101}_{0020}$ \\
    \hline 
        $h^{\Lambda}_{\lambda}$ & $\frac{2}{3}$&$\frac{1}{6}$&$\frac{1}{6}$&$\frac{1}{6}$&$\frac{1}{6}$&$\frac{1}{6}$&$\frac{1}{6}$&$\frac{2}{3}$\\
    \hline
    Label & [$\tau'$] & [$1'$] & [$2'$] & [$3'$] & [$4'$] & [$5'$] & [$6'$]  & [$\eta'$] \\
    \hline \hline
     $\Phi^{\Lambda}_{\lambda}$ & $\Phi^{1100}_{1100}$ & $\Phi^{1100}_{2-110}$&$\Phi^{1100}_{102-1}$ &$\Phi^{1100}_{01-12}$ &$\Phi^{1100}_{021-1}$& $\Phi^{1100}_{0011}$ & $\Phi^{1100}_{-1201}$ &$\Phi^{1100}_{-1120}$ \\
    \hline 
        $h^{\Lambda}_{\lambda}$ & $\frac{1}{8}$&$\frac{1}{8}$&$\frac{5}{8}$&$\frac{5}{8}$&$\frac{1}{8}$&$\frac{5}{8}$&$\frac{1}{8}$&$\frac{5}{8}$\\
    \hline
    Label & [$\tau''$] & [$1''$] & [$2''$] & [$3''$] & [$4''$] & [$5''$] & [$6''$]  & [$\eta''$]   
    \end{tabular}
    \caption{Inequivalent primary fields in $\widehat{\mathfrak{sl}(4)}_2/\widehat{\mathfrak{u}(1)}^3$}
    \label{tab:sl4prifield}
\end{table}
\endgroup

We say the modules with $\Lambda=[2,0,0,0]$  are in the "untwisted sector", those with $\Lambda=[0,1,0,1]$ are in the "$\frac{1}{6}$-twisted sector" and, those with $\Lambda=[1,1,0,0]$ are in the "$\frac{1}{8}$-twisted sectors". 
 
A subset of relevant fusion rules is illustrated in Table \ref{tab:sl4PFfus}, while the fusion rules regarding the $\frac{1}{8}$-twisted sectors are the same as those for $\frac{1}{6}$-twisted sectors with all prime symbols replaced by double prime symbols.
\begingroup
\renewcommand*{\arraystretch}{1.4}
\begin{table}[h!]
    \centering
      \begin{tabular}[c]{c|cccccccccccccccc}
        $\times$  & [0] & [1] & [2] & [3]  & [4] & [5] & [6] & [$\eta$] & [$\tau'$] & [$1'$] & [$2'$]  & [$3'$]  & [$4'$] & [$5'$]  & [$6'$]  & [$\eta'$]\\
        \hline
        $[1]$ & [1] & [0] & [4] & [$\eta$]  & [2] & [6] & [5] & [3] & [$1'$] & [$\tau'$] & [$4'$]  & [$\eta'$]  & [$2'$] & [$6'$]  & [$5'$]  & [$3'$]\\
        $[2]$ & [2] & [4] & [0] & [5]  & [1] & [3] & [$\eta$] & [6] & [$2'$] & [$4'$] & [$\tau'$]  & [$5'$]  & [$1'$] & [$3'$]  & [$\eta'$]  & [$6'$]\\
        $[3]$ & [3] & [$\eta$] & [5] & [0]  & [6] & [2] & [4] & [1] & [$3'$] & [$\eta'$] & [$5'$]  & [$\tau'$]  & [$6'$] & [$2'$]  & [$4'$]  & [$1'$]\\
        $[4]$ & [4] & [2] & [1] & [6]  & [0] & [$\eta$] & [3] & [5] & [$4'$] & [$2'$] & [$1'$]  & [$6'$]  & [$\tau'$] & [$\eta'$]  & [$3'$]  & [$5'$]\\
        $[5]$ & [5] & [6] & [3] & [2]  & [$\eta$] & [0] & [1] & [4] & [$5'$] & [$6'$] & [$3'$]  & [$2'$]  & [$\eta'$] & [$\tau'$]  & [$1'$]  & [$4'$]\\
        $[6]$ & [6] & [5] & [$\eta$] & [4]  & [3] & [1] & [0] & [2] & [$6'$] & [$5'$] & [$\eta'$]  & [$4'$]  & [$3'$] & [$1'$]  & [$\tau'$]  & [$2'$]
    \end{tabular}
    \caption{Subset of fusion rules for $\widehat{\mathfrak{sl}(4)}_2/\widehat{\mathfrak{u}(1)}^3$}
    \label{tab:sl4PFfus}
\end{table}
\endgroup

We again solve for R- and F-matrices of the untwisted sector of  $\widehat{\mathfrak{sl}(4)}_2/\widehat{\mathfrak{u}(1)}^3$ from the pentagon equations \eqref{eq:pent} and hexagon equations \eqref{eq:hex} with the assumptions as in \eqref{eq:Fmat} and \eqref{eq:Rmat}. 

For each $\mathfrak{sl}(3)$-triple $\{i,j,k\}$, the relevant R- and F-matrices should satisfy \eqref{eq:sl3RF}. The F-matrices that involve channels across different $\mathfrak{sl}(3)$-triples are found to satisfy the following relations:
\begin{equation}\label{eq:sl4Fmat}
    \begin{aligned}
&\left(F_{462}^5\right)_{13}=\left(F_{526}^4\right)_{13}=\left(F_{
154}^3\right)_{26}=\left(F_{345}^1\right)_{26}=\left(F_{426}^5\right)_{31}=\left(F_{562}^4\right)_{31}\\=&\left(F_{231}^6\right)_{45}=\left(F_{613}^2\right)_{45}=\left(F_{354}^1\right)_{62}=\left(F_{145}^3\right)_{62}=\left(F_{631}^2\right)_{54}=\left(F_{213}^6\right)_{54},
    \end{aligned}
    \end{equation}
    \begin{align*}
&\left(F_{254}^6\right)_{13}=\left(F_{645}^2\right)_{13}=\left(F_{431}^5\right)_{26}=\left(F_{513}^4\right)_{26}=\left(F_{654}^2\right)_{31}=\left(F_{245}^6\right)_{31}\\=&\left(F_{162}^3\right)_{45}=\left(F_{326}^1\right)_{45}=\left(F_{413}^5\right)_{62}=\left(F_{531}^4\right)_{62}=\left(F_{126}^3\right)_{54}=\left(F_{362}^1\right)_{54}=\left(\left(F_{213}^6\right)_{54}\right)^{-1}.\tag{\ref{eq:sl4Fmat}}
    \end{align*}


The solution to the F-matrices in \eqref{eq:sl4Fmat} suggests that in the generalised commutation relations \eqref{eq:gcommrel} which in this case reads:
    \begin{align}
    \psi^{(i)}_n\psi^{(i)}_m+\psi^{(i)}_m\psi^{(i)}_n=\delta_{m+n,0},&\label{eq:sl4gen-1}\\[0.7ex]
    \psi^{(i)}_n\psi^{(j)}_m+\mu_{ij}\psi^{(j)}_m\psi^{(i)}_n=0,\;\;&\quad \text{if}\;\{i,j\}\in\Xi,\label{eq:sl4gen-2}\\
    \sum_{l\ge0}\binom{l-\frac{1}{2}}{l}\left(\psi^{(i)}_{m-\frac{1}{2}-l}\psi^{(j)}_{n+\frac{1}{2}+l}+\mu_{ij}\psi^{(j)}_{n-l}\psi^{(i)}_{m+l}\right)=c_{ij}\psi^{(i+j)}_{m+n},&\quad\text{if}\;\{i,j\}\not\in\Xi,\label{eq:sl4gen-3}
    \end{align}
    the values of the constants $c_{ij}:=c_{\alpha_i,\alpha_j}$ and $\mu_{ij}:=\mu_{\alpha_i,\alpha_j}$ need to be determined carefully so that the associativity holds. In fact, we found there are even more constraints on these constants beyond the associativity.

From \eqref{eq:sl4gen-3}, we can compute that
\begin{equation}\label{eq:sl4uncprel}
\begin{aligned}
  \psi^{(5)}_{-\frac{1}{2}}\psi^{(4)}_{-\frac{1}{2}}\ket{0}=&\,c^{-1}_{16} \sum_{l\ge 0}\binom{l-\frac{1}{2}}{l}\left(\psi^{(1)}_{-\frac{1}{2}-l}\psi^{(6)}_{l}+\mu_{16}\psi^{(6)}_{-\frac{1}{2}-l}\psi^{(1)}_{l}\right)\psi^{(4)}_{-\frac{1}{2}}\ket{0}\\
  =&\left(c^{-1}_{16}c_{64} \psi^{(1)}_{-\frac{1}{2}}\psi^{(3)}_{-\frac{1}{2}}+ c^{-1}_{16}c_{14}\mu_{16}\psi^{(6)}_{-\frac{1}{2}}\psi^{(2)}_{-\frac{1}{2}}\right)\ket{0}\\
  =&\left(\mu_{13}c^{-1}_{16} c_{64}\psi^{(3)}_{-\frac{1}{2}}\psi^{(1)}_{-\frac{1}{2}}+ c^{-1}_{16}c_{14}\mu_{16}\psi^{(6)}_{-\frac{1}{2}}\psi^{(2)}_{-\frac{1}{2}}\right)\ket{0}.
\end{aligned}
\end{equation}
Similarly we have 
\begin{align*}
  \psi^{(4)}_{-\frac{1}{2}}\psi^{(5)}_{-\frac{1}{2}}\ket{0}
  =&\left(\mu_{13}c^{-1}_{12} c_{25}\psi^{(3)}_{-\frac{1}{2}}\psi^{(1)}_{-\frac{1}{2}}+ \mu_{26}c^{-1}_{12}c_{15}\mu_{12}\psi^{(6)}_{-\frac{1}{2}}\psi^{(2)}_{-\frac{1}{2}}\right)\ket{0}.
\end{align*}
On the other hand, we know $\psi^{(5)}_n\psi^{(4)}_m=\mu_{54}\psi^{(4)}_m\psi^{(5)}_n$ from \eqref{eq:sl4gen-2}, so all together we must have 
\begin{equation}\label{eq:sl4para-0}
    \mu_{54}=\dfrac{c_{14}c_{21}}{\mu_{26}c_{61}c_{15}}=\dfrac{c_{12}c_{64}}{c_{16}c_{25}}.
\end{equation}
With the same spirit we note that
\begin{align*}
  &\psi^{(6)}_{-\frac{1}{2}}\psi^{(2)}_{-\frac{1}{2}}\ket{0}
  =\left(\mu_{13}c^{-1}_{15} c_{52}\psi^{(3)}_{-\frac{1}{2}}\psi^{(1)}_{-\frac{1}{2}}+ c^{-1}_{15}c_{12}\mu_{15}\psi^{(5)}_{-\frac{1}{2}}\psi^{(4)}_{-\frac{1}{2}}\right)\ket{0}\\
  =&\left(\mu_{13}c^{-1}_{15
  } c_{52}\psi^{(3)}_{-\frac{1}{2}}\psi^{(1)}_{-\frac{1}{2}}+ c^{-1}_{15}c_{12}\mu_{15}\left(\mu_{13}c^{-1}_{16} c_{64}\psi^{(3)}_{-\frac{1}{2}}\psi^{(1)}_{-\frac{1}{2}}+ c^{-1}_{16}c_{14}\mu_{16}\psi^{(6)}_{-\frac{1}{2}}\psi^{(2)}_{-\frac{1}{2}}\right)\right)\ket{0}
\end{align*}
which implies
\begin{equation}\label{eq:sl4para-1}
    c^{-1}_{15}c_{12}\mu_{15}c^{-1}_{16}c_{14}\mu_{16}=1\implies c_{12}c_{14}=c_{51}c_{61}
\end{equation}
and
\begin{equation}\label{eq:sl4para-2}
    c_{52}+ c_{12}\mu_{15}c^{-1}_{16} c_{64}=0\implies c_{52}c_{61}=c_{12}c_{64}.
\end{equation}

Now combining \eqref{eq:sl4para-0} - \eqref{eq:sl4para-2} and the associativity conditions, we can solve for compatible choices of the constants $\mu_{ij}$ and $c_{ij}$ and we choose the set of constants as follows:
\begin{equation}\label{eq:sl4para}
\begin{aligned}
    &c_{ij}=\mu_{ij}c_{ji}, \; \mu_{ij}\mu_{ji}=1,\\
    &\mu_{13}=\mu_{26}=1,\; \mu_{54}=-1, \\
    \mu_{12}=\mu_{24}=\mu_{41}=\mu_{15}=&\mu_{56}=\mu_{61}=\mu_{23}=\mu_{35}=\mu_{52}=\mu_{43}=\mu_{64}=\mu_{36}=x^2,\\
    c_{12}=c_{24}=c_{41}=c_{15}=&c_{56}=c_{61}=c_{23}=c_{35}=c_{52}=c_{43}=c_{64}=c_{36}=\dfrac{x}{\sqrt{2}}.
\end{aligned}
\end{equation}
where $x$ is again an 8th root of unity which is chosen to be $e^{-\frac{i\pi }{4}}$.

We would like to find a basis of the Fock space of $\widehat{\mathfrak{sl}(4)}_2/\widehat{\mathfrak{u}(1)}^3$ analogous to that of $\widehat{\mathfrak{sl}(3)}_2/\widehat{\mathfrak{u}(1)}^2$. Naively we would guess that the basis can be formed by states in the form of 
\begin{equation}\label{eq:sl4basis-pseudo}
    \begin{aligned}
      &\psi^{(6)}_{-N_6-\frac{N_1+N_3+N_4+N_5}{2}+\frac{1}{2}-s^{(6)}_{N_6}}\dots\psi^{(6)}_{-\frac{N_1+N_3+N_4+N_5}{2}-\frac{1}{2}-s^{(6)}_{1}}\psi^{(5)}_{-N_5-\frac{N_1+N_2+N_3}{2}+\frac{1}{2}-s^{(5)}_{N_5}}\dots\\&\dots\psi^{(5)}_{-\frac{N_1+N_2+N_3}{2}-\frac{1}{2}-s^{(5)}_{1}}\psi^{(4)}_{-N_4-\frac{N_1+N_2+N_3}{2}+\frac{1}{2}-s^{(4)}_{N_4}}\dots\psi^{(4)}_{-\frac{N_1+N_2+N_3}{2}-\frac{1}{2}-s^{(4)}_{1}}\psi^{(3)}_{-N_3-\frac{N_2}{2}+\frac{1}{2}-s^{(3)}_{N_3}}\dots\\&\dots\psi^{(3)}_{-\frac{N_2}{2}-\frac{1}{2}-s^{(3)}_{1}}\psi^{(2)}_{-N_2-\frac{N_1}{2}+\frac{1}{2}-s^{(2)}_{N_2}}\dots\psi^{(2)}_{-\frac{N_1}{2}-\frac{1}{2}-s^{(2)}_{1}}\psi^{(1)}_{-N_1+\frac{1}{2}-s^{(1)}_{N_1}}\dots\psi^{(1)}_{-\frac{1}{2}-s^{(1)}_{1}}\ket{0}
    \end{aligned}
\end{equation}
with $N_1,N_2,\dots,N_6\ge0$ and $s^{(i)}_N\ge\dots s^{(i)}_2\ge s^{(i)}_1\ge0$ for $i=1,2,\dots,6$. The spanning property of the set of all states in the form of \eqref{eq:sl4basis-pseudo} can be shown simply in the analogy of the proof of the spanning property of \eqref{eq:sl3basis-123}. 

However, as calculated in \eqref{eq:sl4uncprel}, the states $ \psi^{(3)}_{-\frac{1}{2}}\psi^{(1)}_{-\frac{1}{2}}\ket{0}, \psi^{(6)}_{-\frac{1}{2}}\psi^{(2)}_{-\frac{1}{2}}\ket{0}, \psi^{(5)}_{-\frac{1}{2}}\psi^{(4)}_{-\frac{1}{2}}\ket{0}$ already form a linearly dependent set, so \eqref{eq:sl4basis-pseudo} is overcomplete. One of the uncoupled pairs, which we choose to be $\{4,5\}$, can only contribute to the basis when the total energy of $\psi^{(4)}$- and $\psi^{(5)}$-modes is no less than 2. Hence the best we can hope for the basis states in a similar form of \eqref{eq:sl4basis-pseudo} would be those in the form of 
\begin{equation}\label{eq:sl4basis}
    \begin{aligned}
      &\psi^{(6)}_{-N_6-\frac{N_1+N_3+N_4+N_5}{2}+\frac{1}{2}-s^{(6)}_{N_6}}\dots\psi^{(6)}_{-\frac{N_1+N_3+N_4+N_5}{2}-\frac{1}{2}-s^{(6)}_{1}}\psi^{(5)}_{-N_5-N_4-\frac{N_1+N_2+N_3}{2}+\frac{1}{2}-s^{(5)}_{N_5}}\dots\\&\dots\psi^{(5)}_{-N_4-\frac{N_1+N_2+N_3}{2}-\frac{1}{2}-s^{(5)}_{1}}\psi^{(4)}_{-N_4-\frac{N_1+N_2+N_3}{2}+\frac{1}{2}-s^{(4)}_{N_4}}\dots\psi^{(4)}_{-\frac{N_1+N_2+N_3}{2}-\frac{1}{2}-s^{(4)}_{1}}\psi^{(3)}_{-N_3-\frac{N_2}{2}+\frac{1}{2}-s^{(3)}_{N_3}}\dots\\&\dots\psi^{(3)}_{-\frac{N_2}{2}-\frac{1}{2}-s^{(3)}_{1}}\psi^{(2)}_{-N_2-\frac{N_1}{2}+\frac{1}{2}-s^{(2)}_{N_2}}\dots\psi^{(2)}_{-\frac{N_1}{2}-\frac{1}{2}-s^{(2)}_{1}}\psi^{(1)}_{-N_1+\frac{1}{2}-s^{(1)}_{N_1}}\dots\psi^{(1)}_{-\frac{1}{2}-s^{(1)}_{1}}\ket{0}
    \end{aligned}
\end{equation}
with $N_1,N_2,\dots,N_6\ge0$ and $s^{(i)}_N\ge\dots s^{(i)}_2\ge s^{(i)}_1\ge0$ for $i=1,2,\dots,6$.

The lowest energy state in the form of \eqref{eq:sl4basis} with $N_i$ $\psi^{(i)}$-modes is
\begin{align*}
    &\psi^{(6)}_{-N_6-\frac{N_1+N_3+N_4+N_5}{2}+\frac{1}{2}}\dots\psi^{(6)}_{-\frac{N_1+N_3+N_4+N_5}{2}-\frac{1}{2}}\psi^{(5)}_{-N_5-N_4-\frac{N_1+N_2+N_3}{2}+\frac{1}{2}}\dots\\&\dots\psi^{(5)}_{-N_4-\frac{N_1+N_2+N_3}{2}-\frac{1}{2}}\psi^{(4)}_{-N_4-\frac{N_1+N_2+N_3}{2}+\frac{1}{2}}\dots\psi^{(4)}_{-\frac{N_1+N_2+N_3}{2}-\frac{1}{2}}\psi^{(3)}_{-N_3-\frac{N_2}{2}+\frac{1}{2}}\dots\\&\dots\psi^{(3)}_{-\frac{N_2}{2}-\frac{1}{2}}\psi^{(2)}_{-N_2-\frac{N_1}{2}+\frac{1}{2}}\dots\psi^{(2)}_{-\frac{N_1}{2}-\frac{1}{2}}\psi^{(1)}_{-N_1+\frac{1}{2}}\dots\psi^{(1)}_{-\frac{1}{2}}\ket{0}
\end{align*}
whose energy is 
\begin{align*}
    &\frac{1}{2}\left(\sum_{k=1}^{N_1}(2k-1)+\sum_{k=1}^{N_2}(N_1+2k-1)+\sum_{k=1}^{N_3}(N_2+2k-1)+\sum_{k=1}^{N_4}(N_1+N_2+N_3+2k-1)\right.\\
    &\left.+\sum_{k=1}^{N_5}(N_1+N_2+N_3+2N_4+2k-1)+\sum_{k=1}^{N_6}(N_1+N_3+N_4+N_5+2k-1)\right)\\
    =&\frac{1}{2}\left(N_1^2+N_2^2+N_3^2+N_4^2+N_5^2+N_6^2+N_1N_2+N_1N_4+N_1N_5+N_1N_6\right.\\    &\left.+N_2N_3+N_2N_4+N_2N_5+N_3N_4+N_3N_5+N_3N_6+2N_4N_5+N_4N_6+N_5N_6\right),
\end{align*}
or in a compact form,
\[\frac{1}{2}\bold{N}^T\cdot\bold{G}_4\cdot\bold{N},\]
where
\begin{align}\label{eq:G4-def}
\mathbf{G}_4:=\frac{1}{2}\begin{pmatrix}
2&1&0&1&1&1\\
1&2&1&1&1&0\\
0&1&2&1&1&1\\
1&1&1&2&2&1\\
1&1&1&2&2&1\\
1&0&1&1&1&2
\end{pmatrix}.
\end{align}

It is again verified using the dilogarithm\index{dilogarithm} \eqref{eq:dilog} with $\bold{G}=\bold{G}_4$ that the central charge $c$ here is, indeed, $2$ as we expect, which supports our conjectured basis.

The corresponding UCPF \index{universal chiral partition function (UCPF)} in this case reads (up to an overall factor)
\[\sum_{N_i\ge 0}\dfrac{q^{\frac{1}{2}\mathbf{N}^{\mathsf{T}}\cdot\mathbf{G}_4\cdot\mathbf{N}}}{\prod_i (q)_{N_i}},\]
where $\mathbf{N}^{\mathsf{T}}=(N_1,N_2,N_3,N_4,N_5,,N_6)$. If the basis is correct, we should expect
\begin{equation}\label{eq:sl4untwi}
    q^{-\frac{1}{12}}\sum_{N_i\ge 0}\dfrac{q^{\frac{1}{2}\mathbf{N}^{\mathsf{T}}\cdot\mathbf{G}_4\cdot\mathbf{N}}}{\prod_i (q)_{N_i}}=b^{2000}_{2000}+b^{2000}_{0020}+6b^{2000}_{0101}
\end{equation}
where we used the fact that the coset characters $b^{2000}_{\lambda}$ of the modules with primary fields of conformal weight $\frac{1}{2}$ in the untwisted sector are all the same considering \eqref{eq:strfunprop}. 

It is numerically verified, with the help of SageMath and Mathematica, that \eqref{eq:sl4untwi} indeed holds.  However, \eqref{eq:sl4untwi} can not be proved unless we have explicit expressions of $b^{\Lambda}_{\lambda}$ or, in other words, the string functions \index{string function} $c^{\Lambda}_{\lambda}$ of $\widehat{\mathfrak{sl}(4)}_2$. Here we propose their expressions as follows:
\begin{equation}\label{eq:sl4strfun}
    \arraycolsep=1.4pt\def\arraystretch{1.3}
    \begin{array}{lp{1.6cm}l}
         \multicolumn{3}{l}{b^{2000}_{2000}(\tau)+b^{2000}_{0020}(\tau)=\dfrac{\eta(4\tau)^4\eta(6\tau)^8}{\eta(\tau)\eta(2\tau)^4\eta(3\tau)^3\eta(12\tau)^4} +\dfrac{\eta(2\tau)^8\eta(3\tau)\eta(12\tau)^4}{\eta(\tau)^5\eta(4\tau)^4\eta(6\tau)^4},}\\
        b^{2000}_{2000}(\tau)-b^{2000}_{0020}(\tau) = \dfrac{\eta(\tau)^2}{\eta(2\tau)^2}, & & b^{2000}_{0101}(\tau) = \dfrac{\eta(2\tau)^2\eta(6\tau)^2}{\eta(\tau)^3\eta(3\tau)},\\
        b^{0101}_{2000}(\tau) = b^{0101}_{0002}(\tau) = \dfrac{3\eta(6\tau)^3}{\eta(\tau)^2\eta(2\tau)}, & & b^{0101}_{0101}(\tau) = \dfrac{\eta(2\tau)^3\eta(3\tau)^2}{\eta(\tau)^4\eta(6\tau)},\\
        b^{1100}_{1100}(\tau)=\dfrac{\eta(4\tau)^5}{\eta(\tau)^3\eta(8\tau)^2}, & & b^{1100}_{0011}(\tau)=\dfrac{2\eta(2\tau)^2\eta(8\tau)^2}{\eta(\tau)^3\eta(4\tau)}.
    \end{array}
\end{equation}

 The derivation of \eqref{eq:sl4strfun} and the proof of \eqref{eq:sl4untwi} require one of the most important observations in this project that will be presented in Chapter \ref{Chapter4}, so we leave them till then. However, we can confirm these expressions despite the derivation using the important Theorem \ref{thm:strfunmodtrans} given in \parencite{KACPETERSON}.
\begin{proof}[Proof of \eqref{eq:sl4strfun}]
On one hand, using Theorem \ref{thm:strfunmodtrans}(1) we know that $b^{\Lambda}_{\lambda}$'s transform as follows:
\begin{equation}\label{eq:sl4strfuntrans}
\begin{aligned}
    b^{2000}_{2000}\left(-\frac{1}{\tau}\right)-b^{2000}_{0020}\left(-\frac{1}{\tau}\right)&=2\left(b^{1100}_{1100}(\tau)+b^{1100}_{0011}(\tau)\right)\\
    b^{2000}_{2000}\left(-\frac{1}{\tau}\right)+b^{2000}_{0020}\left(-\frac{1}{\tau}\right)&=\dfrac{1}{2\sqrt{3}}\left(b^{2000}_{2000}(\tau)+b^{2000}_{0020}(\tau) + 6 b^{2000}_{0101}(\tau) + 6b^{0101}_{0101}(\tau) +  2b^{0101}_{2000}(\tau)\right)\\
    b^{0101}_{2000}\left(-\frac{1}{\tau}\right)=b^{0101}_{0002}\left(-\frac{1}{\tau}\right)&=\dfrac{1}{2\sqrt{3}}\left(b^{2000}_{2000}(\tau)+b^{2000}_{0020}(\tau) + 6 b^{2000}_{0101}(\tau) -3 b^{0101}_{0101}(\tau) -  b^{0101}_{2000}(\tau)\right)\\
    b^{2000}_{0101}\left(-\frac{1}{\tau}\right)&=\dfrac{1}{4\sqrt{3}}\left(b^{2000}_{2000}(\tau)+b^{2000}_{0020}(\tau) - 2 b^{2000}_{0101}(\tau) -2 b^{0101}_{0101}(\tau) + 2 b^{0101}_{2000}(\tau)\right)\\
    b^{0101}_{0101}\left(-\frac{1}{\tau}\right)&=\dfrac{1}{2\sqrt{3}}\left(b^{2000}_{2000}(\tau)+b^{2000}_{0020}(\tau) - 2 b^{2000}_{0101}(\tau) + b^{0101}_{0101}(\tau) - b^{0101}_{2000}(\tau)\right)\\
    b^{1100}_{1100}\left(-\frac{1}{\tau}\right)&=\dfrac{1}{4}\left(b^{2000}_{2000}(\tau)-b^{2000}_{0020}(\tau) + 2 b^{1100}_{1100}(\tau) -2 b^{1100}_{0011}(\tau)\right)\\
     b^{1100}_{0011}\left(-\frac{1}{\tau}\right)&=\dfrac{1}{4}\left(b^{2000}_{2000}(\tau)-b^{2000}_{0020}(\tau) - 2 b^{1100}_{1100}(\tau) +2 b^{1100}_{0011}(\tau)\right)
\end{aligned}
\end{equation}
On the other hand, modular transformations of the Dedekind eta functions can be easily computed using the property of $\eta(\tau)$ \eqref{eq:etamodtrans}. One readily observes that the right-hand sides of expressions in \eqref{eq:sl4strfun} transform exactly as in \eqref{eq:sl4strfuntrans}.

Now, using Theorem \ref{thm:strfunmodtrans}(2) and Sturm's bound for modular forms (see, for example, Theorem 6.4.7 in \parencite{murty2015problems}), we know that comparing coefficients of the analytical expressions of $b^{\Lambda}_{\lambda}$ with the numerical expressions of $\eta(\tau)^{3}c^{\Lambda}_{\lambda}$, computed from the definition of string functions, up to the order of $q^{100}$ is (more than) sufficient to conclude that $\eta(\tau)^{-3}b^{\Lambda}_{\lambda}=c^{\Lambda}_{\lambda}$. It is verified, with the help of SageMath and Mathematica, that the coefficients all match. Hence the expressions in \eqref{eq:sl4strfun} are confirmed. 
\end{proof}
With \eqref{eq:sl4untwi} proved, the only thing left to confirm is the spanning property of our conjectured basis. It is again done inductively just as the $\widehat{\mathfrak{sl}(3)}$ case, but it is unsurprisingly much more involved due to the presence of dependence for the states involving uncoupled pairs such as \eqref{eq:sl4uncprel}.
\begin{proof}[Proof of the spanning property of \eqref{eq:sl4basis}]
We say a state in the form of \eqref{eq:sl4basis-pseudo} is of pair-length $n+m$ if it is of length $n$ and $\min\{N_4,N_5\}=m$. We say a state of pair-length $n+m$ is well-ordered if it is $0$ or a linear combination of states in the form of \eqref{eq:sl4basis} of pair-length less or equal than $n+m$.

Let $P(n)$ be the statement that any state of pair-length $n$ can be well-ordered. $P(1)$ is obvious. Now suppose $P(n)$ for all $n<N$, we intend to prove $P(N)$. 

We have the following facts about \eqref{eq:sl4gen-1} - \eqref{eq:sl4gen-3}:
\begin{enumerate}[label=\textbf{F$'$.\arabic*}]
    \item They will not change the number of $\psi^{(i)}$-modes for each $i=1,\dots,6$ without changing the length of the state.\label{fact:sl4-basis-1}
    \item They will not change the total mode number of a state. \label{fact:sl4-basis-2}  
\end{enumerate}

By the inductive hypothesis, we can assume that $\ket{\psi}=\psi^{(i_N)}_{m_N}\ket{\psi'}$ where $\ket{\psi'}:=\psi^{(i_{N-1})}_{m_{N-1}}\cdots\psi^{(i_{N-N'})}_{m_{N-N'}}\ket{0}$ is a state in the form of \eqref{eq:sl4basis} with $N_i'$ $\psi^{(i)}$-modes. Note that if $i_N\notin\{4,5\}$ or it is the case that $i_N=5$ (resp. $i_N=4$) and $N'_4\le N'_5$ (resp. $N'_4\ge N'_5$), then $\ket{\psi'}$ is of length $N'$ and pair-length $N-1$, while if $N'_4> N'_5$ (resp. $N'_4< N'_5$) and $i_N=5$ (resp. $i_N=4$), then $\ket{\psi'}$ is of pair-length $N-2$.

The modification of the proof focuses on the following additional situations that could happen for the $\widehat{\mathfrak{sl}(4)}$ case, compared to the $\widehat{\mathfrak{sl}(3)}$ case: 
\begin{enumerate}[label=\textbf{D.\arabic*}]
    \item The right-hand-side of \eqref{eq:sl4gen-3} can not always be handled by the inductive hypothesis because the right-hand-side of \eqref{eq:sl4gen-3} can produce a state of pair-length $N$. This can only happen when using \eqref{eq:sl4gen-3} to interchange the first two modes in a state $\psi^{(i)}_{m_N}\psi^{(j)}_{m_{N-1}}\ket{\psi''}$ whose $\{i,j\}\in\big\{\{1,2\},\{3,6\}\big\}$ (resp. $\big\{\{2,3\},\{1,6\}\big\}$), and $\ket{\psi''}$ has $N''_4$ $\psi^{(4)}$-modes, $N''_5$ $\psi^{(5)}$-modes with $N''_4<N''_5$ (resp. $N''_4>N''_5$).\label{diff-1}
    \item When applying $P(n)$ with $n<N$, we can not always assume the states of pair-length $n$ in the resulting well-ordered expression has the same number of $\psi^{(i)}$-modes for each $i=1,\dots,6$, because we need to do operations as what have been done in \eqref{eq:sl4uncprel}. However, because of \ref{fact:sl4-basis-1}, we can carefully control our operations so that there are only two possible ways to produce a pair-length $n$ in the resulting well-ordered expression:
    \begin{enumerate}
        \item Using \eqref{eq:sl4gen-3} to interchange a $\psi^{(i)}$-mode and a $\psi^{(j)}$-mode with $\{i,j\}\in\big\{\{1,2\},\{3,6\}\big\}$ (resp. $\big\{\{2,3\},\{1,6\}\big\}$) for a state has more (resp. less) $\psi^{(5)}$-modes than $\psi^{(4)}$-modes. \label{diff-2-1}
        \item Reducing the number of $\{4,5\}$-pairs by expanding a $\psi^{(5)}$-mode in terms of $\psi^{(2)}$-modes and  $\psi^{(3)}$-modes using \eqref{eq:sl4gen-3}
\[\psi^{(5)}_{m}=c_{23} \sum_{l\ge0}\binom{l-\frac{1}{2}}{l}\left(\psi^{(2)}_{m+\epsilon-\frac{1}{2}-l}\psi^{(3)}_{-\epsilon+\frac{1}{2}+l}+\mu_{23}\psi^{(3)}_{-\epsilon-l}\psi^{(2)}_{m+\epsilon+l}\right)\tag{$\diamonddot$}\]
where $\epsilon$ is some integer or half-integer, depending on which module the $\psi^{(4)}$-mode or $\psi^{(5)}$-mode acts on.\label{diff-2-2}
    \end{enumerate}\label{diff-2}
    Note that neither of the above would increase the number of $\psi^{(6)}$-modes.
    \item For a state in the form of  $\psi^{(i)}_{m_{N}}\cdots\psi^{(i)}_{m_{N-p+1}}\ket{\psi'}$ with $i=4$ (resp. $i=5$) and $\ket{\psi'}$ of pair-length $N-p$, for some $p>0$, when applying $P(N-p)$ to $\ket{\psi'}$, say a state $\ket{\psi''}$ in the resulting well-ordered expression of pair-length $N-p-q$, for some $q>0$, has $N''_4$ $\psi^{(4)}$-modes and $N''_5$ $\psi^{(5)}$-modes. If $N''_5-N''_4=q$ (resp. $N''_4-N''_5=q$), then $\psi^{(i)}_{m_{N}}\cdots\psi^{(i)}_{m_{N-p+1}}\ket{\psi''}$ is of pair-length $N$. \label{diff-3}
\end{enumerate}
Suppose $i_N=i$ and $i_{N-1}=j$, we have the following cases:
\begin{enumerate}[label=\textbf{C$'$.\arabic*}]
\item If $i=j\notin \{4,5\}$, then the proof is similar to \ref{case:sl3-1}, where \eqref{eq:sl4gen-3} is not used (explicitly), so we do not need to consider \ref{diff-1}. Furthermore, if
\begin{enumerate}
    \item $i=1,2,3$, then since $N'_4=N'_5=0$ by assumption, none of \ref{diff-2-1} and \ref{diff-2-2} could happen.
    \item $i=6$, then by using \eqref{eq:sl4gen-1} repeatedly, we can apply $P(N-N'_6)$ on the  state $\psi^{(6)}_{m_{N}}\ket{\psi''}$ with $\ket{\psi''}$ containing only $N'_j$ $\psi^{(j)}$-modes for $j=1,\dots,5$. Although any of \ref{diff-2} could happen when applying $P(N-N'_6)$, they would not change the total number of $\psi^{(1)}-,\psi^{(3)}-,\psi^{(4)}-,$ $\psi^{(5)}-$modes. Hence, with the fact that $m_{N-N'_6}\le-\frac{N'_1+N'_3+N'_4+N'_5}{2}-\frac{1}{2}$ by assumption and that \ref{diff-2} would not increase the number of $\psi^{(6)}-$modes, we know that in the resulting well-ordered expression of $\psi^{(6)}_{m_{N}}\ket{\psi''}$, any term with no $\psi^{(6)}-$modes is well-ordered, while any other terms of pair-length $N-N'_6$ can be well-ordered following \ref{case:sl3-1-4}. \label{case:sl4-1-2}
\end{enumerate}
\item $(i,j)\in\{(3,1),(6,2)\}$, then the proof is similar to \ref{case:sl3-1} with \eqref{eq:sl3com-sim-1} replaced by \eqref{eq:sl4gen-2}. Since $N'_4=N'_5=0$ by assumption, none of \ref{diff-2-1} or \ref{diff-2-2} could happen.
\item For some $\{i,j\}\notin\Xi$, the proof can be done based on \ref{case:sl3-2}. We consider the following cases:
\begin{enumerate}
  \item $(i,j)\in\{(2,1),(3,2),(6,1),(6,3)\}$, then since $N'_4=N'_5=0$ by assumption, so none of \ref{diff-1} and \ref{diff-2} could happen.
        
        \item $i\in\{4,5\}$ and $j\le 3$,  then it is obvious that \ref{diff-1} cannot happen. When applying $P(n)$ to ($\meddiamond$), since either $N'_4=0$ or $N'_5=0$, we know \ref{diff-2-2} cannot happen, while if \ref{diff-2-1} happens, then we would have terms in the form of $\psi^{(j)}_{m_{N-1}-\frac{1}{2}-l_1}\psi^{(5)}_{m'_{N''}}\psi^{(4)}_{m'_{N''-1}}\ket{\psi''}$ with $\ket{\psi''}$ of pair-length $N-4$ and having $N''_k$ $\psi^{(k)}$-modes for $k=1,2,3$ only. We see $m'_{N''}\le -2-\frac{N''_1+N''_2+N''_3}{2}+\frac{1}{2}$, so by \eqref{eq:sl4gen-3}, we have
        \begin{align*}
    &\psi^{(j)}_{m_{N-1}-\frac{1}{2}-l_1}\psi^{(5)}_{m'_{N''}}\psi^{(4)}_{m'_{N''-1}}\ket{\psi''}\\
    \sim&\sum_{r_1\ge 1}\psi^{(j)}_{m_{N-1}-\frac{1}{2}-l_1-r_1}\psi^{(5)}_{m'_{N''}+r_1}\psi^{(4)}_{m'_{N''-1}}\ket{\psi''}\\ 
    +& \sum_{r_1\ge 0} \psi^{(5)}_{m'_{N''}-\frac{1}{2}-r_1}\psi^{(j)}_{m_{N-1}-l_1+r_1}\psi^{(4)}_{m'_{N''-1}}\ket{\psi''}.\tag{$\filleddiamond'$}
\end{align*}
    Note that \ref{diff-1} cannot happen in this step. The rest follows as in \ref{case:sl3-2b}. While applying \ref{case:sl3-2b}, if \ref{diff-2-1}  happens or \ref{diff-3} happens without $\psi^{(6)}-$modes involved when applying  $P(N-1)$ to ($\filleddiamond'$), the argument is still true, since they would not increase the total of twice the number of $\psi^{(5)}-,\psi^{(4)}-$modes and the number of $\psi^{(1)}-,\psi^{(2)}-,\psi^{(3)}-$modes for $\psi^{(j)}_{m_{N-1}-l_1+r_1}\psi^{(4)}_{m'_{N''-1}}\ket{\psi''}$. If \ref{diff-3} happens with $\psi^{(6)}$-modes involved, then we use \eqref{eq:sl4gen-3} and do a similar process as above, where if \ref{diff-2-1} happens again, then it leads to \ref{case:sl4-1-2}.
\end{enumerate}\label{case:sl4-3}

\item If $i=1$ and $j=3$, then by using \eqref{eq:sl4gen-2} and then $P(N-1)$ we can convert it to cases with $i_N=3$ and $i_{N-1}\le 3$, all of which have been done above.

    \item If $i\in\{4,5\}$ and $j=6$, then by using \eqref{eq:sl4gen-3}, we have 
     \begin{align*}
                 &\psi^{(i)}_{m_N}\psi^{(6)}_{m_{N-1}}\psi^{(i_{N-2})}_{m_{N-2}}\cdots \psi^{(i_{N-N'})}_{m_{N-N'}}\ket{0}\\
           \sim&\sum_{l_1\ge 1}\psi^{(i)}_{m_N-l_1}\psi^{(6)}_{m_{N-1}+l_1}\psi^{(i_{N-2})}_{m_{N-2}}\cdots \psi^{(i_{N-N'})}_{m_{N-N'}}\ket{0}\tag{$\medcircle'$}\\
           +&\sum_{l_1\ge 0}\psi^{(6)}_{m_{N-1}-\frac{1}{2}-l_1}\psi^{(i)}_{m_N+\frac{1}{2}+l_1}\psi^{(i_{N-2})}_{m_{N-2}}\cdots \psi^{(i_{N-N'})}_{m_{N-N'}}\ket{0}.\tag{$\meddiamond'$}
            \end{align*}
Note that \ref{diff-1} cannot happen at this step. With a similar argument as in \ref{case:sl4-1-2}, we see that ($\meddiamond'$) can be well-ordered by applying $P(N-1)$ even if \ref{diff-2} happens. Therefore $\ket{\psi}$ can be well-ordered using ($\medcircle'$) repeatedly, as argued in \ref{case:sl3-2b}.

\item  For some $\{i,j\}\notin\Xi$, the proof can be done based on \ref{case:sl3-3}, for which we only need to deal with \ref{diff-1}. We consider the following cases:
\begin{enumerate}
    \item If $i\in\{1,2,3,6\}$ and $j\in\{4,5\}$, then \ref{diff-1} cannot happen as $(i,j)\notin\big\{(1,2),(3,6),(2,3),(1,6)\big\}$.
    
    \item If $(i,j)\in\{(1,2),(2,3)\}$, then since $N'_4=N'_5=0$, \ref{diff-1} cannot happen.
\end{enumerate}

\item If $i=2$ and $j=6$, then by using \eqref{eq:sl4gen-2} and then $P(N-1)$ we can convert it to cases with $i_N=6$, all of which have been done above.

\item If $i=4$ and $j=5$, denote $\ket{\psi''}:=\psi^{(i_{N-2})}_{m_{N-2}}\cdots \psi^{(i_{N-N'})}_{m_{N-N'}}\ket{0}$, and then by using ($\diamonddot$) in \ref{diff-2-2}, we have
                \[\psi^{(4)}_{m_N}\psi^{(5)}_{m_{N-1}}\ket{\psi''}
                \sim \sum_l \psi^{(4)}_{m_N}\psi^{(2)}_{m_{N-1}+\epsilon-\frac{1}{2}-l}\psi^{(3)}_{-\epsilon+\frac{1}{2}+l}\ket{\psi''}+\psi^{(4)}_{m_N}\psi^{(3)}_{-\epsilon-l}\psi^{(2)}_{m_{N-1}+\epsilon+l}\ket{\psi''}.\]
            We can choose a large enough $\epsilon$ so that $\psi^{(2)}_{m_{N-1}+\epsilon+l}\ket{\psi''}=0$ for all $l$ by \ref{fact:sl4-basis-2}. For the remaining terms, by using \eqref{eq:sl4gen-3} we have
            \begin{align*}                
                &\psi^{(4)}_{m_N}\psi^{(2)}_{m_{N-1}+\epsilon-\frac{1}{2}-l}\psi^{(3)}_{-\epsilon+\frac{1}{2}+l}\ket{\psi''}\\
                \sim& \sum_{l_1\ge 1}\psi^{(4)}_{m_N-l_1}\psi^{(2)}_{m_{N-1}+\epsilon-\frac{1}{2}-l+l_1}\psi^{(3)}_{-\epsilon+\frac{1}{2}+l}\ket{\psi''}\\
                +&\sum_{l_1\ge 0}\psi^{(2)}_{m_{N-1}+\epsilon-1-l-l_1}\psi^{(4)}_{m_N+\frac{1}{2}+l_1}\psi^{(3)}_{-\epsilon+\frac{1}{2}+l}\ket{\psi''}\tag{$\meddiamond''$}.
            \end{align*}
            Note that \ref{diff-1} cannot happen at this step. Now, by consulting the cases with $i_N=2$, all of which have been done above, we see that ($\meddiamond''$) can be well-ordered and therefore $\ket{\psi}$ can be well-ordered as argued in \ref{case:sl3-2b}.\label{case:sl4-8}

\item If $i=5$ and $j=4$, then by using \eqref{eq:sl4gen-2}, we convert it to case \ref{case:sl4-8}. \label{case:sl4-9}

\item If $i=j=5$ and 
        \begin{enumerate}
            \item  $N'_4=0$, the proof is based on \ref{case:sl3-1-4}. By using \eqref{eq:sl4gen-1} repeatedly, we can apply $P(n)$ with some $n<N$ on the state $\psi^{(5)}_{m_{N}}\ket{\psi''}$ with $\ket{\psi''}$ containing only $N'_j$ $\psi^{(j)}$-modes for $j=1,\dots,3$, so \ref{diff-2-2} cannot happen. Although \ref{diff-2-1} could happen, any resulting term, say, with $N''_j$ $\psi^{(j)}$-modes, will satisfy $N''_6=0$ and $(N''_1+N''_2+N''_3)/2+N''_4\le (N'_1+N'_2+N'_3)/2$, so the rest of \ref{case:sl3-1-4} still follows.
            \item $N'_4\ne 0$, by using \eqref{eq:sl4gen-2}, we convert it to case \ref{case:sl4-9}. 
        \end{enumerate}

\item If $i=j=4$, then the proof is again based on \ref{case:sl3-1}. By using \eqref{eq:sl4gen-1} repeatedly, we can apply $P(N-1)$ to $\psi^{(4)}_{m_N}\ket{\psi''}$ with $\ket{\psi''}$ contains no $\psi^{(4)}-,\psi^{(5)}-,\psi^{(6)}-$modes, so \ref{diff-2-2} cannot happen. If \ref{diff-2-1} or \ref{diff-3} happens, we would end up with the cases where $i_N=4$ and $i_{N-1}\ge 5$ because of \ref{fact:sl4-basis-1}, which have both been done above. Otherwise, the proof follows \ref{case:sl3-1-4}. 

\item $i=1$ (resp. $i=3$) and $j=6$, then the proof is again based on \ref{case:sl3-3}. If \ref{diff-1} happens, we need to consult the cases with $i_N=5$ (resp. $i_N=4$), all of which have been done above.
\end{enumerate}
\end{proof}
Similarly, we can find a basis for the $\frac{1}{6}$-twisted sector where the states are in the form of
\begin{equation}\label{eq:sl4basistwi-1}
    \begin{aligned}
      &\psi^{(6)}_{-N_6-\frac{N_1+N_3+N_4+N_5}{2}+1-s^{(6)}_{N_6}}\dots\psi^{(6)}_{-\frac{N_1+N_3+N_4+N_5}{2}-s^{(6)}_{1}}\psi^{(5)}_{-N_5-N_4-\frac{N_1+N_2+N_3}{2}+1-s^{(5)}_{N_5}}\dots\\
      &\dots\psi^{(5)}_{-N_4-\frac{N_1+N_2+N_3}{2}-s^{(5)}_{1}}\psi^{(4)}_{-N_4-\frac{N_1+N_2+N_3}{2}+1-s^{(4)}_{N_4}}\dots\psi^{(4)}_{-\frac{N_1+N_2+N_3}{2}-s^{(4)}_{1}}\psi^{(3)}_{-N_3-\frac{N_2}{2}+\frac{1}{2}-s^{(3)}_{N_3}}\dots\\
      &\dots\psi^{(3)}_{-\frac{N_2}{2}-\frac{1}{2}-s^{(3)}_{1}}\psi^{(2)}_{-N_2-\frac{N_1}{2}+1-s^{(2)}_{N_2}}\dots\psi^{(2)}_{-\frac{N_1}{2}-s^{(2)}_{1}}\psi^{(1)}_{-N_1+\frac{1}{2}-s^{(1)}_{N_1}}\dots\psi^{(1)}_{-\frac{1}{2}-s^{(1)}_{1}}\ket{1'}
    \end{aligned}
\end{equation}
with $N_1,N_2,\dots,N_6\ge0$ and $s^{(i)}_N\ge\dots s^{(i)}_2\ge s^{(i)}_1\ge0$ for $i=1,2,\dots,6$, where $\ket{1'}$ is the parefermionic primary state in correspondence to $\Phi^{0101}_{12-10}$ that satisfies $L^{(i)}_0\ket{1'}=0$ for $i=1,3$ and $L^{(i)}_0\ket{1'}=\frac{1}{16}\ket{1'}$ if $i=2,4,5,6$.

The lowest energy state in the form of \eqref{eq:sl4basistwi-1} with $N_i$ $\psi^{(i)}$-modes, $1\le i\le 6$, is
\begin{align*}
    &\psi^{(6)}_{-N_6-\frac{N_1+N_3+N_4+N_5}{2}+1}\dots\psi^{(6)}_{-\frac{N_1+N_3+N_4+N_5}{2}}\psi^{(5)}_{-N_5-N_4-\frac{N_1+N_2+N_3}{2}+1}\dots\\&\dots\psi^{(5)}_{-N_4-\frac{N_1+N_2+N_3}{2}}\psi^{(4)}_{-N_4-\frac{N_1+N_2+N_3}{2}+1}\dots\psi^{(4)}_{-\frac{N_1+N_2+N_3}{2}}\psi^{(3)}_{-N_3-\frac{N_2}{2}+\frac{1}{2}}\dots\\&\dots\psi^{(3)}_{-\frac{N_2}{2}-\frac{1}{2}}\psi^{(2)}_{-N_2-\frac{N_1}{2}+1}\dots\psi^{(2)}_{-\frac{N_1}{2}}\psi^{(1)}_{-N_1+\frac{1}{2}}\dots\psi^{(1)}_{-\frac{1}{2}}\ket{1'}
\end{align*}
whose energy is 
\[\frac{1}{2}\bold{N}^T\cdot\bold{G}_4\cdot\bold{N}-\frac{1}{2}(N_2+N_4+N_5+N_6).\]
Thus the corresponding UCPF is expected to satisfy
\begin{equation}\label{eq:sl4twi-1}
    q^{\frac{1}{6}-\frac{1}{12}}\sum_{N_i\ge 0}\dfrac{q^{\frac{1}{2}\mathbf{N}^{\mathsf{T}}\cdot\mathbf{G}_4\cdot\mathbf{N}-\frac{1}{2}(N_2+N_4+N_5+N_6)}}{\prod_i (q)_{N_i}}=2b^{0101}_{2000}+6b^{0101}_{0101},
\end{equation}
which is proved in the same manner of the proof of \eqref{eq:sl4untwi}. 

For the $\frac{1}{8}$-twisted sector, a basis is formed by the states in the form of
\begin{equation}\label{eq:sl4basistwi-2}
    \begin{aligned}
      &\psi^{(6)}_{-N_6-\frac{N_1+N_3+N_4+N_5}{2}+\frac{1}{2}-s^{(6)}_{N_6}}\dots\psi^{(6)}_{-\frac{N_1+N_3+N_4+N_5}{2}-\frac{1}{2}-s^{(6)}_{1}}\psi^{(5)}_{-N_5-N_4-\frac{N_1+N_2+N_3}{2}+1-s^{(5)}_{N_5}}\dots\\&\dots\psi^{(5)}_{-N_4-\frac{N_1+N_2+N_3}{2}-s^{(5)}_{1}}\psi^{(4)}_{-N_4-\frac{N_1+N_2+N_3}{2}+\frac{1}{2}-s^{(4)}_{N_4}}\dots\psi^{(4)}_{-\frac{N_1+N_2+N_3}{2}-\frac{1}{2}-s^{(4)}_{1}}\psi^{(3)}_{-N_3-\frac{N_2}{2}+\frac{1}{2}-s^{(3)}_{N_3}}\dots\\&\dots\psi^{(3)}_{-\frac{N_2}{2}-\frac{1}{2}-s^{(3)}_{1}}\psi^{(2)}_{-N_2-\frac{N_1}{2}+1-s^{(2)}_{N_2}}\dots\psi^{(2)}_{-\frac{N_1}{2}-s^{(2)}_{1}}\psi^{(1)}_{-N_1+1-s^{(1)}_{N_1}}\dots\psi^{(1)}_{-s^{(1)}_{1}}\ket{1''}
    \end{aligned}
\end{equation}
with $N_1,N_2,\dots,N_6\ge0$ and $s^{(i)}_N\ge\dots s^{(i)}_2\ge s^{(i)}_1\ge0$ for $i=1,2,\dots,6$. where $\ket{1''}$ is the parefermionic primary state in correspondence to $\Phi^{1100}_{1100}$ that satisfies $L^{(i)}_0\ket{1'}=0$ for $i=3,4,6$ and $L^{(i)}_0\ket{1''}=\frac{1}{16}\ket{1''}$ if $i=1,2,5$.

The lowest energy state in the form of \eqref{eq:sl4basistwi-2} with $N_i$ $\psi^{(i)}$-modes 
is
\begin{align*}
    &\psi^{(6)}_{-N_6-\frac{N_1+N_3+N_4+N_5}{2}+\frac{1}{2}}\dots\psi^{(6)}_{-\frac{N_1+N_3+N_4+N_5}{2}-\frac{1}{2}}\psi^{(5)}_{-N_5-N_4-\frac{N_1+N_2+N_3}{2}+1}\dots\\&\dots\psi^{(5)}_{-N_4-\frac{N_1+N_2+N_3}{2}}\psi^{(4)}_{-N_4-\frac{N_1+N_2+N_3}{2}+\frac{1}{2}}\dots\psi^{(4)}_{-\frac{N_1+N_2+N_3}{2}-\frac{1}{2}}\psi^{(3)}_{-N_3-\frac{N_2}{2}+\frac{1}{2}}\dots\\&\dots\psi^{(3)}_{-\frac{N_2}{2}-\frac{1}{2}}\psi^{(2)}_{-N_2-\frac{N_1}{2}+1}\dots\psi^{(2)}_{-\frac{N_1}{2}}\psi^{(1)}_{-N_1+1}\dots\psi^{(1)}_{0}\ket{1''}
\end{align*}
whose energy is 
\[\frac{1}{2}\bold{N}^T\cdot\bold{G}_4\cdot\bold{N}-\frac{1}{2}(N_1+N_2+N_5),\]
so the UCPF in this case is expected to satisfy
\begin{equation}\label{eq:sl4twi-2}
    q^{\frac{1}{8}-\frac{1}{12}}\sum_{N_i\ge 0}\dfrac{q^{\frac{1}{2}\mathbf{N}^{\mathsf{T}}\cdot\mathbf{G}_4\cdot\mathbf{N}-\frac{1}{2}(N_1+N_2+N_5)}}{\prod_i (q)_{N_i}}=4b^{1100}_{1100}+4b^{1100}_{0011}.
\end{equation}
The proof of \eqref{eq:sl4twi-2} uses a different approach from those of \eqref{eq:sl4untwi} and \eqref{eq:sl4twi-1}, which does not rely on any result in Chapter \ref{Chapter4}, but we leave the sketch of this proof to Chapter \ref{Chapter4} on purpose.

We shall investigate the decomposition of the modules of $\widehat{\mathfrak{sl}(4)}_2/\widehat{\mathfrak{u}(1)}^3$ in terms of the modules of minimal models. As stated in \eqref{eq:CFFMINcc}, we expect that the modules $\mathcal{L}^{\Lambda}_{\lambda}$ can be decomposed to the sum of $\mathcal{L}\left(\frac{1}{2},0\right)\otimes \mathcal{L}\left(\frac{7}{10},0\right)\otimes \mathcal{L}\left(\frac{4}{5},0\right)$-modules in the form of $$\mathcal{L}\left(\frac{1}{2},h^{(1)}_{r,s}\right)\otimes \mathcal{L}\left(\frac{7}{10},h^{(2)}_{r',s'}\right)\otimes \mathcal{L}\left(\frac{4}{5},h^{(3)}_{r'',s''}\right).$$
We choose to deconstruct the total energy-momentum tensor of $\widehat{\mathfrak{sl}(4)}_2/\widehat{\mathfrak{u}(1)}^3$ as
\begin{align*}
   &\quad L^{(\frac{1}{2})}=L^{(1)},&L^{(\frac{7}{10})}=-\frac{1}{5}L^{(1)}+\frac{4}{5}\left(L^{(2)}+L^{(4)}\right),
\end{align*}
\[L^{(\frac{4}{5})}=-\frac{2}{15}\left(L^{(1)}+L^{(2)}+L^{(4)}\right)+\frac{2}{3}\left(L^{(3)}+L^{(5)}+L^{(6)}\right),\]
where the fractional superscripts indicate their central charges.

\label{sl4decomp}
Comparing the coefficients of the coset characters $b^{\Lambda}_{\lambda}$ and the characters of minimal modules (up to a sufficiently large order of $q$), we observe the following identities:
\begingroup
\renewcommand*{\arraystretch}{1.5}
\begin{align*}
b^{2000}_{2000}=&\text{ch}\textstyle\left[\mathcal{L}\left(\frac{1}{2},0\right)\otimes \mathcal{L}\left(\frac{7}{10},0\right)\otimes \mathcal{L}\left(\frac{4}{5},0\right)\boldsymbol{\oplus} \mathcal{L}\left(\frac{1}{2},0\right)\otimes \mathcal{L}\left(\frac{7}{10},\frac{3}{5}\right)\otimes \mathcal{L}\left(\frac{4}{5},\frac{7}{5}\right)\right.\\
&\left.\textstyle\boldsymbol{\oplus} \mathcal{L}\left(\frac{1}{2},\frac{1}{2}\right)\otimes \mathcal{L}\left(\frac{7}{10},\frac{3}{2}\right)\otimes \mathcal{L}\left(\frac{4}{5},0\right)\boldsymbol{\oplus} \mathcal{L}\left(\frac{1}{2},\frac{1}{2}\right)\otimes \mathcal{L}\left(\frac{7}{10},\frac{1}{10}\right)\otimes \mathcal{L}\left(\frac{4}{5},\frac{7}{5}\right)\right]\\
   b^{2000}_{0020}= &\text{ch}\textstyle\left[\mathcal{L}\left(\frac{1}{2},0\right)\otimes \mathcal{L}\left(\frac{7}{10},\frac{3}{5}\right)\otimes \mathcal{L}\left(\frac{4}{5},\frac{2}{5}\right)\boldsymbol{\oplus} \mathcal{L}\left(\frac{1}{2},\frac{1}{2}\right)\otimes \mathcal{L}\left(\frac{7}{10},\frac{1}{10}\right)\otimes \mathcal{L}\left(\frac{4}{5},\frac{2}{5}\right)\right.\\
   &\left.\textstyle\boldsymbol{\oplus}\mathcal{L}\left(\frac{1}{2},0\right)\otimes \mathcal{L}\left(\frac{7}{10},0\right)\otimes \mathcal{L}\left(\frac{4}{5},3\right)\boldsymbol{\oplus} \mathcal{L}\left(\frac{1}{2},\frac{1}{2}\right)\otimes \mathcal{L}\left(\frac{7}{10},\frac{3}{2}\right)\otimes \mathcal{L}\left(\frac{4}{5},3\right)\right],\\ 
   b^{2000}_{0101}= &\text{ch}\textstyle\left[\mathcal{L}\left(\frac{1}{2},\frac{1}{16}\right)\otimes \mathcal{L}\left(\frac{7}{10},\frac{7}{16}\right)\otimes \mathcal{L}\left(\frac{4}{5},0\right)\boldsymbol{\oplus} \mathcal{L}\left(\frac{1}{2},\frac{1}{16}\right)\otimes \mathcal{L}\left(\frac{7}{10},\frac{3}{80}\right)\otimes \mathcal{L}\left(\frac{4}{5},\frac{7}{5}\right)\right]\\=&\text{ch}\textstyle\left[\mathcal{L}\left(\frac{1}{2},\frac{1}{16}\right)\otimes \mathcal{L}\left(\frac{7}{10},\frac{3}{80}\right)\otimes \mathcal{L}\left(\frac{4}{5},\frac{2}{5}\right)\boldsymbol{\oplus} \mathcal{L}\left(\frac{1}{2},\frac{1}{16}\right)\otimes \mathcal{L}\left(\frac{7}{10},\frac{7}{16}\right)\otimes \mathcal{L}\left(\frac{4}{5},3\right)\right],\\
    b^{0101}_{2000}=&\text{ch}\textstyle\left[\mathcal{L}\left(\frac{1}{2},0\right)\otimes \mathcal{L}\left(\frac{7}{10},0\right)\otimes \mathcal{L}\left(\frac{4}{5},\frac{2}{3}\right)\boldsymbol{\oplus} \mathcal{L}\left(\frac{1}{2},0\right)\otimes \mathcal{L}\left(\frac{7}{10},\frac{3}{5}\right)\otimes \mathcal{L}\left(\frac{4}{5},\frac{1}{15}\right)\right.\\
    &\left.\textstyle\boldsymbol{\oplus}\mathcal{L}\left(\frac{1}{2},\frac{1}{2}\right)\otimes \mathcal{L}\left(\frac{7}{10},\frac{1}{10}\right)\otimes \mathcal{L}\left(\frac{4}{5},\frac{1}{15}\right)\boldsymbol{\oplus} \mathcal{L}\left(\frac{1}{2},\frac{1}{2}\right)\otimes \mathcal{L}\left(\frac{7}{10},\frac{3}{2}\right)\otimes \mathcal{L}\left(\frac{4}{5},\frac{2}{3}\right)\right],\\
    b^{0101}_{0101}=&\text{ch}\textstyle\left[\mathcal{L}\left(\frac{1}{2},\frac{1}{16}\right)\otimes \mathcal{L}\left(\frac{7}{10},\frac{3}{80}\right)\otimes \mathcal{L}\left(\frac{4}{5},\frac{1}{15}\right)\boldsymbol{\oplus} \mathcal{L}\left(\frac{1}{2},\frac{1}{16}\right)\otimes \mathcal{L}\left(\frac{7}{10},\frac{7}{16}\right)\otimes \mathcal{L}\left(\frac{4}{5},\frac{2}{3}\right)\right],\\
    b^{1100}_{1100}=&\text{ch}\textstyle\left[\mathcal{L}\left(\frac{1}{2},0\right)\otimes \mathcal{L}\left(\frac{7}{10},0\right)\otimes \mathcal{L}\left(\frac{4}{5},\frac{1}{8}\right)\boldsymbol{\oplus} \mathcal{L}\left(\frac{1}{2},\frac{1}{2}\right)\otimes \mathcal{L}\left(\frac{7}{10},\frac{3}{5}\right)\otimes \mathcal{L}\left(\frac{4}{5},\frac{1}{40}\right)\right.\\
    &\left.\textstyle\boldsymbol{\oplus}\mathcal{L}\left(\frac{1}{2},\frac{1}{2}\right)\otimes \mathcal{L}\left(\frac{7}{10},\frac{1}{10}\right)\otimes \mathcal{L}\left(\frac{4}{5},\frac{21}{40}\right)\boldsymbol{\oplus} \mathcal{L}\left(\frac{1}{2},0\right)\otimes \mathcal{L}\left(\frac{7}{10},\frac{3}{2}\right)\otimes \mathcal{L}\left(\frac{4}{5},\frac{13}{8}\right)\right]\\=&\text{ch}\textstyle\left[\mathcal{L}\left(\frac{1}{2},\frac{1}{16}\right)\otimes \mathcal{L}\left(\frac{7}{10},\frac{3}{80}\right)\otimes \mathcal{L}\left(\frac{4}{5},\frac{1}{40}\right)\boldsymbol{\oplus} \mathcal{L}\left(\frac{1}{2},\frac{1}{16}\right)\otimes \mathcal{L}\left(\frac{7}{10},\frac{7}{16}\right)\otimes \mathcal{L}\left(\frac{4}{5},\frac{13}{8}\right)\right],\\
    b^{1100}_{0011}=&\text{ch}\textstyle\left[\mathcal{L}\left(\frac{1}{2},0\right)\otimes \mathcal{L}\left(\frac{7}{10},\frac{3}{5}\right)\otimes \mathcal{L}\left(\frac{4}{5},\frac{1}{40}\right)\boldsymbol{\oplus} \mathcal{L}\left(\frac{1}{2},\frac{1}{2}\right)\otimes \mathcal{L}\left(\frac{7}{10},\frac{1}{10}\right)\otimes \mathcal{L}\left(\frac{4}{5},\frac{1}{40}\right)\right.\\
    &\left.\textstyle\boldsymbol{\oplus} \mathcal{L}\left(\frac{1}{2},0\right)\otimes \mathcal{L}\left(\frac{7}{10},0\right)\otimes \mathcal{L}\left(\frac{4}{5},\frac{13}{8}\right)\boldsymbol{\oplus} \mathcal{L}\left(\frac{1}{2},\frac{1}{2}\right)\otimes \mathcal{L}\left(\frac{7}{10},\frac{3}{2}\right)\otimes \mathcal{L}\left(\frac{4}{5},\frac{13}{8}\right)\right]\\
   =&\text{ch}\textstyle\left[\mathcal{L}\left(\frac{1}{2},\frac{1}{16}\right)\otimes \mathcal{L}\left(\frac{7}{10},\frac{3}{80}\right)\otimes \mathcal{L}\left(\frac{4}{5},\frac{21}{40}\right)\boldsymbol{\oplus} \mathcal{L}\left(\frac{1}{2},\frac{1}{16}\right)\otimes \mathcal{L}\left(\frac{7}{10},\frac{7}{16}\right)\otimes \mathcal{L}\left(\frac{4}{5},\frac{1}{8}\right)\right].
\end{align*}   
\endgroup
Keeping in mind that we also have the following identities:
\begingroup
\renewcommand*{\arraystretch}{1.5}
\begin{equation*}
    \begin{aligned}
    \text{ch}\textstyle\left[\mathcal{L}\left(\frac{1}{2},\frac{1}{16}\right)\otimes \left(\frac{7}{10},\frac{7}{16}\right)\right]=&\;\text{ch}\textstyle\left[\mathcal{L}\left(\frac{1}{2},\frac{1}{2}\right)\otimes \mathcal{L}\left(\frac{7}{10},0\right)\boldsymbol{\oplus} \mathcal{L}\left(\frac{1}{2},0\right)\otimes \mathcal{L}\left(\frac{7}{10},\frac{3}{2}\right)\right],\\ 
    \text{ch}\textstyle\left[\mathcal{L}\left(\frac{1}{2},\frac{1}{16}\right)\otimes \mathcal{L}\left(\frac{7}{10},\frac{3}{80}\right)\right]=&\;\text{ch}\textstyle\left[\mathcal{L}\left(\frac{1}{2},0\right)\otimes \mathcal{L}\left(\frac{7}{10},\frac{1}{10}\right)\boldsymbol{\oplus} \mathcal{L}\left(\frac{1}{2},\frac{1}{2}\right)\otimes \mathcal{L}\left(\frac{7}{10},\frac{3}{5}\right)\right],\\
    \text{ch}\textstyle\left[\mathcal{L}\left(\frac{1}{2},\frac{1}{16}\right)\otimes \mathcal{L}\left(\frac{7}{10},\frac{7}{16}\right)\right]=&\;\text{ch}\textstyle\left[\mathcal{L}\left(\frac{1}{2},\frac{1}{2}\right)\otimes \mathcal{L}\left(\frac{7}{10},0\right)\boldsymbol{\oplus} \mathcal{L}\left(\frac{1}{2},0\right)\otimes \mathcal{L}\left(\frac{7}{10},\frac{3}{2}\right)\right],\\
    \text{ch}\textstyle\left[\mathcal{L}\left(\frac{1}{2},\frac{1}{16}\right)\otimes \mathcal{L}\left(\frac{7}{10},\frac{3}{80}\right)\right]=& \;\text{ch}\textstyle\left[\mathcal{L}\left(\frac{1}{2},0\right)\otimes \mathcal{L}\left(\frac{7}{10},\frac{1}{10}\right)\boldsymbol{\oplus} \mathcal{L}\left(\frac{1}{2},\frac{1}{2}\right)\otimes \mathcal{L}\left(\frac{7}{10},\frac{3}{5}\right)\right],
\end{aligned}
\end{equation*}
\endgroup
we see that, in terms of minimal model modules, there are 8 different ways to decompose $b^{2000}_{0101}$, 4 different ways to decompose $b^{0101}_{0101}$, and 5 different ways to decompose each of $b^{1100}_{1100}$ and $b^{1100}_{0011}$.

One can find all the highest weight vectors of relevant $\mathcal{L}\left(\frac{1}{2},0\right)\otimes \mathcal{L}\left(\frac{7}{10},0\right)\otimes \mathcal{L}\left(\frac{4}{5},0\right)$-modules  in the Fock space of $\widehat{\mathfrak{sl}(4)}_2/\widehat{\mathfrak{u}(1)}^3$. It would be rather tedious to list all of them,
but we attach Table \ref{tab:sl4hwv} for partial results. 
\begingroup
\renewcommand*{\arraystretch}{2}
\begin{center}
\begin{longtable}{c|c}
      Module of minimal models & Highest weight vector(s)\\
      \hline
      $\mathcal{L}\left(\frac{1}{2},0\right)\otimes\mathcal{L}\left(\frac{7}{10},0\right)\otimes\mathcal{L}\left(\frac{4}{5},0\right)$&$\ket{0}$\\
      $\mathcal{L}\left(\frac{1}{2},\frac{1}{2}\right)\otimes \mathcal{L}\left(\frac{7}{10},\frac{3}{2}\right)\otimes \mathcal{L}\left(\frac{4}{5},0\right)$ & $\left(\psi^{(4)}_{-\frac{3}{2}}\psi^{(4)}_{-\frac{1}{2}}-\psi^{(2)}_{-\frac{3}{2}}\psi^{(2)}_{-\frac{1}{2}}\right)\ket{0}$\\
      $\mathcal{L}\left(\frac{1}{2},\frac{1}{2}\right)\otimes \mathcal{L}\left(\frac{7}{10},\frac{1}{10}\right)\otimes \mathcal{L}\left(\frac{4}{5},\frac{7}{5}\right)$ & $\left(\frac{1}{7}\psi^{(2)}_{-\frac{3}{2}}\psi^{(2)}_{-\frac{1}{2}}-\frac{1}{7}\psi^{(4)}_{-\frac{3}{2}}\psi^{(4)}_{-\frac{1}{2}}-\psi^{(5)}_{-\frac{3}{2}}\psi^{(5)}_{-\frac{1}{2}}+\psi^{(6)}_{-\frac{3}{2}}\psi^{(6)}_{-\frac{1}{2}}\right)\ket{0}$\\
$\mathcal{L}\left(\frac{1}{2},0\right)\otimes \mathcal{L}\left(\frac{7}{10},\frac{3}{5}\right)\otimes \mathcal{L}\left(\frac{4}{5},\frac{2}{5}\right)$ & $\left(2\psi^{(6)}_{-\frac{1}{2}}\psi^{(2)}_{-\frac{1}{2}}-\psi^{(3)}_{-\frac{1}{2}}\psi^{(1)}_{-\frac{1}{2}}\right)\ket{0}$\\
$\mathcal{L}\left(\frac{1}{2},\frac{1}{2}\right)\otimes \mathcal{L}\left(\frac{7}{10},\frac{1}{10}\right)\otimes \mathcal{L}\left(\frac{4}{5},\frac{2}{5}\right)$ & $\psi^{(3)}_{-\frac{1}{2}}\psi^{(1)}_{-\frac{1}{2}}\ket{0}$\\
$\mathcal{L}\left(\frac{1}{2},\frac{1}{2}\right)\otimes \mathcal{L}\left(\frac{7}{10},0\right)\otimes\mathcal{L}\left(\frac{4}{5},0\right)$ & $\psi^{(1)}_{-\frac{1}{2}}\ket{0}$\\
$\mathcal{L}\left(\frac{1}{2},0\right)\otimes \mathcal{L}\left(\frac{7}{10},\frac{1}{10}\right)\otimes \mathcal{L}\left(\frac{4}{5},\frac{2}{5}\right)$ & $\psi^{(3)}_{-\frac{1}{2}}\ket{0}$\\
$\mathcal{L}\left(\frac{1}{2},\frac{1}{16}\right)\otimes\mathcal{L}\left(\frac{7}{10},\frac{7}{16}\right)\otimes\mathcal{L}\left(\frac{4}{5},0\right)$ & $\psi^{(2)}_{-\frac{1}{2}}\ket{0},\psi^{(4)}_{-\frac{1}{2}}\ket{0}$\\
$\mathcal{L}\left(\frac{1}{2},\frac{1}{16}\right)\otimes \mathcal{L}\left(\frac{7}{10},\frac{3}{80}\right)\otimes \mathcal{L}\left(\frac{4}{5},\frac{2}{5}\right)$ & $\psi^{(5)}_{-\frac{1}{2}}\ket{0},\psi^{(6)}_{-\frac{1}{2}}\ket{0}$\\
$\mathcal{L}\left(\frac{1}{2},0\right)\otimes \mathcal{L}\left(\frac{7}{10},\frac{3}{2}\right)\otimes \mathcal{L}\left(\frac{4}{5},0\right)$ & $\left(\frac{-1+i}{4}\psi^{(1)}_{-\frac{3}{2}}+\psi^{(4)}_{-1}\psi^{(2)}_{-\frac{1}{2}}\right)\ket{0}$\\
$\mathcal{L}\left(\frac{1}{2},\frac{1}{16}\right)\otimes \mathcal{L}\left(\frac{7}{10},\frac{3}{80}\right)\otimes \mathcal{L}\left(\frac{4}{5},\frac{7}{5}\right)$ & $\left(\frac{-2+2i}{7}\psi^{(2)}_{-\frac{3}{2}}-\frac{i}{7}\psi^{(4)}_{-1}\psi^{(1)}_{-\frac{1}{2}}+\psi^{(5)}_{-1}\psi^{(3)}_{-\frac{1}{2}}\right)\ket{0}$\\
$\mathcal{L}\left(\frac{1}{2},0\right)\otimes \mathcal{L}\left(\frac{7}{10},\frac{1}{10}\right)\otimes \mathcal{L}\left(\frac{4}{5},\frac{7}{5}\right)$ & $\left(\frac{3+3i}{2}\psi^{(3)}_{-\frac{3}{2}}+\psi^{(5)}_{-1}\psi^{(2)}_{-\frac{1}{2}}+\psi^{(6)}_{-1}\psi^{(4)}_{-\frac{1}{2}}\right)\ket{0}$\\
$\mathcal{L}\left(\frac{1}{2},\frac{1}{2}\right)\otimes \mathcal{L}\left(\frac{7}{10},\frac{3}{5}\right)\otimes \mathcal{L}\left(\frac{4}{5},\frac{2}{5}\right)$ & $\left(\psi^{(6)}_{-1}\psi^{(4)}_{-\frac{1}{2}}-\psi^{(5)}_{-1}\psi^{(2)}_{-\frac{1}{2}}\right)\ket{0}$\\
$\mathcal{L}\left(\frac{1}{2},\frac{1}{2}\right)\otimes \mathcal{L}\left(\frac{7}{10},\frac{3}{5}\right)\otimes \mathcal{L}\left(\frac{4}{5},\frac{7}{5}\right)$ & $\begin{array}{l}
    \left(-\frac{i}{7}\psi^{(1)}_{-\frac{5}{2}}+\frac{1-i}{28}\psi^{(4)}_{-2}\psi^{(2)}_{-\frac{1}{2}}+\frac{1-i}{14}\psi^{(4)}_{-1}\psi^{(2)}_{-\frac{3}{2}}\right.\\\left.+\frac{1-i}{4}\psi^{(6)}_{-1}\psi^{(5)}_{-\frac{3}{2}}+\psi^{(3)}_{-\frac{3}{2}}\psi^{(3)}_{-\frac{1}{2}}\psi^{(1)}_{-\frac{1}{2}}\right)\ket{0}
\end{array}$\\
$\mathcal{L}\left(\frac{1}{2},0\right)\otimes \mathcal{L}\left(\frac{7}{10},\frac{1}{10}\right)\otimes \mathcal{L}\left(\frac{4}{5},\frac{1}{15}\right)$ & $\ket{1'},  \psi_0^{\left(6\right)}\psi_0^{\left(2\right)}\ket{1'}$\\
$\mathcal{L}(\frac{1}{2},\frac{1}{16})\otimes \mathcal{L}(\frac{7}{10},\frac{3}{80})\otimes \mathcal{L}(\frac{4}{5},\frac{1}{15})$ & $\psi_0^{\left(2\right)}\ket{1'},\psi_0^{\left(4\right)}\ket{1'},\psi_0^{\left(5\right)}\ket{1'},\psi_0^{\left(6\right)}\ket{1'}$\\
$\mathcal{L}\left(\frac{1}{2},\frac{1}{2}\right)\otimes \mathcal{L}\left(\frac{7}{10},\frac{3}{5}\right)\otimes \mathcal{L}\left(\frac{4}{5},\frac{1}{15}\right)$ & $\left(\psi_{-1}^{(4)}\psi_{0}^{(4)}-\psi_{-1}^{(2)}\psi_{0}^{(2)}\right)\ket{1'}$ \\
$\mathcal{L}\left(\frac{1}{2},\frac{1}{2}\right)\otimes \mathcal{L}\left(\frac{7}{10},0\right)\otimes \mathcal{L}\left(\frac{4}{5},\frac{2}{3}\right)$ &  $\left(\frac{1}{3}\psi_{-1}^{(2)}\psi_{0}^{(2)}-\frac{1}{3}\psi_{0}^{(4)}-\psi_{-1}^{(5)}\psi_{0}^{(5)}+\psi_{-1}^{(6)}\psi_{0}^{6)}\right)\ket{1'}$\\
$\mathcal{L}\left(\frac{1}{2},\frac{1}{16}\right)\otimes \mathcal{L}\left(\frac{7}{10},\frac{7}{16}\right)\otimes \mathcal{L}\left(\frac{4}{5},\frac{2}{3}\right)$& $\left( \frac{1+i}{3}\psi_{-\frac{1}{2}}^{(2)}\psi_{-\frac{1}{2}}^{(1)}-\frac{1+i}{2}\psi_{-\frac{1}{2}}^{(6)}\psi_{-\frac{1}{2}}^{(3)}+\psi_{-\frac{1}{2}}^{(6)}\psi_{-\frac{1}{2}}^{(5)}\psi_{0}^{(2)}\right)\ket{1'}$\\
$\mathcal{L}\left(\frac{1}{2},\frac{1}{2}\right)\otimes \mathcal{L}\left(\frac{7}{10},\frac{1}{10}\right)\otimes \mathcal{L}\left(\frac{4}{5},\frac{1}{15}\right)$ & $ \psi_{-\frac{1}{2}}^{(1)}\ket{1'}$ \\
$\mathcal{L}\left(\frac{1}{2},0\right)\otimes \mathcal{L}\left(\frac{7}{10},\frac{3}{5}\right)\otimes \mathcal{L}\left(\frac{4}{5},\frac{1}{15}\right)$ & $\left(\frac{i-1}{4}\psi_{-\frac{1}{2}}^{(1)}+\psi_{-\frac{1}{2}}^{(4)}\psi_{0}^{(2)}\right)\ket{1'}$\\
$\mathcal{L}\left(\frac{1}{2},0\right)\otimes \mathcal{L}\left(\frac{7}{10},0\right)\otimes \mathcal{L}\left(\frac{4}{5},\frac{2}{3}\right)$ & $ \left(\frac{i-1}{6}\psi_{-\frac{1}{2}}^{(1)}\ket{1'}-\frac{1}{3}\psi_{-\frac{1}{2}}^{(4)}\psi_{0}^{(2)}+\psi_{-\frac{1}{2}}^{(6)}\psi_{0}^{(5)}\right)\ket{1'}$\\
$\mathcal{L}\left(\frac{1}{2},0\right)\otimes \mathcal{L}\left(\frac{7}{10},0\right)\otimes \mathcal{L}\left(\frac{4}{5},\frac{1}{8}\right)$ & $\psi_{0}^{(5)}\ket{1''}$\\
$\mathcal{L}\left(\frac{1}{2},0\right)\otimes \mathcal{L}\left(\frac{7}{10},\frac{1}{10}\right)\otimes \mathcal{L}\left(\frac{4}{5},\frac{1}{40}\right)$ & $ \psi_{0}^{(2)}\ket{1''}$\\
$\mathcal{L}\left(\frac{1}{2},\frac{1}{16}\right)\otimes \mathcal{L}\left(\frac{7}{10},\frac{3}{80}\right)\otimes \mathcal{L}\left(\frac{4}{5},\frac{1}{40}\right)$ & $\ket{1''},\psi_{0}^{(1)}\ket{1''}$\\
$\mathcal{L}\left(\frac{1}{2},\frac{1}{16}\right)\otimes \mathcal{L}\left(\frac{7}{10},\frac{7}{16}\right)\otimes \mathcal{L}\left(\frac{4}{5},\frac{1}{8}\right)$ & $\left(\frac{-1-i}{2}\psi^{(3)}_{-\frac{1}{2}}+ \psi^{(5)}_{-\frac{1}{2}}\psi^{(2)}_{0}\right)\ket{1''}$\\
$\mathcal{L}\left(\frac{1}{2},\frac{1}{2}\right)\otimes \mathcal{L}\left(\frac{7}{10},0\right)\otimes \mathcal{L}\left(\frac{4}{5},\frac{1}{8}\right)$ & $\left(\frac{i-1}{2}\psi^{(6)}_{-\frac{1}{2}}+ \psi^{(5)}_{-\frac{1}{2}}\psi^{(1)}_{0}\right)\ket{1''}$\\
$\mathcal{L}\left(\frac{1}{2},\frac{1}{16}\right)\otimes \mathcal{L}\left(\frac{7}{10},\frac{3}{80}\right)\otimes \mathcal{L}\left(\frac{4}{5},\frac{21}{40}\right)$ & $\left(\psi^{(6)}_{-\frac{1}{2}}\psi^{(2)}_{0}-\psi^{(3)}_{-\frac{1}{2}}\psi^{(1)}_{0}\right)\ket{1''},\left(\frac{1+i}{2}\psi^{(3)}_{-\frac{1}{2}}+ \psi^{(5)}_{-\frac{1}{2}}\psi^{(2)}_{0}\right)\ket{1''}$ \\
$\mathcal{L}\left(\frac{1}{2},0\right)\otimes \mathcal{L}\left(\frac{7}{10},\frac{1}{10}\right)\otimes \mathcal{L}\left(\frac{4}{5},\frac{21}{40}\right)$ & $\left(\frac{1-i}{2}\psi^{(6)}_{-\frac{1}{2}}+ \psi^{(5)}_{-\frac{1}{2}}\psi^{(1)}_{0}\right)\ket{1''}$\\
$\mathcal{L}\left(\frac{1}{2},\frac{1}{2}\right)\otimes \mathcal{L}\left(\frac{7}{10},\frac{1}{10}\right)\otimes \mathcal{L}\left(\frac{4}{5},\frac{1}{40}\right)$ &  $\left(\frac{i-1}{2}\psi^{(4)}_{-\frac{1}{2}}+ \psi^{(2)}_{-\frac{1}{2}}\psi^{(1)}_{0}\right)\ket{1''}$\\
$\mathcal{L}\left(\frac{1}{2},0\right)\otimes \mathcal{L}\left(\frac{7}{10},\frac{3}{5}\right)\otimes \mathcal{L}\left(\frac{4}{5},\frac{1}{40}\right)$ & $\left(\frac{1-i}{2}\psi^{(4)}_{-\frac{1}{2}}+ \psi^{(2)}_{-\frac{1}{2}}\psi^{(1)}_{0}\right)\ket{1''}$\\
    \caption{Virasoro highest weight vectors in $\widehat{\mathfrak{sl}(4)}_2/\widehat{\mathfrak{u}(1)}^3$ \\ (partial results)}
    \label{tab:sl4hwv}
\end{longtable}
\end{center}
\endgroup


\chapter{Coupled Free Fermions: The Lattice Construction} 

\label{Chapter4} 

As briefly mentioned in the previous chapter, we expect more constructions of coupled free fermions\index{coupled free fermion}. A natural candidate is the lattice construction \index{lattice construction} based on root lattices of Lie algebras, which can indeed give us a family of coupled free fermion CFTs by a scaling of $1/\sqrt{2}$, as we will describe in this chapter. Surprisingly, we observe a connection between a CFT from this family and one from the coset family. Before the discussion of coupled free fermions, we first review some studies on lattice CFT constructed from root lattices scaled by $\sqrt{2}$.

In terms of notations, we reserve the conventional A-D-E notations of simple Lie algebras for their root lattices. For example, $A_n$ denotes the root lattice of $\mathfrak{sl}(n+1)$.

\section{Motivation: $V_{\sqrt{2}A_n}$}\label{sec:latsqrt2}

In addition to massive exploration of the lattice CFTs $V_{X_n}$ ($X=A,D,E$), there are also many interests in the scaled lattice CFTs $V_{\sqrt{2}X_n}$. One of the most important reasons is that according to \eqref{eq:latOPE}, each root of $X_n$ will contribute a spin-2 field in $V_{\sqrt{2}X_n}$, which often intrigues physicists for the important roles of spin-2 fields in gravity theories \parencite{BOULWARE1975193,FierzPauli}. Another reason is that the famous Moonshine VOA is constructed from the Leech lattice which contains no elements of length square 2, which is a property that any $\sqrt{2}X_n$ also holds. 

We note that $V_{\sqrt{2}X_n}$ have central charges no less than 1. To study such CFTs, it is often helpful to decompose the theory into those with central charges less than 1, as (rational) CFTs with central charges less than 1 are well-studied, and especially the minimal models are fully classified. For CFTs containing primary fields of conformal dimension 2, a general method of deconstruction\index{lattice construction!deconstruction} is stated in \parencite{Bae2021}. 

Let $\varphi(z)$ be a Virasoro primary field of conformal dimension 2, with respect to a energy-momentum tensor $T(z)$ of central charge $c$, which satisfies the OPE
\[\varphi(z)\varphi(w)\sim \dfrac{1}{(z-w)^4}+\dfrac{\frac{4}{c}T(w)+b\varphi(w)}{(z-w)^2}+\dfrac{\frac{1}{2}\partial\left(\frac{4}{c}T(w)+b\varphi(w)\right)}{z-w},\]
then the energy-momentum tensor $T(z)$ can be deconstructed as $T(z)=t^+(z)+t^-(z)$ with $t^{\pm}(z)=\alpha_{\pm}T(z)+\beta_{\pm}\varphi(z)$ of central charges $$c_{\pm}=\dfrac{c}{2}\left(1\pm\left(1+\dfrac{32}{b^2c}\right)^{-\frac{1}{2}}\right),$$ where
\begin{align*}
    &\alpha_{\pm}= \dfrac{1}{2}\left(1\pm\left(1+\dfrac{32}{b^2c}\right)^{-\frac{1}{2}}\right), &\beta_{\pm}=\mp \dfrac{2}{b}\left(1+\dfrac{32}{b^2c}\right)^{-\frac{1}{2}}\left(1+\dfrac{32}{b^2c}\right)^{-\frac{1}{2}},
\end{align*}
if $b\ne 0$, and 
\begin{align*}
    &\alpha_\pm=\frac{1}{2},&\beta_\pm=\mp\frac{1}{2}\sqrt{\frac{c}{2}},
\end{align*}
if $b=0$.

One can further decompose $t^+(z)$ if there is a primary field of conformal dimension 2 with respect to $t^+(z)$. For $V_{\sqrt{2}X_n}$, this nested deconstruction is always possible until $T(z)$ is deconstructed to a sum of $n+1$ mutually commuting energy-momentum tensors. It is proved in \parencite{Dong1996AssociativeSO} that the energy-momentum tensor of $V_{\sqrt{2}A_n}$ can be deconstructed as (in terms of central charges) 
\[n=\dfrac{2n}{n+3}+\sum_{i=1}^{n}\left(1-\dfrac{6}{(i+2)(i+3)}\right).\]

More specifically, it is shown in \parencite{LAM2004614} that as a module of $\mathcal{L}(c(1),0)\otimes\cdots\otimes\mathcal{L}(c(n),0)\otimes W(n,0)$, $V_{\sqrt{2}A_n}$ can be decomposed as follows:
\[V_{\sqrt{2}A_n}\cong\underset{\substack{0\le k_j\le j+1 \\ j=1,\dots,n\\ k_j\;\mathrm{even}}}{\bigoplus}\mathcal{L}(c(1),h^{(1)}_{k_0+1,k_1+1})\otimes\cdots\otimes\mathcal{L}(c(n),h^{(1)}_{k_{n-1}+1,k_n+1})\otimes W(n,k_n),\]
where $c(m)$ and $h^{(m)}_{r,s}$ are central charges and conformal dimensions of unitary minimal models as given in \eqref{eq:mincc} and \eqref{eq:minwt} respectively, $W(n,k)$ is the highest weight module of $\widehat{\mathfrak{sl}(2)}_{n+1}$ of weight $(n+1-k)\Lambda_0+k\Lambda_1$. It is also known that $W(n,0)$ is isomorphic to a parafermion algebra of central charge $\frac{2n}{n+3}$ constructed in \parencite{ZF1985}. 

\subsubsection*{Example: Decomposition of $V_{\sqrt{2}A_2}$}

The difficulty of decomposing $V_{\sqrt{2}A_n}$ is that for generic $n$, it contains the subalgebra $W(n,0)$ whose central charge is not one of those of minimal model series. However, in the case of $A_2$, this issue does not occur and the decomposition is easier. We quote some explicit results on the decomposition of $V_{\sqrt{2}A_2}$-modules from \parencite{KITAZUME2000893} here, as they provide important motivations for us revealing the connection between two coupled free fermion constructions.

Let $\{\alpha_1,\alpha_2\}$ be a set of fundamental roots of $A_2$ with an inner product $(\;,\;)$. Write $\Tilde{\alpha}_i=\sqrt{2}\alpha_i$. Denote the dual of $\sqrt{2}A_2$ by $(\sqrt{2}A_2)^*$ . Then $(\sqrt{2}A_2)^*$ has a basis given by $\{\left(\Tilde{\alpha}_1-2\Tilde{\alpha}_2\right)/6, \left(-\Tilde{\alpha}_1+2\Tilde{\alpha}_2\right)/6\}$ and it can be seen that $(\sqrt{2}A_2)^*/\sqrt{2}A_2$ has 12 cosets. Hence $V_{\sqrt{2}A_2}$ has 12 irreducible modules given by $\{V_{\gamma+\sqrt{2}A_2}\,|\,\gamma\in(\sqrt{2}A_2)^*/\sqrt{2}A_2\}$.

Note that the central charge of $V_{\sqrt{2}A_2}$ is again $2=\frac{1}{2}+\frac{7}{10}+\frac{4}{5}$, that is, the sum of the first three values of central charges of discrete series of unitary minimal models. It is claimed in \parencite{KITAZUME2000893} that the $V_{\sqrt{2}A_2}$-modules can be decomposed into the minimal model\index{minimal model} modules.

We shall focus on the decomposition of 6 irreducible modules which are $V^0:=V_{\sqrt{2}A_2},\;V^1:=V_{\left(-\Tilde{\alpha}_1+\Tilde{\alpha}_2\right)/3+\sqrt{2}A_2},$ $ V^2:=V_{\left(\Tilde{\alpha}_1-\Tilde{\alpha}_2\right)/3+\sqrt{2}A_2}, \;V^a:=V_{\Tilde{\alpha}_1/2+\sqrt{2}A_2},\; V^b:=V_{\Tilde{\alpha}_2/2+\sqrt{2}A_2},\; V^c:=V_{\left(\Tilde{\alpha}_1+\Tilde{\alpha}_2\right)/2+\sqrt{2}A_2}$. The decomposition of other modules can be obtained by fusion products between $\{V^1,V^2\}$ and $\{V^a,V^b,V^c\}$ using the fusion rules computed in \parencite{Miy2001}. We summarise the results as follows:
\begingroup
\renewcommand*{\arraystretch}{2}
\begin{equation}\label{eq:2A2decomp}
\begin{aligned}
    V^0\cong&\textstyle \mathcal{L}\left(\frac{1}{2},0\right)\otimes \mathcal{L}\left(\frac{7}{10},0\right)\otimes \mathcal{L}\left(\frac{4}{5},0\right)\boldsymbol{\oplus} \mathcal{L}\left(\frac{1}{2},0\right)\otimes \mathcal{L}\left(\frac{7}{10},\frac{3}{5}\right)\otimes \mathcal{L}\left(\frac{4}{5},\frac{7}{5}\right)\\
    \boldsymbol{\oplus} &\textstyle\mathcal{L}\left(\frac{1}{2},\frac{1}{2}\right)\otimes \mathcal{L}\left(\frac{7}{10},\frac{3}{2}\right)\otimes \mathcal{L}\left(\frac{4}{5},0\right)\boldsymbol{\oplus} \mathcal{L}\left(\frac{1}{2},\frac{1}{2}\right)\otimes \mathcal{L}\left(\frac{7}{10},\frac{1}{10}\right)\otimes \mathcal{L}\left(\frac{4}{5},\frac{7}{5}\right)\\
    \boldsymbol{\oplus}&\textstyle\mathcal{L}\left(\frac{1}{2},0\right)\otimes \mathcal{L}\left(\frac{7}{10},\frac{3}{5}\right)\otimes \mathcal{L}\left(\frac{4}{5},\frac{2}{5}\right)\boldsymbol{\oplus} \mathcal{L}\left(\frac{1}{2},\frac{1}{2}\right)\otimes \mathcal{L}\left(\frac{7}{10},\frac{1}{10}\right)\otimes \mathcal{L}\left(\frac{4}{5},\frac{2}{5}\right)\\
   \boldsymbol{\oplus} & \textstyle\mathcal{L}\left(\frac{1}{2},0\right)\otimes \mathcal{L}\left(\frac{7}{10},0\right)\otimes \mathcal{L}\left(\frac{4}{5},3\right)\boldsymbol{\oplus} \mathcal{L}\left(\frac{1}{2},\frac{1}{2}\right)\otimes \mathcal{L}\left(\frac{7}{10},\frac{3}{2}\right)\otimes \mathcal{L}\left(\frac{4}{5},3\right),\\
     V^1\cong V^2\cong&\textstyle \mathcal{L}\left(\frac{1}{2},0\right)\otimes \mathcal{L}\left(\frac{7}{10},0\right)\otimes \mathcal{L}\left(\frac{4}{5},\frac{2}{3}\right)\boldsymbol{\oplus} \mathcal{L}\left(\frac{1}{2},0\right)\otimes \mathcal{L}\left(\frac{7}{10},\frac{3}{5}\right)\otimes \mathcal{L}\left(\frac{4}{5},\frac{1}{15}\right)\\
     \boldsymbol{\oplus}&\textstyle \mathcal{L}\left(\frac{1}{2},\frac{1}{2}\right)\otimes \mathcal{L}\left(\frac{7}{10},\frac{1}{10}\right)\otimes \mathcal{L}\left(\frac{4}{5},\frac{1}{15}\right)\boldsymbol{\oplus} \mathcal{L}\left(\frac{1}{2},\frac{1}{2}\right)\otimes \mathcal{L}\left(\frac{7}{10},\frac{3}{2}\right)\otimes \mathcal{L}\left(\frac{4}{5},\frac{2}{3}\right),\\
      V^a\cong V^b\cong&\textstyle \mathcal{L}\left(\frac{1}{2},\frac{1}{16}\right)\otimes \mathcal{L}\left(\frac{7}{10},\frac{7}{16}\right)\otimes \mathcal{L}\left(\frac{4}{5},0\right)\boldsymbol{\oplus} \mathcal{L}\left(\frac{1}{2},\frac{1}{16}\right)\otimes \mathcal{L}\left(\frac{7}{10},\frac{3}{80}\right)\otimes \mathcal{L}\left(\frac{4}{5},\frac{7}{5}\right)\\
      \boldsymbol{\oplus}&\textstyle\mathcal{L}\left(\frac{1}{2},\frac{1}{16}\right)\otimes \mathcal{L}\left(\frac{7}{10},\frac{3}{80}\right)\otimes \mathcal{L}\left(\frac{4}{5},\frac{2}{5}\right)\boldsymbol{\oplus} \mathcal{L}\left(\frac{1}{2},\frac{1}{16}\right)\otimes \mathcal{L}\left(\frac{7}{10},\frac{7}{16}\right)\otimes \mathcal{L}\left(\frac{4}{5},3\right),\\
      V^c\cong &\textstyle\mathcal{L}\left(\frac{1}{2},\frac{1}{2}\right)\otimes \mathcal{L}\left(\frac{7}{10},0\right)\otimes \mathcal{L}\left(\frac{4}{5},0\right)\boldsymbol{\oplus} \mathcal{L}\left(\frac{1}{2},\frac{1}{2}\right)\otimes \mathcal{L}\left(\frac{7}{10},\frac{3}{5}\right)\otimes \mathcal{L}\left(\frac{4}{5},\frac{7}{5}\right)\\
      \boldsymbol{\oplus} &\textstyle\mathcal{L}\left(\frac{1}{2},0\right)\otimes \mathcal{L}\left(\frac{7}{10},\frac{3}{2}\right)\otimes \mathcal{L}\left(\frac{4}{5},0\right)\boldsymbol{\oplus} \mathcal{L}\left(\frac{1}{2},0\right)\otimes \mathcal{L}\left(\frac{7}{10},\frac{1}{10}\right)\otimes \mathcal{L}\left(\frac{4}{5},\frac{7}{5}\right)\\
    \boldsymbol{\oplus}&\textstyle\mathcal{L}\left(\frac{1}{2},\frac{1}{2}\right)\otimes \mathcal{L}\left(\frac{7}{10},\frac{3}{5}\right)\otimes \mathcal{L}\left(\frac{4}{5},\frac{2}{5}\right)\boldsymbol{\oplus} \mathcal{L}\left(\frac{1}{2},0\right)\otimes \mathcal{L}\left(\frac{7}{10},\frac{1}{10}\right)\otimes \mathcal{L}\left(\frac{4}{5},\frac{2}{5}\right)\\
    \boldsymbol{\oplus}&\textstyle \mathcal{L}\left(\frac{1}{2},\frac{1}{2}\right)\otimes \mathcal{L}\left(\frac{7}{10},0\right)\otimes \mathcal{L}\left(\frac{4}{5},3\right)\boldsymbol{\oplus}  \mathcal{L}\left(\frac{1}{2},0\right)\otimes \mathcal{L}\left(\frac{7}{10},\frac{3}{2}\right)\otimes \mathcal{L}\left(\frac{4}{5},3\right).
\end{aligned}
\end{equation}
\endgroup
where $\cong$ means isomorphic as modules of $\mathcal{L}\left(\frac{1}{2},0\right)\otimes \mathcal{L}\left(\frac{7}{10},0\right)\otimes \mathcal{L}\left(\frac{4}{5},0\right)$.

We notice that the decomposition in \eqref{eq:2A2decomp} overlaps with the decomposition of $\widehat{\mathfrak{sl}(4)}_2/\widehat{\mathfrak{u}(1)}^3$-modules on page \pageref{sl4decomp}. In terms of characters, we have the following \nopagebreak[3] identities:
\begin{align*}
\mathrm{ch}\left(V^0\right)=&b^{2000}_{2000}+b^{2000}_{0020}\\
    \text{ch}\left(V^1\right)=\text{ch}\left(V^2\right)=&b^{0101}_{0020}\\
    \text{ch}\left(V^a\right)=\text{ch}(V^b)=\text{ch}\left(V^c\right)=&2b_{0101}^{0020}
\end{align*}
It can also be checked that characters of other 6 modules are directly related to $b^{0101}_{0101}$. They exhaust the $\widehat{\mathfrak{sl}(4)}_2/\widehat{\mathfrak{u}(1)}^3$-modules in the untwisted sector and $\frac{1}{6}$-twisted sector. 

Naturally, one would like to look for characters of some structure closely related to $V_{\sqrt{2}A_2}$ that has a connection with characters of $\widehat{\mathfrak{sl}(4)}_2/\widehat{\mathfrak{u}(1)}^3$-modules in $\frac{1}{8}$-twisted sector. It turns out that one needs to consider the $\mathbb{Z}_2$-orbifold of $V_{\sqrt{2}A_2}$ whose $\mathbb{Z}_2$-twisted modules have characters identical to $b^{1100}_{1100}+b^{1100}_{0011}$.

All of the above suggests that lattice construction from $A_2$ has a potential connection to our coupled free fermions CFT $\widehat{\mathfrak{sl}(4)}_2/\widehat{\mathfrak{u}(1)}^3$.


\section{Investigation: $V_{\frac{1}{\sqrt{2}}A_n}$}\label{sec:lat1/sqrt2}

Motivated by scaled lattice CFTs $V_{\sqrt{2}A_n}$, in order to construct another family of coupled free fermions, we consider $V_{\frac{1}{\sqrt{2}}A_n}$, where each root of $A_n$ contributes a field of conformal dimension $\frac{1}{2}$, suggesting a fermion. For convenience, we redefine the notation of $\psi^{\alpha}$, for each $\alpha\in A_n$, in \eqref{eq:latfield} to be 
\[\psi^{\alpha}(z)=:e^{i\frac{\alpha}{\sqrt{2}}\cdot\Phi(z)}:\]
and then according to \eqref{eq:latOPE}, the non-trivial OPEs between the generating fields $\psi^{\alpha}$ with $\alpha\in\Delta$ read as follows:
    \begin{align}
       \psi^{\alpha}(z)\overline{\psi^{\alpha}}(w)&\sim \dfrac{1}{z-w},\label{eq:latfreefer}\\
       \psi^{\alpha}(z)\psi^{\beta}(w)&\sim \dfrac{c_{\alpha,\beta}\psi^{\alpha+\beta}(w)}{(z-w)^{1/2}},\;\text{if}\;\;\alpha+\beta\in\Delta.\label{eq:latcpfer}
    \end{align}
Here we used the observation that $\psi^{-\alpha}(z)=\overline{\psi^{\alpha}}(z)$, where $\overline{\psi^{\alpha}}$ is the complex conjugate of $\psi^{\alpha}$. Now, \eqref{eq:latfreefer} suggests that each $\psi^{\alpha}$ is a free complex fermion and \eqref{eq:latcpfer} suggests that the complex fermions are coupled to each other, so we indeed have coupled free fermions in $V_{\frac{1}{\sqrt{2}}A_n}$ as desired.

We shall derive an energy-momentum tensor $L(z)$ in terms of coupled free fermions as in \eqref{eq:CFFEM}. First, by expanding $e^{\frac{i}{\sqrt{2}}\alpha\cdot\Phi(z)}e^{-\frac{i}{\sqrt{2}}\alpha\cdot\Phi(z)}$, we see that
\[:\psi^{\alpha}(z)\overline{\psi^{\alpha}}(z):=\frac{i}{\sqrt{2}}\alpha\cdot\partial\Phi(z).\]
Then we have
\[L^{\alpha}(z):=-\frac{1}{2}:\psi^{\alpha}(z)\partial\overline{\psi^{\alpha}}(z):=\frac{1}{2}(\frac{i}{\sqrt{2}}\alpha\cdot\partial\Phi)(z)(\frac{i}{\sqrt{2}}\alpha\cdot\partial\Phi)(z).\]
We construct the total energy-momentum tensor\index{coupled free fermion!energy-momentum tensor} as
\[L(z)=\delta\sum_{\alpha\in\Delta_{+}}L^{\alpha}(z)=\frac{1}{2}\delta\sum_{\alpha\in\Delta_{+}}(\frac{i}{\sqrt{2}}\alpha\cdot\partial\Phi)(z)(\frac{i}{\sqrt{2}}\alpha\cdot\partial\Phi)(z)\]
where $\delta$ is a to-be-determined overall factor by requiring $L(z)$ to satisfy the desired OPEs, as described in the following.

With this expression of $L(z)$, the conformal weight $h_{\beta}$ of any field $\psi^{\beta}$ can be read from the OPE $L(z)\psi^{\beta}(w)$, which gives
\[h_{\beta}=\frac{1}{4}\delta\sum_{\alpha\in\Delta_+}(\beta^T\alpha)(\alpha^T\beta)=\frac{1}{4}\delta\sum_{\alpha\in\Delta_+}\beta^T(\alpha\alpha^T)\beta=\frac{1}{4}\delta h^{\vee}|\beta|^2,\]
where we used the fact
\[\sum_{\alpha\in\Delta_+}\alpha\alpha^T=h^{\vee}\mathbf{1}.\]
On the other hand, we know by construction that the conformal weight of $\psi^{\beta}$ is $\frac{1}{2}|\beta|^2$, so we have $\frac{1}{2}|\beta|^2=\frac{1}{4}\delta h^{\vee}|\beta|^2$, which gives
\[\delta=\frac{2}{h^{\vee}}=\frac{2}{n+1}.\]
Hence we conclude that 
\[L(z)=\frac{2}{n+1}\sum_{\alpha\in\Delta_{+}}L^{\alpha}(z)=-\frac{1}{n+1}\sum_{\alpha\in\Delta_{+}}:\psi^{\alpha}(z)\partial\overline{\psi^{\alpha}}(z):.\]

Since $\frac{1}{\sqrt{2}}A_n$ is not integral, $\frac{1}{\sqrt{2}}A_n\not\subset \left(\frac{1}{\sqrt{2}}A_n\right)^*$ and we cannot expect to classify the modules by the cosets of $\frac{1}{\sqrt{2}}A_n$ in its dual as we did for $V_{\sqrt{2}A_n}$. However, we notice that $\frac{1}{\sqrt{2}}A_n\subset \frac{1}{2}\left(\frac{1}{\sqrt{2}}A_n\right)^*=\frac{1}{\sqrt{2}}A^*_n$, so we may expect the modules corresponding to the cosets in $\left(\frac{1}{\sqrt{2}}A^*_n\right)/\left(\frac{1}{\sqrt{2}}A_n\right)$. If this is the case, for $\gamma\in\left(\frac{1}{\sqrt{2}}A^*_n\right)/\left(\frac{1}{\sqrt{2}}A_n\right)$, the character\index{lattice construction!character} would be given by
\begin{equation}\label{eq:latch-scaled}
    \dfrac{1}{\eta(\tau)^{n}}\sum_{\nu\in\gamma+\frac{1}{\sqrt{2}}A_n}q^{\frac{1}{2}|\nu|^2}.
\end{equation}

We can first test our conjecture on $V_{\frac{1}{\sqrt{2}}A_1}$. Let $\alpha$ denote the root of $A_1$. It is obvious that there is only one complex free fermion $\psi^{\alpha}$ in $V_{\frac{1}{\sqrt{2}}A_1}$. We can rewrite the complex free ferimion $\psi^{\alpha}$ in terms of two real free fermions by setting 
\begin{align*}
    &\chi^{\alpha}(z)=\frac{1}{\sqrt{2}}\left(\psi^{\alpha}(z)+\overline{\psi^{\alpha}}(z)\right),& \xi^{\alpha}(z)=\frac{1}{i\sqrt{2}}\left(\psi^{\alpha}(z)-\overline{\psi^{\alpha}}(z)\right),
\end{align*}
for which it can be seen that
\begin{equation*}
    \begin{tabular}{ccc}
      $\chi^{\alpha}(z)\chi^{\alpha}(w)\sim\dfrac{1}{z-w}$, &$\xi^{\alpha}(z)\xi^{\alpha}(w)\sim\dfrac{1}{z-w}$, &$\chi^{\alpha}(z)\xi^{\alpha}(w)\sim0$.
    \end{tabular}
\end{equation*}
Therefore we should expect $V_{\frac{1}{\sqrt{2}}A_1}$-modules corresponding to the NS and R sectors of two (uncoupled) free fermions. 

On the other hand, we know $\frac{1}{\sqrt{2}}A_1^*=\mathbb{Z}\left\{\frac{1}{2\sqrt{2}}\alpha\right\}$ and therefore we expect two modules $V_{\frac{1}{\sqrt{2}}A_1}$ and $V_{\frac{1}{2\sqrt{2}}\alpha+\frac{1}{\sqrt{2}}A_1}$, whose characters are, according to \eqref{eq:latch-scaled},
\begin{equation}\label{eq:A1latch}
  \begin{aligned}
  \mathrm{ch}\left[V_{\frac{1}{\sqrt{2}}A_1}\right](\tau)&=\dfrac{1}{\eta(\tau)}\sum_{n\in\mathbb{Z}}q^{\frac{1}{4}|n\alpha|^2}=\dfrac{1}{\eta(\tau)}\sum_{n\in\mathbb{Z}}q^{\frac{1}{2}n^2},\\
   \mathrm{ch}\left[V_{\frac{1}{2\sqrt{2}}\alpha+\frac{1}{\sqrt{2}}A_1}\right](\tau)&=\dfrac{1}{\eta(\tau)}\sum_{n\in\mathbb{Z}}q^{\frac{1}{4}\left|\left(n+\frac{1}{2}\right)\alpha\right|^2}=\dfrac{1}{\eta(\tau)}\sum_{n\in\mathbb{Z}}q^{\frac{1}{2}\left(n+\frac{1}{2}\right)^2}.
  \end{aligned}  
\end{equation}
Comparing \eqref{eq:A1latch} and characters of fermions given in \eqref{eq:freeferch}, we note that
\begin{align*}
     \textstyle\mathrm{ch}\left[V_{\frac{1}{\sqrt{2}}A_1}\right]&=\left( \chi_0+\chi_{\frac{1}{2}}\right)^2,\\
     \textstyle\mathrm{ch}\left[V_{\frac{1}{\sqrt{2}}A_1}\right]&=\left( \chi_{\frac{1}{16}}\right)^2,
\end{align*}
which can be proved by the Jacobi triple identity \eqref{eq:JacobitriId} and \eqref{eq:JacobitriId-1} respectively.

We shall proceed to investigate $V_{\frac{1}{\sqrt{2}}A_2}$. Let $\{\alpha_1,\alpha_2\}$ denote the set of simple roots of $A_2$ and set $\alpha_3:=\alpha_1+\alpha_2$. Then we have three complex fermions $\psi^{\alpha_1}$, $\psi^{\alpha_2}$, $\psi^{\alpha_3}$ in $V_{\frac{1}{\sqrt{2}}A_2}$. We can also rewrite them in terms of real fermions by setting, for $i=1,2,3$,
\begin{align}\label{eq:A2realfer}
    &\chi^{i}(z)=\frac{1}{\sqrt{2}}\left(\psi^{\alpha_i}(z)+\overline{\psi^{\alpha_i}}(z)\right),&\xi^{i}(z)=\frac{1}{i\sqrt{2}}\left(\psi^{\alpha_i}(z)-\overline{\psi^{\alpha_i}}(z)\right),
\end{align}
for which the OPEs are calculated to be
\begingroup
\renewcommand*{\arraystretch}{2}
\begin{equation}\label{eq:A2latOPE}
    \begin{array}{ccc}
            \chi^{i}(z)\chi^{i}(w)\sim \dfrac{1}{z-w},& \xi^{i}(z)\xi^{i}(w)\sim \dfrac{1}{z-w},&\chi^{i}(z)\xi^{i}(w)\sim0,\;\;\mathrm{for}\;i=1,2,3;\\
         \chi^{1}(z)\chi^{2}(w)\sim \dfrac{\frac{1}{\sqrt{2}}c^{1}_{2}\chi^{3}(w)}{(z-w)^{1/2}},& \chi^{1}(z)\chi^{3}(w)\sim\dfrac{\frac{1}{\sqrt{2}}c^{1}_{-3}\chi^{2}(w)}{(z-w)^{1/2}},&\chi^{2}(z)\chi^{3}(w)\sim \dfrac{\frac{1}{\sqrt{2}}c^{2}_{-3}\chi^{1}(w)}{(z-w)^{1/2}},\\
         \chi^{1}(z)\xi^{2}(w)\sim\dfrac{\frac{1}{\sqrt{2}}c^{1}_{2}\xi^{3}(w)}{(z-w)^{1/2}},&\chi^{1}(z)\xi^{3}(w)\sim \dfrac{\frac{1}{\sqrt{2}}c^{1}_{-3}\xi^{2}(w)}{(z-w)^{1/2}},&\chi^{2}(z)\xi^{3}(w)\sim \dfrac{\frac{1}{\sqrt{2}}c^{2}_{-3}\xi^{1}(w)}{(z-w)^{1/2}},\\
         \xi^{1}(z)\chi^{2}(w)\sim\dfrac{\frac{1}{\sqrt{2}}c^{1}_{2}\xi^{3}(w)}{(z-w)^{1/2}},&\xi^{1}(z)\chi^{3}(w)\sim \dfrac{\frac{-1}{\sqrt{2}}c^{1}_{-3}\xi^{2}(w)}{(z-w)^{1/2}},&\xi^{2}(z)\chi^{3}(w)\sim \dfrac{\frac{-1}{\sqrt{2}}c^{2}_{-3}\xi^{1}(w)}{(z-w)^{1/2}},\\
         \xi^{1}(z)\xi^{2}(w)\sim \dfrac{\frac{-1}{\sqrt{2}}c^{1}_{2}\chi^{3}(w)}{(z-w)^{1/2}},&\xi^{1}(z)\xi^{3}(w)\sim \dfrac{\frac{1}{\sqrt{2}}c^{1}_{-3}\chi^{2}(w)}{(z-w)^{1/2}},&\xi^{2}(z)\xi^{3}(w)\sim \dfrac{\frac{1}{\sqrt{2}}c^{2}_{-3}\chi^{1}(w)}{(z-w)^{1/2}},
    \end{array}
\end{equation}
\endgroup
where $c^{\pm i}_{\pm j}:=c_{\pm\alpha_i,\pm\alpha_j}$ and we have used the fact $c_{i,j}=c_{-i,-j}$, as the choice of the positive for roots is arbitrary.

We notice that if the values of $c_{i,j}$ are 8th roots of unity with the $\pm$ signs chosen carefully, then the OPEs in \eqref{eq:A2latOPE} are exactly the same as the OPEs for generating fields in $\widehat{\mathfrak{sl}(4)}_2/\widehat{\mathfrak{u}(1)}^3$ as given by \eqref{eq:freefer} and \eqref{eq:couplefer} with parameters given in \eqref{eq:sl4para}. Explicitly, we have following correspondence:
\begin{equation}\label{eq:A2sl4corrsp}
    \begin{array}{cccccc}
    \chi^1\leftrightarrow \psi^{(1)}, & \chi^2\leftrightarrow \psi^{(2)}, & \chi^3\leftrightarrow \psi^{(4)}, & \xi^1\leftrightarrow \psi^{(3)},& \xi^2\leftrightarrow \psi^{(6)}, & \xi^3\leftrightarrow \psi^{(5)}.
    \end{array}
\end{equation}

Hence we expect that the $V_{\frac{1}{\sqrt{2}}A_2}$-modules are related to $\widehat{\mathfrak{sl}(4)}_2/\widehat{\mathfrak{u}(1)}^3$-modules in one way or another. We may investigate their relations by looking at the characters.

We first note that $\frac{1}{\sqrt{2}}A_2^*$ has a basis $\left\{\frac{1}{3\sqrt{2}}(\alpha_1-\alpha_2),\frac{1}{\sqrt{2}}\alpha_1\right\}$, so $\left(\frac{1}{\sqrt{2}}A_2^*\right)/\left(\frac{1}{\sqrt{2}}A_2\right)$ has three cosets that can be represented by 
\begin{equation*}
    \begin{array}{ccc}
       \gamma_0:=0,  & \gamma_1:=\frac{1}{3\sqrt{2}}(\alpha_1-\alpha_2),&\gamma_2:=\frac{2}{3\sqrt{2}}(\alpha_1-\alpha_2).
    \end{array}
\end{equation*}
The characters of $V_{\gamma_i+\frac{1}{\sqrt{2}}A_2}$ are then given by 
\begin{align}
    \mathrm{ch}\left[V_{\frac{1}{\sqrt{2}}A_2}\right](\tau)&=\eta(\tau)^{-2}\sum_{m,n\in\mathbb{Z}}q^{\frac{1}{2}(m^2+n^2-mn)},\label{eq:A2latuntwi}\\
    \mathrm{ch}\left[V_{\gamma_1+\frac{1}{\sqrt{2}}A_2}\right](\tau)=\mathrm{ch}\left[V_{\gamma_2+\frac{1}{\sqrt{2}}A_2}\right](\tau)&=\eta(\tau)^{-2}q^{\frac{2}{3}}\sum_{m,n\in\mathbb{Z}}q^{\frac{1}{2}(m^2+n^2-mn+2m)}.\label{eq:A2lattwi}
\end{align}
Comparing \eqref{eq:A2latuntwi} with \eqref{eq:sl4untwi} and \eqref{eq:A2lattwi} with \eqref{eq:sl4twi-1}, we find that
\begin{align}
    \mathrm{ch}\left[V_{\frac{1}{\sqrt{2}}A_2}\right](\tau)&=q^{-\frac{1}{12}}\sum_{N_i\ge 0}\dfrac{q^{\frac{1}{2}\mathbf{N}^{\mathsf{T}}\cdot\mathbf{G}_4\cdot\mathbf{N}}}{\prod_i (q)_{N_i}},\label{eq:A2UCPF-1}\\
     \mathrm{ch}\left[V_{\gamma_1+\frac{1}{\sqrt{2}}A_2}\right](\tau)&=\frac{1}{2}q^{\frac{1}{6}-\frac{1}{12}}\sum_{N_i\ge 0}\dfrac{q^{\frac{1}{2}\mathbf{N}^{\mathsf{T}}\cdot\mathbf{G}_4\cdot\mathbf{N}-\frac{1}{2}(N_2+N_4+N_5+N_6)}}{\prod_i (q)_{N_i}},\label{eq:A2UCPF-2}
\end{align}
which support our conjecture on the connection between $V_{\frac{1}{\sqrt{2}}A_2}$ and $\widehat{\mathfrak{sl}(4)}_2/\widehat{\mathfrak{u}(1)}^3$.


\section{$\mathbb{Z}_2$-orbifold: $V_{\frac{1}{\sqrt{2}}A_2}^{\mathbb{Z}_2}$ }

Ideally, we expect that the $\frac{1}{8}$-twisted sector of $\widehat{\mathfrak{sl}(4)}_2/\widehat{\mathfrak{u}(1)}^3$ could also be recovered from the modules of $V_{\frac{1}{\sqrt{2}}A_2}$. It is seen that the shifted lattices cannot lead us any further. Motivated by the decomposition of $V^{\mathbb{Z}_2}_{\sqrt{2}A_2}$ and the fact that the conformal dimension of the highest weight state in the $\mathbb{Z}_2$-orbifold\index{orbifold construction!Z@$\mathbb{Z}_2$-orbifold} from the lattice $\frac{1}{\sqrt{2}}A_2$ would be $\frac{d}{16}=\frac{1}{8}$, we shall consider the $\mathbb{Z}_2$-orbifold of $V_{\frac{1}{\sqrt{2}}A_2}$, denoted by $V_{\frac{1}{\sqrt{2}}A_2}^{\mathbb{Z}_2}$.

 Suppose that we have a symmetry which acts by taking the bosons $\Phi=(\phi^1,\phi^2)^T$ to $-\Phi$. We follow the orbifold construction reviewed in Section \ref{sec:CFTs} with relaxed locality conditions whenever necessary and calculate the characters\index{orbifold construction!character} of $V_{\frac{1}{\sqrt{2}}A_2}^{\mathbb{Z}_2}$ according to \eqref{eq:orbch-1} and \eqref{eq:orbch-2}. The first equation in  \eqref{eq:orbch-1} gives \eqref{eq:A2latuntwi} and \eqref{eq:A2lattwi} for the untwisted sector and $\frac{1}{6}$-twisted sector respectively. The reflection map $\theta$ in the second equation of \eqref{eq:orbch-1} gives an extra $-1$ for each simple root, so it gives a factor of $(-1)^{m+n}$ to the summations in \eqref{eq:A2latuntwi} and \eqref{eq:A2lattwi} and we have
\begin{align}
   \mathrm{ch}\left[V_{\frac{1}{\sqrt{2}}A_2}^{\mathbb{Z}_2}\right]_{+-}(\tau)&=\eta(\tau)^{-2}\sum_{m,n\in\mathbb{Z}}(-1)^{m+n}q^{\frac{1}{2}(m^2+n^2-mn)},\label{eq:A2orbch-1}\\
    \mathrm{ch}\left[V_{\gamma_i+\frac{1}{\sqrt{2}}A_2}^{\mathbb{Z}_2}\right]_{+-}(\tau)&= \eta(\tau)^{-2}q^{\frac{2}{3}}\sum_{m,n\in\mathbb{Z}}(-1)^{m+n}q^{\frac{1}{2}(m^2+n^2-mn+2m)},\quad \mathrm{for}\; i=1,2.\label{eq:A2orbch-2}   
\end{align}

Geometrically, we think of the orbifold of $\left(A_2/\sqrt{2}\right)/\mathbb{Z}_2$. According to \parencite{ItoKuniZ2string,Bagger1986}, the $\mathbb{Z}_2$-twisted states are located at the fixed points of the orbifold. Hence, in this case, the $\mathbb{Z}_2$-twisted characters are given by the squares of the characters of a free boson in the anti-periodic sector, which is given in \eqref{eq:orbch-boson}, that is,
\begin{align}
      \mathrm{ch}\left[V_{\frac{1}{\sqrt{2}}A_2}^{\mathbb{Z}_2}\right]_{-+}:=[\mathrm{ch}_{-+}(\tau)]^2&=\eta(\tau)^2\eta(\tau/2)^{-2},\label{eq:A2orbch-3} \\
     \mathrm{ch}\left[V_{\frac{1}{\sqrt{2}}A_2}^{\mathbb{Z}_2}\right]_{--}:=[\mathrm{ch}_{--}(\tau)]^2&=\eta(2\tau)^2\eta(\tau/2)^2\eta(\tau)^{-4}.\label{eq:A2orbch-4} 
\end{align}
Comparing the $V_{\frac{1}{\sqrt{2}}A_2}^{\mathbb{Z}_2}$-characters \eqref{eq:A2orbch-1} - \eqref{eq:A2orbch-4} with $\widehat{\mathfrak{sl}(4)}_2$-string functions in \eqref{eq:sl4strfun}, we note that
\begin{align}
     \mathrm{ch}\left[V_{\frac{1}{\sqrt{2}}A_2}^{\mathbb{Z}_2}\right]_{+-}&=b^{2000}_{2000}+b^{2000}_{0020}-2b^{2000}_{0101},\label{eq:A2Tunshf}\\
     \mathrm{ch}\left[V_{\gamma_i+\frac{1}{\sqrt{2}}A_2}^{\mathbb{Z}_2}\right]_{+-}&=b^{0101}_{2000}-b^{0101}_{0101}.\label{eq:A2Tshf}\\
   \mathrm{ch}\left[V_{\frac{1}{\sqrt{2}}A_2}^{\mathbb{Z}_2}\right]_{-+}&=b^{1100}_{1100}+b^{1100}_{0011},\label{eq:A2orb}\\
   \mathrm{ch}\left[V_{\frac{1}{\sqrt{2}}A_2}^{\mathbb{Z}_2}\right]_{--}&=b^{1100}_{1100}-b^{1100}_{0011},\label{eq:A2Torb}
\end{align}
which suggest a strong connection between $V_{\frac{1}{\sqrt{2}}A_2}^{\mathbb{Z}_2}$ and $\widehat{\mathfrak{sl}(4)}_2/\widehat{\mathfrak{u}(1)}^3$.

Furthermore, by \eqref{eq:A2orb}, we note that the UCPF of the $\frac{1}{8}$-twisted sectors of $\widehat{\mathfrak{sl}(4)}_2/\widehat{\mathfrak{u}(1)}^3$ is related to the orbifold character by 
\[q^{\frac{1}{8}-\frac{1}{12}}\sum_{N_i\ge 0}\dfrac{q^{\frac{1}{2}\mathbf{N}^{\mathsf{T}}\cdot\mathbf{G}_4\cdot\mathbf{N}-\frac{1}{2}(N_1+N_2+N_5)}}{\prod_i (q)_{N_i}}=4\;\mathrm{ch}\left[V_{\frac{1}{\sqrt{2}}A_2}^{\mathbb{Z}_2}\right]_{-+},\]
where the factor 4 interestingly agrees with the number of fixed points of the $\mathbb{Z}_2$ action on $A_2/\sqrt{2}$.


\section{Proofs of $q$-identities connecting $V_{\frac{1}{\sqrt{2}}A_n}^{\mathbb{Z}_2}$ and $\widehat{\mathfrak{sl}(4)}_2/\widehat{\mathfrak{u}(1)}^3$}

Here we explain the procedure of obtaining the proofs without knowing the explicit expressions of $\widehat{\mathfrak{sl}(4)}_2$-string functions a priori: Assuming the validity of all of the identities between $\widehat{\mathfrak{sl}(4)}_2/\widehat{\mathfrak{u}(1)}^3$-UCPFs and $\widehat{\mathfrak{sl}(4)}_2$-string functions, namely \eqref{eq:sl4untwi}, \eqref{eq:sl4twi-1}, \eqref{eq:sl4twi-2}, we propose some expressions of $\widehat{\mathfrak{sl}(4)}_2$-string functions using these identities. Then once the modular properties of these expressions are shown to be the same as $\widehat{\mathfrak{sl}(4)}_2$-string functions (which has been presented in Chapter \ref{Chapter3}), we will have these identities as well as \eqref{eq:A2orbch-1} - \eqref{eq:A2orbch-4} proved with the logic described in \parencite{BCH}. Therefore our main task in this section becomes to determine explicit expressions of $\widehat{\mathfrak{sl}(4)}_2$-string functions.

We notice that \eqref{eq:A2UCPF-1} and \eqref{eq:A2UCPF-2} can simplify 
the left-hand sides of \eqref{eq:sl4untwi} and \eqref{eq:sl4twi-1} as they will reduce the number of summation variables from 6 to 2, so we shall prove  \eqref{eq:A2UCPF-1} and \eqref{eq:A2UCPF-2} first. 

To make the motivation more clear, we rewrite the statistical interaction matrix $\mathbf{G}_4$ under the basis $\{\chi^1,\chi^2,\chi^3,\xi^1,\xi^2,\xi^3\}$ in terms of the real fermions \eqref{eq:A2realfer} of $V_{\frac{1}{\sqrt{2}}A_n}$. Then \eqref{eq:A2UCPF-2} converts to 
\begin{equation}\label{eq:A2UCPF-2-1}
   \dfrac{2q^{\frac{2}{3}}}{\eta(\tau)^{2}}\sum_{m,n\in\mathbb{Z}}q^{\frac{1}{2}(m^2+n^2-mn+2m)}=q^{\frac{1}{12}}\sum_{N_i\ge 0}\dfrac{q^{\frac{1}{2}\mathbf{N}^{\mathsf{T}}\cdot\mathbf{G}_4\cdot\mathbf{N}-\frac{1}{2}(N_2+N_3+N_5+N_6)}}{\prod_i (q)_{N_i}},
\end{equation}
where $\mathbf{G}_4$ is now redefined according to \eqref{eq:A2sl4corrsp} as
\begin{align}\label{eq:G4-def-1}
\mathbf{G}_4:=\frac{1}{2}\begin{pmatrix}
2&1&1&0&1&1\\
1&2&1&1&0&1\\
1&1&2&1&1&2\\
0&1&1&2&1&1\\
1&0&1&1&2&1\\
1&1&2&1&1&2
\end{pmatrix}.
\end{align}
Consider the following identity:
\begin{theorem}\label{th:VOA-z}
	Let $\mathbf{G}_4$ be as in \eqref{eq:G4-def-1} and $\mathbf{N}=(N_1,N_2,\dots,N_6)^T$. Then for $\delta_1,\delta_2\in\{0,1\}$,
	\begin{align}
		&\sum_{N_i\ge 0} \frac{z_1^{N_1+N_3-N_4-N_6}z_2^{N_2+N_3-N_5-N_6}q^{\mathbf{N}^{\mathsf{T}}\cdot \mathbf{G}_4\cdot \mathbf{N}-2\delta_1(N_4+N_6)-2\delta_2(N_5+N_6)} }{\prod_i(q^2)_{N_i}}\notag\\
		&\qquad = \frac{1}{(q^2)_\infty^2}\sum_{M_1,M_2\in\mathbb{Z}} z_1^{M_1} z_2^{M_2}q^{M_1^2+M_2^2-M_1M_2}(1+\delta_1 q^{2M_1}) (1+\delta_2 q^{2M_2}).\label{eq:VOA-z-1}
	\end{align}
\end{theorem}
\noindent Here the $z$-factors are assigned by the following rules:
\begin{equation*}
    \begin{array}{cccccc}
    \chi^1\leftrightarrow z_1, & \chi^2\leftrightarrow z_2, & \chi^3\leftrightarrow z_1z_2, & \xi^1\leftrightarrow z_1^{-1},& \xi^2\leftrightarrow z_2^{-1}, & \xi^3\leftrightarrow z_1^{-1}z_2^{-1}.
    \end{array}
\end{equation*}
Then we see that \eqref{eq:A2UCPF-1} is the special case of \eqref{eq:VOA-z-1} with  $\delta_1=\delta_2=0$ and $(q,z_1,z_2)\mapsto (q^{\frac{1}{2}},1,1)$ and so is \eqref{eq:A2UCPF-2-1} with $\delta_1=0$, $\delta_2=1$ and $(q,z_1,z_2)\mapsto (q^{\frac{1}{2}},1,q^{-\frac{1}{2}})$ followed by the change of variables $(M_1,M_2)\mapsto (M_1-M_2,-M_2-1)$.

To prove Theorem \ref{th:VOA-z} we need the following results:
\begin{lemma}
	For any integers $M,N\ge 0$,
	\begin{align}\label{eq:q-CV-spe}
	\sum_{n\ge 0}\frac{q^{n^2-(M+N)n}}{(q)_n(q)_{M-n}(q)_{N-n}} = \frac{q^{-MN}}{(q)_M(q)_N}.
	\end{align}
\end{lemma}
\begin{proof}
    See \parencite[Lemma D.3]{Bouwknegt1997SpinonDA}.
\end{proof}
\begin{lemma}
	Let $\delta\in\{0,1\}$. For any integer $N$,
	\begin{align}
	\sum_{n\ge 0}\frac{q^{n^2+n(N-\delta)}}{(q)_n (q)_{n+N}}&=\frac{1+\delta q^N}{(q)_\infty}.\label{eq:q-Gauss-spe}
	\end{align}
\end{lemma}
\begin{proof}
    See \parencite[App. A]{BCH}.
\end{proof}
\begin{proof}[Proof of Theorem \ref{th:VOA-z}]
We begin with the following change of variables
\begin{align*}
	\left\{
	\begin{aligned}
	M_1&=N_1+N_3-N_4-N_6,\\
	M_2&=N_2+N_3-N_5-N_6,\\
	M_3&=N_4+N_6,\\
	M_4&=N_5+N_6,
	\end{aligned}
	\right.
	\quad\Longleftrightarrow\quad
	\left\{
	\begin{aligned}
	N_1&=M_1+M_3-N_3,\\
	N_2&=M_2+M_4-N_3,\\
	N_4&=M_3-N_6,\\
	N_5&=M_4-N_6.
	\end{aligned}
	\right.
\end{align*}
For convienience, we also write 
\[M_0:=M_1^2+M_2^2+2M_3^2+2M_4^2+M_1M_2+2M_1M_3+2M_1M_4+2M_2M_3+2M_2M_4+4M_3M_4\]
Then the summation on the left-hand side of \eqref{eq:VOA-z-1} is rewritten over indices $N_3,N_6$ and $M_1,M_2,M_3,M_4$:
\begin{align*}
	\eqref{eq:VOA-z-1}&=\sum_{M_1,M_2\in\mathbb{Z}}z_1^{M_1} z_2^{M_2} \sum_{M_3,M_4\ge 0} q^{-2\delta_1 M_3-2\delta_2 M_4+M_0}\\
	&\quad\times\sum_{N_3\ge 0}\frac{q^{2N_3^2-2N_3(M_1+M_2+M_3+M_4)}}{(q^2)_{N_3}(q^2)_{M_1+M_3-N_3}(q^2)_{M_2+M_4-N_3}}\sum_{N_6\ge 0}\frac{q^{2N_6^2-2N_6(M_3+M_4)}}{(q^2)_{N_6}(q^2)_{M_3-N_6}(q^2)_{M_4-N_6}}\\
	\text{\tiny (by \eqref{eq:q-CV-spe})}&=\sum_{M_1,M_2\in\mathbb{Z}} z_1^{M_1} z_2^{M_2} \sum_{M_3,M_4\ge 0} q^{-2\delta_1 M_3-2\delta_2 M_4+M_0}\frac{q^{-2(M_1+M_3)(M_2+M_4)}}{(q^2)_{M_1+M_3}(q^2)_{M_2+M_4}} \frac{q^{-2M_3M_4}}{(q^2)_{M_3}(q^2)_{M_4}}\\
	\text{\tiny (by \eqref{eq:q-CV-spe})}&=\sum_{M_1,M_2\in\mathbb{Z}}z_1^{M_1} z_2^{M_2} q^{M_1^2+M_2^2-M_1M_2} \sum_{M_3\ge 0}\frac{q^{2M_3^2+2(M_1-\delta_1)M_3}}{(q^2)_{M_3}(q^2)_{M_1+M_3}}\sum_{M_4\ge 0}\frac{q^{2M_4^2+2(M_2-\delta_2)M_4}}{(q^2)_{M_4}(q^2)_{M_2+M_4}}\\
	\text{\tiny (by \eqref{eq:q-Gauss-spe})}&=\frac{1}{(q^2)_\infty^2}\sum_{M_1,M_2\in\mathbb{Z}} z_1^{M_1} z_2^{M_2}q^{M_1^2+M_2^2-M_1M_2}(1+\delta_1 q^{2M_1}) (1+\delta_2 q^{2M_2}),
\end{align*}
confirming the desired result. 
\end{proof}

Now, combining \eqref{eq:sl4untwi} and \eqref{eq:sl4twi-1} with \eqref{eq:A2UCPF-1} and \eqref{eq:A2UCPF-2} respectively, we have
\begin{align}
   b^{2000}_{2000}(\tau)+b^{2000}_{0020}(\tau)+6b^{2000}_{0101}(\tau)&=\eta(\tau)^{-2}\sum_{m,n\in\mathbb{Z}}q^{\frac{1}{2}(m^2+n^2-mn)},\label{eq:A2sl4ch-1}\\
    b^{0101}_{2000}(\tau)+3b^{0101}_{0101}(\tau)&= \eta(\tau)^{-2}q^{\frac{2}{3}}\sum_{m,n\in\mathbb{Z}}q^{\frac{1}{2}(m^2+n^2-mn+2m)}.\label{eq:A2sl4ch-2}  
\end{align}
Although we did not know the explicit expressions of $\widehat{\mathfrak{sl}(4)}_2$-string functions a priori, we know that they are very likely to be (linear combinations of) product(s) of Dedekind eta functions because of their modular transformation properties given in Theorem \ref{thm:strfunmodtrans}. Consider the following lemma:
\begin{lemma}\label{le:double-sum-rep}
	We have
	\begin{align}\label{eq:double-sum-rep}
		\sum_{m,n=-\infty}^\infty z_1^m z_2^n q^{\frac{m^2}{2}+\frac{n^2}{2}-\frac{mn}{2}}&= (-z_1^2 z_2q^{\frac{3}{2}},-z_1^{-2} z_2^{-1} q^{\frac{3}{2}},q^3;q^3)_\infty (-z_2q^{\frac{1}{2}},-z_2^{-1}q^{\frac{1}{2}},q;q)_\infty\notag\\
		&\quad+z_1q^{\frac{1}{2}}(-z_1^2 z_2q^{3},-z_1^{-2} z_2^{-1},q^3;q^3)_\infty (-z_2,-z_2^{-1}q,q;q)_\infty,
	\end{align}
 where the $q$-Pochhammer symbols are defined for $n\in\mathbb{N}\cup\{\infty\}$:
\begin{align*}
	(a;q)_n&:=\prod_{j=0}^{n-1} (1-aq^{j} ),\\
	(a_1, a_2, \ldots, a_r;q)_{n} &:= (a_1;q)_n (a_2;q)_n \cdots (a_r;q)_n.
\end{align*}
\end{lemma}
\begin{proof}
    Consider the parity of $m$ and use the Jacobi triple identity \eqref{eq:JacobitriId}. See \parencite[Lemma 20]{BCH} for details.
\end{proof}
We see that Lemma \ref{le:double-sum-rep} in fact handles \eqref{eq:A2sl4ch-1}, \eqref{eq:A2sl4ch-2}, \eqref{eq:A2Tunshf} and \eqref{eq:A2Tshf} all at once. Together with some basic algebraic manipulations of the q-Pochhammer symbols, we find that
\begin{align*}    b^{2000}_{2000}(\tau)+b^{2000}_{0020}(\tau)+6b^{2000}_{0101}(\tau)&=\frac{\eta(\tau)^3\eta(3\tau)^5}{\eta(\tau/2)^2\eta(3\tau/2)^2\eta(2\tau)^2\eta(6\tau)^2}+\frac{4q^{\frac{1}{2}}\eta(4\tau)^2\eta(6\tau)^2}{\eta(\tau)^3\eta(3\tau)},\\
   b^{0101}_{2000}(\tau)+3b^{0101}_{0101}(\tau)&=\dfrac{3\eta(3\tau/2)^3}{\eta(\tau/2)\eta(\tau)^{2}},\\
   b^{2000}_{2000}(\tau)+b^{2000}_{0020}(\tau)-2b^{2000}_{0101}(\tau)&=\dfrac{\eta(\tau/2)^2\eta(3\tau/2)^2}{\eta(\tau)^{3}\eta(3\tau)},\\
   b^{0101}_{2000}(\tau)-b^{0101}_{0101}(\tau)&=\dfrac{\eta(\tau/2)^3 \eta(3\tau)^2}{\eta(\tau)^{4} \eta(3\tau/2)}.
\end{align*}
At this stage it is clear that we can write $b^{2000}_{2000}+b^{2000}_{0020}$, $b^{2000}_{0101}$, $ b^{0101}_{2000}$ and $b^{0101}_{0101}$ as linear combinations of products of Dedekind eta functions. Then by consulting many known identities of Dedekind eta functions, which are summarised in \parencite[Lemma 13,14]{BCH}, and comparing with the coefficient expansions of these $\widehat{\mathfrak{sl}(4)}_2$-string functions up to a large power of $q$, we obtain their expressions as given in \eqref{eq:sl4strfun}.

Next, we want to obtain an expression of \eqref{eq:sl4twi-2} in terms of Dedekind eta functions. Unfortunately, in this case, we have no shortcut as above to reduce the number of summation variables, so we have to find a different method to achieve it. The core here is the following lemma for an auxiliary series constructed by S. Chern.
\begin{lemma}\label{le:T-Exp}
Define
	\begin{align*}
N_0:=&N_1^2+N_2^2+N_3^2+N_4^2+N_5^2+N_1N_2+N_1N_3+N_1N_5+N_2N_3\\+&N_2N_4+N_3N_4+N_3N_5+N_4N_5-N_1-N_2,\\
	T_M:=&\sum_{\substack{N_1,N_2,N_3,N_4,N_5\ge 0\\N_1+N_2+2N_3+N_4+N_5=M}}\frac{z_1^{N_1+N_3+N_4}z_2^{N_2+N_3+N_5}q^{N_0}}{\prod_i(q^2)_{N_i}}.
	\end{align*}
	Then for $M\ge 0$,
	\begin{align*}
	T_M=\frac{\big(z_1+z_2\big)z_2^{M-1}q^{\binom{M}{2}}(-z_1 z_2^{-1}q^{2-M};q^2)_M}{\big(1+z_1 z_2^{-1}q^M\big)(q)_M}.
	\end{align*}
 and thus
    \[\sum_{M\ge 0}(-z_1z_2q^{M};q^2)_\infty T_M=\sum_{M\ge 0}(-z_1z_2q^{2M};q^2)_\infty T_{2M}+\sum_{M\ge 0}(-z_1z_2q^{2M+1};q^2)_\infty T_{2M+1}.\]
\end{lemma}
\begin{proof}
   See \parencite[Lemma 22]{BCH}.
\end{proof}
With this lemma we are able to prove the following result:
\begin{theorem}\label{th:VOA-1}
Let $\mathbf{G}_4$ be as in \eqref{eq:G4-def-1} and $\mathbf{N}=(N_1,N_2,\dots,N_6)^T$. Then
	\begin{align}
		&\sum_{N_i\ge 0} \frac{z_1^{N_1+N_3+N_4+N_6} z_2^{N_2+N_3+N_5+N_6} q^{\mathbf{N}^{\mathsf{T}}\cdot \mathbf{G}_4\cdot \mathbf{N}-(N_1+N_2+N_6)}}{\prod_i(q^2)_{N_i}}\notag\\
		=&\frac{(-z_1z_2;q)_\infty}{(q;q^2)_\infty}\sum_{M\ge 0}\frac{(-1)^M q^{M^2} (z_1^2;q^2)_M(z_2^2;q^2)_M}{(q^2)_M (-z_1 z_2;q)_{2M}}\notag\\
		+&\frac{(-z_1z_2;q)_\infty}{(q;q^2)_\infty}\sum_{M\ge 0}\frac{(-1)^M \big(z_1+z_2\big)q^{M^2+2M} (z_1^2;q^2)_M(z_2^2;q^2)_M}{(q^2)_M (-z_1 z_2;q)_{2M+1}},\label{eq:VOA-1-3}
	\end{align}
\end{theorem}
\begin{proof}
    See \parencite[Theorem 9]{BCH}.
\end{proof}
Now using \eqref{eq:VOA-1-3} with $z_1=z_2=1$ and $q\mapsto q^{\frac{1}{2}}$, we can conclude that 
\begin{equation*}
    \sum_{N_i\ge 0} \frac{q^{\mathbf{N}^{\mathsf{T}}\cdot \mathbf{G}_4\cdot \mathbf{N}-(N_1+N_2+N_6)}}{\prod_i(q^2)_{N_i}}=b^{1100}_{1100}(\tau)+b^{1100}_{0011}(\tau)=\frac{\eta(4\tau)^5}{\eta(\tau)^3 \eta(8\tau)^2} + \frac{2q^{\frac{1}{2}}\eta(2\eta)^2 \eta(8\tau)^2}{\eta(\tau)^3 \eta (4\tau)}.
\end{equation*}
Since we know from \eqref{eq:strfunprefac} that there is a $q^{\frac{1}{2}}$-shift between $b^{1100}_{1100}$ and $b^{1100}_{0011}$, it is easy to observe the expressions for each of them.

With $b^{1100}_{1100}$ and $b^{1100}_{0011}$ known, \eqref{eq:A2orb} and \eqref{eq:A2Torb} are simply recourses of a known identity given by \parencite[Eq. (1.9.4)]{Hir2007}: 
\[\frac{\eta(4\tau)^5}{\eta(\tau)^3 \eta(8\tau)^2} + \frac{2q^{\frac{1}{2}}\eta(2\eta)^2 \eta(8\tau)^2}{\eta(\tau)^3 \eta (4\tau)}=\eta(\tau)^2\eta(\tau/2)^{-2}.\]
Considering the modular transformations of $b^{1100}_{1100}+b^{1100}_{0011}$ as given in \eqref{eq:sl4strfuntrans} and that of $\eta(\tau)^2\eta(\tau/2)^{-2}$ by \eqref{eq:etamodtrans}, it is implied that we must have
\[b^{2000}_{2000}(\tau)-b^{2000}_{0020}(\tau) = \eta(\tau)^2\eta(2\tau)^{-2},\]
which completes \eqref{eq:sl4strfun}.

By Theorem \ref{th:VOA-z} and \ref{th:VOA-1}, we are also able to obtain some UCPF-like expressions for other orbifold characters, which we summarise in Table \ref{tab:CharId} along with all identities\footnote{up to overall factors, with $\mathbf{G}_4$ as in \eqref{eq:G4-def-1}.} we have found in Chapter \ref{Chapter3} and previous sections of this chapter.
\begingroup
\renewcommand*{\arraystretch}{2}
\begin{table}[h!]
    \centering
     \begin{equation*}
        \begin{array}{c|c|c}
         A_2/\sqrt{2}  &  \widehat{\mathfrak{sl}(4)}_2/\widehat{\mathfrak{u}(1)}^{3} & \text{UCPF(-like)}\\
        \hline
      \mathrm{ch}\left[V_{\frac{1}{\sqrt{2}}A_2}\right] &  b^{2000}_{2000}+b^{2000}_{0020}+6b^{2000}_{0101}&\sum\limits_{N_i\ge 0}\dfrac{q^{\frac{1}{2}\mathbf{N}^{\mathsf{T}}\cdot\mathbf{G}_4\cdot\mathbf{N}}}{\prod_i (q)_{N_i}}\\
          \mathrm{ch}\left[V_{\gamma_1+\frac{1}{\sqrt{2}}A_2}\right]  &b^{0101}_{2000}+3b^{0101}_{0101}& \sum\limits_{N_i\ge 0}\dfrac{q^{\frac{1}{2}\mathbf{N}^{\mathsf{T}}\cdot\mathbf{G}_4\cdot\mathbf{N}-\frac{1}{2}(N_2+N_3+N_5+N_6)}}{\prod_i (q)_{N_i}}\\
       \mathrm{ch}\left[V_{\frac{1}{\sqrt{2}}A_2}^{\mathbb{Z}_2}\right]_{-+}&  b^{1100}_{1100}+b^{1100}_{0011}&\sum\limits_{N_i\ge 0} \dfrac{q^{\frac{1}{2}\mathbf{N}^{\mathsf{T}}\cdot \mathbf{G}_4\cdot \mathbf{N}-\frac{1}{2}(N_1+N_2+N_6)}}{\prod_i(q)_{N_i}}\\
         \mathrm{ch}\left[V_{\frac{1}{\sqrt{2}}A_2}^{\mathbb{Z}_2}\right]_{+-}&b^{2000}_{2000}+b^{2000}_{0020}-2b^{2000}_{0101}&\sum\limits_{N_i\ge 0}\dfrac{(-1)^{N_1+N_2+N_4+N_5}q^{\frac{1}{2}\mathbf{N}^{\mathsf{T}}\cdot\mathbf{G}_4\cdot\mathbf{N}}}{\prod_i (q)_{N_i}}\\
         \mathrm{ch}\left[V_{\gamma_i+\frac{1}{\sqrt{2}}A_2}^{\mathbb{Z}_2}\right]_{+-} & b^{0101}_{2000}-b^{0101}_{0101}&\sum\limits_{N_i\ge 0}\dfrac{(-1)^{N_2+N_3+N_5+N_6}q^{\frac{1}{2}\mathbf{N}^{\mathsf{T}}\cdot\mathbf{G}_4\cdot\mathbf{N}-\frac{1}{2}(N_2+N_3+N_5+N_6)}}{\prod_i (q)_{N_i}}\\
        \mathrm{ch}\left[V_{\frac{1}{\sqrt{2}}A_2}^{\mathbb{Z}_2}\right]_{--}&b^{1100}_{1100}-b^{1100}_{0011}&\sum\limits_{N_i\ge 0} \dfrac{(-1)^{N_1+N_2+N_6}q^{\frac{1}{2}\mathbf{N}^{\mathsf{T}}\cdot \mathbf{G}_4\cdot \mathbf{N}-\frac{1}{2}(N_1+N_2+N_6)}}{\prod_i(q)_{N_i}}
    \end{array}
\end{equation*}
    \caption{Summary of character identities}
    \label{tab:CharId}
\end{table}
\endgroup
 
\chapter{Concluding Remarks} 

\label{Chapter5}

In Chapter \ref{Chapter1}, we reviewed various constructions of CFTs and tools to study the representations of CFTs.

In Chapter \ref{Chapter2}, we noticed that one can extract some string functions of $\widehat{\mathfrak{sl}(n+1)}_{n+1}$ from characters of free fermion theories that are equivalent to the WZW models with underlying algebra $\widehat{\mathfrak{sl}(n+1)}$. It seems very feasible to prove the conjectures using some dimensional results of the space of modular forms, so besides computing more explicit expressions, the next task will be looking for proofs of the two conjectures there from either the number theory or the physics perspective. 

In Chapter \ref{Chapter3}, we introduced the notion of coupled free fermions and focused on the structure of the class of coupled free fermion CFTs from the coset construction $\hat{\mathfrak{g}}_k/\widehat{\mathfrak{u}(1)}^n$. In particular, we studied in detail the examples: $\widehat{\mathfrak{sl}(3)}_2/\widehat{\mathfrak{u}(1)}^2$ and $\widehat{\mathfrak{sl}(4)}_2/\widehat{\mathfrak{u}(1)}^3$. We analysed their representation spaces and chiral vertex operators with extensive use of generalised commutation relations. We found specific bases for them so that more information about the exclusion statistics of coupled free fermions can be revealed via the universal chiral partition functions. We explicitly decomposed their modules into Virasoro modules of minimal models. We would like to proceed to look for exclusion statistic matrices in the UCPFs for $\widehat{\mathfrak{sl}(n+1)}_2/\widehat{\mathfrak{u}(1)}^n$ with $n\ge 4$. Although the algorithm of finding UCPFs exists \parencite{Bou2000,BH2000}, besides many worked examples included in \parencite{BouwknegtUCPF}, it was mentioned there that for $\widehat{\mathfrak{sl}(n+1)}$, $n\ge 4$, the algorithm becomes extremely cumbersome, and it seems that in the coupled free fermion case, the problem is even more subtle.

In Chapter \ref{Chapter4}, we noticed another class of coupled free fermion CFTs which are from the lattice construction based on $\frac{1}{\sqrt{2}}X_n$, with $X_n$ the root lattice of any simply-laced Lie algebra. We paid special attention to the example of the $\frac{1}{\sqrt{2}}A_2$ lattice model because of its unexpected connection to the $\widehat{\mathfrak{sl}(4)}_2/\widehat{\mathfrak{u}(1)}^3$ coset model. We later found out that more precisely speaking, it is the $\mathbb{Z}_2$-orbifold of $\frac{1}{\sqrt{2}}A_2$-lattice model that is intimately connected to $\widehat{\mathfrak{sl}(4)}_2/\widehat{\mathfrak{u}(1)}^3$ by studying the modules of $\frac{1}{\sqrt{2}}A_2$-lattice model and its orbifold. At this stage, we can conclude that the orbifold of the $\frac{1}{\sqrt{2}}A_2$ lattice model and the $\widehat{\mathfrak{sl}(4)}_2/\widehat{\mathfrak{u}(1)}^3$ coset model can be projected onto sub-sectors of each other. We obtained full correspondence between the characters of the $\mathbb{Z}_2$-orbifold of the $\frac{1}{\sqrt{2}}A_2$ lattice model and the characters of $\widehat{\mathfrak{sl}(4)}_2/\widehat{\mathfrak{u}(1)}^3$, which allowed us to prove the UCPFs of $\widehat{\mathfrak{sl}(4)}_2/\widehat{\mathfrak{u}(1)}^3$ and compute explicit expressions of string functions of $\widehat{\mathfrak{sl}(4)}_2$. Given the success of proving these character identities, we are counting on producing such proofs for more and more generalised Rogers-Ramanujan identities for Lie algebras in general \parencite{Gepner2014yva} as well as some twisted affine algebras \parencite{Gepner:2022yme,Genish:2017sor}. 

It should be mentioned that we have not yet discussed the irreducibility of the modules of the $\frac{1}{\sqrt{2}}A_2$ lattice model that we have used. To determine the irreducibility, we may have to consult mathematical formulations of non-local CFTs involving fractional powers in OPEs, such as the theory of generalised VOA \parencite{dong2012generalized}. Another crucial step is to figure out the underlying reason for the connection between the $\frac{1}{\sqrt{2}}A_2$ lattice model and the $\widehat{\mathfrak{sl}(4)}_2/\widehat{\mathfrak{u}(1)}^3$ coset model from the view of Lie algebras. We also wonder if this connection implies any correspondence for physical systems. Moreover, some bizarre identities between fermionic characters of VOAs and 3-manifolds characters have been proposed lately \parencite{CHE2022} and we would like to explore whether coupled free fermions could be related to 3-manifolds in a similar spirit.

Throughout the thesis, we have seen that in the context of coupled free fermion, a number of character identities have been established. Additionally, an unexpected connection integrating the coset construction, lattice construction, and orbifold construction has been uncovered. These findings highlight the importance of investigating coupled free fermions and signal that they deserve more attention from researchers. In light of our discoveries, we emphasize the need to classify all coupled free fermion CFTs and further explore their potential applications in other areas. With a better understanding of coupled free fermions, there is potential to unlock new insights into other fields such as condensed matter physics and quantum computing. We would like to invite researchers to continue investigating this fascinating object and push the boundaries of our knowledge in this field.


\appendix 


\chapter{Appendix A: Affine Lie algebras and string functions} 

\label{AppendixA} 

Let $\mathfrak{g}$ be a complex finite-dimensional semi-simple Lie algebra with Cartan matrix $A$ and Killing form $\kappa(\;,\;):\mathfrak{g}\times \mathfrak{g}\mapsto \mathbb{C}$ defined by $$\kappa(X,Y)=\Tr(\mathrm{ad}(X)\mathrm{ad}(Y)),$$ where $\mathrm{ad}:\mathfrak{g}\mapsto\mathfrak{gl}(\mathfrak{g})$ is the adjoint representation of $\mathfrak{g}$. Let $\mathfrak{h}$ be the Cartan subalgebra of $\mathfrak{g}$ and $\Delta$ be the set of roots in the dual of $\mathfrak{h}$, denoted by $\mathfrak{h}^*$. Fix a set of positive roots $\Delta_+$. Let $\theta$ be the highest root and $\rho$ be the Weyl vector defined by $$\rho:=\frac{1}{2}\displaystyle\sum_{\alpha\in\Delta_+}\alpha.$$\index{affine Lie algebra! weyl vector}Let $\Delta_0$ be the set of simple roots $\{\alpha_i\mid 1\le i\le n:=\rank(\mathfrak{g})\}$ and $\langle\;,\;\rangle: \mathfrak{h}^*\times\mathfrak{h}^*\mapsto \mathbb{C}$ be the bilinear form on $\mathfrak{h}^*$ such that $A_{ij}=\dfrac{2\langle\alpha_i,\alpha_j\rangle}{\langle\alpha_i,\alpha_i\rangle}.$ 
Let $$(\;,\;):=\dfrac{1}{2h^{\vee}}\kappa(\;,\;)$$ be the normalised Killing form, where $h^{\vee}$ is the dual Coxeter number defined by $$h^{\vee}:=1+\langle \theta,\rho\rangle.$$\index{affine Lie algebra! dual Coxeter number}Identify $\mathfrak{h}$ with $\mathfrak{h}^*$ via $(\;,\;)$. Note that the normalisation is fixed so that $|\theta|^2=(\theta,\theta)=2$.

Set $\hat{\mathfrak{g}}:=(\mathbb{C}[t,t^{-1}]\otimes\mathfrak{g})\oplus\mathbb{C}\hat{k}\oplus\mathbb{C}L_0$ and define the bracket on $\hat{\mathfrak{g}}$ as follows:
\begin{align}
    &[t^m\otimes X,t^n\otimes Y]=t^{m+n}\otimes [X,Y]+m\delta_{m+n,0}(X,Y)\hat{k},\label{eq:KMcomm}\\
    &[L_0,t^n\otimes X]=nt^n\otimes X,\notag\\
    &[\hat{k},X]=0,\notag
\end{align}
for all $X,Y\in \hat{\mathfrak{g}}$.
Then $\hat{\mathfrak{g}}$ is called the affine Lie algebra\index{affine Lie algebra} associated to $\mathfrak{g}$. 

Let $\hat{\mathfrak{h}}=\mathfrak{h}\oplus\mathbb{C}\hat{k}\oplus\mathbb{C}L_0$ and $\hat{\mathfrak{h}}^*$ be its dual. Define the elements $\delta$ and $\Lambda_0$ of $\hat{\mathfrak{h}}^*$ by
\begin{align*}
  &\delta|_{\mathfrak{h}\oplus\mathbb{C}\hat{k}}=0,&\delta(L_0)=1,\\&\Lambda_0|_{\mathfrak{h}\oplus\mathbb{C}L_0}=0,&\Lambda_0(\hat{k})=1.
    \end{align*}
Then $\{\alpha_1,\cdots,\alpha_l,\
\delta,\Lambda_0\}$ forms a basis of $\hat{\mathfrak{h}}^*$ and the bilinear form $(\;,\;)$ can be extended to $\hat{\mathfrak{h}}^*$ by setting
\begin{align*}
    &(\alpha_i,\delta)=(\alpha_i,\Lambda_0)=0,\;(1\le i\le l),\\
    &(\delta,\delta)=(\Lambda_0,\Lambda_0)=0,\\
    &(\delta,\Lambda_0)=1.
    \end{align*}
In addition, we also define $\alpha_0:=\delta-\theta$. The set of simple roots\index{affine Lie algebra!simple root} of $\hat{\mathfrak{g}}$ is then $$\hat{\Delta}_0=\{\alpha_i|0\le i\le n\}.$$ The set of roots\index{affine Lie algebra!root} of $\hat{\mathfrak{g}}$ is 
\[\hat{\Delta}=\{\alpha+n\delta| n\in\mathbb{Z},\alpha\in\Delta\}\cup\{n\delta| n\in\mathbb{Z},n\ne 0\}\]
and the set of positive roots is
\[\hat{\Delta}_+=\Delta_+\cup\{\alpha+n\delta|n>0,\alpha\in\Delta\}.\]
The triangular decomposition of $\mathfrak{g}$ is given by $\mathfrak{g}=\hat{\mathfrak{n}}_-\oplus\hat{\mathfrak{h}}\oplus\hat{\mathfrak{n}}_+$
where 
\[\hat{\mathfrak{n}}_{\pm}=\bigoplus_{\alpha\in\hat{\Delta}_{\pm}}\hat{\mathfrak{g}}_{\alpha}\]
where $\hat{\mathfrak{g}}_{\alpha}=\{x\in\hat{\mathfrak{g}}|[h,x]=\alpha(h)x,\;\forall h\in\hat{\mathfrak{h}}^*\}$ is the root space of $\alpha$ and $\hat{\Delta}_{-}=-\hat{\Delta}_{+}$.

The set of fundamental weights $\{\omega_i\in\hat{\mathfrak{h}}|0\le i\le n\}$ is defined to be the basis of $\hat{\mathfrak{h}}$ dual to the set of simple roots $\hat{\Delta}_{0}=\{\alpha_i\in\hat{\mathfrak{h}}^*|0\le i\le n\}$. 

Set $\hat{P}=\{\lambda\in\hat{\mathfrak{h}}^*|\lambda(L_0)=0,\lambda(\omega_i)\in\mathbb{Z},0\le i\le n\}$ to be affine weights, $\hat{P}_+=\{\lambda\in\hat{P}|\lambda(\omega_i)\ge0,0\le i\le n\}$ to be dominant affine weights and $\hat{P}_+^k=\{\lambda\in\hat{P}_+|\lambda(\hat{k})=k\}$ to be dominant affine weights\index{affine Lie algebra! dominant weight} at level $k$. Affine weights are often given in terms of Dynkin labels\index{affine Lie algebra! Dynkin label} under the form $\lambda=[\lambda(\omega_0),\lambda(\omega_1),\dots,\lambda(\omega_n)]$.

For any fixed $\Lambda\in\hat{P}_+$, there exists a unique irreducible highest weight $\hat{\mathfrak{g}}$-module\index{affine Lie algebra! highest weight module}  $\mathcal{L}(\Lambda)$ for which there exists a non-zero vector $v\in\mathcal{L}(\Lambda)$ such that
\[\begin{tabular}{ccc}
    $\hat{\mathfrak{n}}_+\cdot v=0$ &and& $h\cdot v=\Lambda(h)v,\;\forall h\in\hat{\mathfrak{h}}^*$.
\end{tabular}\]
For $\lambda\in\hat{P}$, set $\mathcal{L}(\Lambda)_{\lambda}:=\{v\in\mathcal{L}(\Lambda)|h(v)=\lambda(h)v,\;\forall h\in\hat{\mathfrak{h}}\}$. If $\dim\mathcal{L}(\Lambda)_{\lambda}\ne0$, then $\lambda$ is called a weight of $\mathcal{L}(\Lambda)$. A weight $\lambda$ of $\mathcal{L}(\Lambda)$ is maximal if $\lambda+\delta$ is not a weight of $L$. The character of $\mathcal{L}(\Lambda)$ is defined as
\begin{equation}\label{eq:affch}
    \text{ch}\left[\mathcal{L}(\Lambda)\right]=\sum_{\lambda}\text{mult}_{\Lambda}(\lambda)e^{\lambda},
\end{equation}
where $\text{mult}_{\Lambda}(\lambda)=\dim\mathcal{L}(\Lambda)_{\lambda}$ is the multiplicity of $\lambda$ in $\mathcal{L}(\Lambda)$.
The multiplicities can be tracked by the Freudenthal recursion formula:
\[[|\lambda+\hat{\rho}|^2-|\lambda'+\hat{\rho}|^2]\text{mult}_{\lambda}(\lambda')=2\sum_{\alpha\in\hat{\Delta}_+}\sum_{p=1}^{\infty}(\lambda'+p\alpha,\alpha)\text{mult}_{\lambda}(\lambda'+p\alpha).\]

Given a dominant weight $\Lambda\in\hat{P}_+^k$ and a maximal weight $\lambda$ of $\mathcal{L}(\Lambda)$, the so-called string functions $c^{\Lambda}_{\lambda}$ are defined in \parencite{KACPETERSON} as 
\begin{equation*}
    c^{\Lambda}_{\lambda}(\tau)=q^{s_{\Lambda}(\lambda)}\sum_{n\ge0}\text{mult}_{\Lambda}(\lambda-n\delta)q^n,
\end{equation*}
where $q=e^{2\pi i \tau}$ and
\begin{equation*}
s_{\Lambda}(\lambda)=s_{\Lambda}-\dfrac{|\lambda|^2}{2k},
\end{equation*}
where
\[s_{\Lambda}=\dfrac{|\Lambda+\hat{\rho}|^2}{2(k+h^{\vee})}-\dfrac{|\hat{\rho}|^2}{2h^{\vee}}.\]
 Furthermore, for any 
$\mu\in\hat{\mathfrak{h}}^*$ such that $\mu-\gamma\in\mathbb{C}\delta$, set $c^{\Lambda}_{\lambda}=c^{\Lambda}_{\mu}$. In total, string functions have the property that 
\begin{equation}\label{eq:strfunprop}
    c^{\Lambda+a\delta}_{w(\lambda)+k\gamma+b\delta}=c^{\Lambda}_{\lambda},\;\text{for}\; w\in W,\gamma\in M,a,b\in\mathbb{C},
\end{equation}
where $M$ is the dual root lattice of $\hat{\mathfrak{g}}$ and $W$ is the weyl group of $\mathfrak{g}$.

It is known that the character of $\mathcal{L}(\Lambda)$ can be expressed in terms of string functions \eqref{eq:strfun} as follows:
\begin{equation}\label{eq:affch-str}
\text{ch}\left[\mathcal{L}(\Lambda)\right]=q^{-s_{\Lambda}}\sum_{\lambda\in \hat{P}\;\text{mod}\;(kM+\mathbb{C}\delta)}c^{\Lambda}_{\lambda}\Theta_{\lambda|_{\mathfrak{h}^*},k},    
\end{equation}
where $\Theta_{\mu,k}$ are theta functions:
\[\Theta_{\mu,k}\:=\sum_{\gamma\in M+\mu/k}\exp\left(k(\Lambda_0+\gamma-\frac{1}{2}|\gamma|^2\delta)\right),\;\text{for}\;\mu\in\mathfrak{h}^*.\]
The character can also be computed by the Weyl-Kac character formula
\begin{equation}\label{eq:weylkac}
    \text{ch}\left[\mathcal{L}(\Lambda)\right]=\dfrac{\sum_{w\in\hat{W}}\det(w)e^{w(\Lambda+\rho)}}{\sum_{w\in\hat{W}}\det(w)e^{w\rho}}=\dfrac{\sum_{w\in\hat{W}}\det(w)e^{w(\Lambda+\rho)}}{e^{\rho}\prod_{\alpha\in\hat{\Delta}_+}(1-e^{-\alpha})^{\mathrm{mult}(\alpha)}},
\end{equation}
where $\hat{W}$ is the Weyl group of $\hat{\mathfrak{g}}$.


\printbibliography[heading=bibintoc]


\printindex


\end{document}